\documentclass[12pt,reqno]{amsart}
\usepackage{amsaddr} 

\usepackage[utf8]{inputenc}
\usepackage[T1]{fontenc}
\usepackage[a4paper,margin=1in]{geometry}
\IfFileExists{lmodern.sty}{\usepackage{lmodern}}{}
\usepackage[protrusion=true,expansion=true]{microtype}

\usepackage{amssymb,mathtools,amsthm}
\usepackage{bm}
\usepackage{bbm} 
\numberwithin{equation}{section}

\usepackage{graphicx}
\usepackage{booktabs}
\usepackage{enumitem}
\setlist[enumerate]{label=(\roman*),itemsep=0.25em,topsep=0.25em}

\usepackage[numbers,sort&compress]{natbib}
\usepackage[dvipsnames]{xcolor}
\usepackage[colorlinks=true,allcolors=MidnightBlue]{hyperref}
\usepackage[nameinlink,capitalize,noabbrev]{cleveref}
\hypersetup{
	pdftitle={Strategic Play and Home Advantage: Coaches' Tactical Impact in Serie A},
	pdfauthor={Francesco Angelini; Massimiliano Castellani; Gery A. D\'\i az Rubio; Simone Giannerini; Greta Goracci},
	pdfsubject={Statistics / Mathematics},
	pdfkeywords={sports statistics, home advantage, coaching strategies, GLM, logit, ordered logit, winning probability}
}

\usepackage[english]{babel}
\usepackage{csquotes}

\usepackage{amsmath, amssymb, amsfonts, amsthm, mathtools}
\usepackage{graphicx, float, subcaption, adjustbox, tabularx, longtable, multirow, multicol}
\usepackage{enumitem, enumerate, booktabs, url, listings, siunitx}
\usepackage{pdflscape, lscape, textcomp, xcolor, framed, alltt, algorithm, algorithmicx, algpseudocode}
\usepackage{fancyhdr, lipsum, etoolbox}
\usepackage{bm, dsfont, array, psfrag, epstopdf, pifont, bbding, wasysym, fontawesome5}
\usepackage{mathrsfs, manyfoot}

\usepackage{placeins}     
\usepackage{needspace}    
\usepackage{caption}      
\usepackage[toc]{appendix}   
\usepackage{hyperref}        

\newtheorem{model}{Model}

\DeclareMathOperator{\E}{E}
\DeclareMathOperator{\Pbb}{\mathbb{P}}

\DeclareMathOperator{\Var}{Var}

\newcommand{\eps}{\varepsilon}

\newcommand{\bk}{\mathbf{k}}

\newcommand{\bR}{\mathbf{R}}

\newcommand{\bZ}{\mathbf{Z}}
\newcommand{\bX}{\mathbf{X}}
\newcommand{\bs}{\mathbf{s}}
\newcommand{\bz}{\mathbf{z}}
\newcommand{\bc}{\mathbf{c}}

\newcommand*{\bw}{\mathbf{w}}
\newcommand*{\bW}{\mathbf{W}}

\newcommand{\bpsi}{\boldsymbol{\psi}}

\newcommand{\btheta}{\boldsymbol{\theta}}
\newcommand{\bEta}{\boldsymbol{\eta}}
\newcommand{\bBeta}{\boldsymbol{\beta}}

\newcommand{\balpha}{\boldsymbol{\alpha}}

\newcommand*{\bphi}{\boldsymbol{\phi}}

\newcommand*{\bgamma}{\boldsymbol{\gamma}}

\newcommand{\cC}{\mathcal{C}}
\newcommand{\bcI}{\boldsymbol{\mathcal{I}}}
\newcommand{\bcT}{\boldsymbol{\mathcal{T}}}
\newcommand{\bcN}{\boldsymbol{\mathcal{N}}}
\newcommand{\bS}{\mathbf{S}}
\newcommand{\bY}{\mathbf{Y}}
\newcommand{\cT}{\mathcal{T}}

\theoremstyle{plain}

\theoremstyle{definition}

\theoremstyle{remark}

\title[Strategic Play and Home Advantage]{Strategic Play and Home Advantage:\\Coaches' Tactical Impact in Serie A}

\author[F. Angelini]{Francesco Angelini\vspace*{-0.5cm}}
\address{Department of Statistical Sciences ``Paolo Fortunati'', University of Bologna, Via delle Belle Arti 41, 40126 Bologna, Italy\vspace*{-0.4cm}}
\email{francesco.angelini7@unibo.it}

\author[M. Castellani]{Massimiliano Castellani\vspace*{-0.5cm}}
\address{Department of Statistical Sciences ``Paolo Fortunati'', University of Bologna, Via delle Belle Arti 41, 40126 Bologna, Italy\vspace*{-0.4cm}}
\email{m.castellani@unibo.it}

\author[G. A. D\'\i az Rubio]{Gery A. D\'\i az Rubio\vspace*{-0.5cm}}
\address{Department of Statistical Sciences ``Paolo Fortunati'', University of Bologna, Via delle Belle Arti 41, 40126 Bologna, Italy\vspace*{-0.4cm}}
\email{geryandre.diazrubio2@unibo.it}
\thanks{Corresponding author: G. A. D\'\i az Rubio.}

\author[S. Giannerini]{Simone Giannerini\vspace*{-0.5cm}}
\address{Department of Economics and Statistics, University of Udine, via Tomadini 30/a, 33100 Udine, Italy\vspace*{-0.4cm}}
\email{simone.giannerini@uniud.it}

\author[G. Goracci]{Greta Goracci\vspace*{-0.5cm}}
\address{Faculty of Economics and Management, University of Bozen--Bolzano, Piazza Universit\`a 1, 39100 Bozen--Bolzano, Italy}
\email{greta.goracci@unibz.it}

\thanks{Funding: No funding to declare.}

\date{August, 2025} 

\subjclass[2020]{Primary 62-F07; Secondary 62-J12; Other: 62-F40, 62-P99}
\keywords{sports statistics, home advantage, coaching strategies, GLM, logit, ordered logit, winning probability}

\makeatletter\let\ps@firstpage\ps@plain\makeatother

\begin{document}
	
	
	\begin{abstract}
		We analyze how coaching strategies affect goal difference and home win probabilities using hand-coded Serie A match commentary (2011/12--2013/14). Our dataset captures in-game dynamics, referee actions, and team behavior. Applying generalized linear, logit, and proportional-odds models with robust and bootstrap standard errors, we uncover stable effects across model averaging. Aggressive opening tactics consistently boost performance, while technical actions like crosses and goal-kicks show distinct patterns. Home advantage remains significant after full control. Our approach reveals the economic logic of real-time coaching, offering a novel, data-driven method to study decision-making under uncertainty in competitive environments.
	\end{abstract}
	
	\maketitle
			
	\thanks{JEL Classification: C25; L83; Z29.}
	
	\section{Introduction}\label{sec:1_INTRO}
	The influence of coaches on sports outcome has gained considerable attention in the economics and sports statistics literature, both from a theoretical and an empirical point of view. The impact of coaches' strategies and tactics on team performance may vary considerably and is highlighted in numerous studies, often borrowing from management theory and leveraging technological advancements in monitoring team sports \citep{FRY21}. Indeed, coaches, like managers, operate in varying contexts that affect their teams' success, where their strategic decisions play a crucial role in performance \citep{BRU03}. 
	\par
	Another phenomenon, home advantage (HA), is widely recognized as a critical factor influencing match outcomes.\footnote{The `soccernomics' literature \citep{kuper2009} suggests that playing in a home stadium has a positive effect on team performance, significantly increasing the probability of winning \citep{JAM10}.} Several studies suggest that home team fans can influence match outcomes by intimidating referees, disrupting communication among opposing players, boosting the motivation of home players, inducing errors in away players, and reducing their effort, ultimately favoring the home team \citep[e.g.,][]{COU92,CAR05}. HA stems from psychological and contextual factors, including crowd influence, familiarity with the environment, and reduced travel fatigue. These elements create conditions that amplify home teams' capacity to exert dominance or adapt effectively.
	\par
	Both home advantage (HA) and the role of coaches significantly influence team performance, yet their interplay remains underexplored. From a theoretical perspective, this synthesis aligns with resource allocation theory \citep{Bower2018} and contingency theory \citep{northouse2025}.\footnote{Game theory, often applied in strategic interactions \citep{madden2011}, highlights how decision-makers (e.g., coaches) anticipate and respond to their opponents' strategies. When combined with HA, game theory suggests that coaches can exploit the predictable advantages of being at home (e.g., referee bias, crowd influence) or adapt their tactics away to neutralize these factors.} Resource allocation theory suggests that contextual advantages, such as HA, can enhance performance if effectively utilized. Coaches act as allocators of these resources, adapting strategies to capitalize on HA's psychological and environmental benefits. In turn, contingency theory models emphasize the importance of contextual awareness in decision-making, as leaders (coaches) adjust their styles based on external conditions. For instance, coaches might adopt aggressive tactics at home, leveraging crowd support, whereas, in away matches, focus on defensive strategies to counteract HA for the opposing team. 
	\par
	While it is evident that HA and coaches' decisions are two essential ingredients of performance determinants in soccer, this crucial interplay has received little attention in existing research. This paper aims to bridge such gap, offering a significant leap forward in understanding the complex dynamics of soccer matches. For this purpose, we build a unique hand-collected dataset comprising 157,985 event observations from the Italian Serie A league during the 2011/12 to 2013/14 seasons, sourced from original, non-lemmatized soccer commentary. To the best of our knowledge, this is the first attempt to create such a data set, which holds a great potential in a wide range of research problems linked to soccer and team sports analysis. Furthermore, we study different aspects of the phenomenon, namely, home-away goal differences, home win probability, and home resulting points. These  correspond to three modeling approaches within the generalized linear models framework, combined with non-standard statistical approaches, some of which leverage upon recent results in the mathematical statistics literature, such as model averaging and bootstrap standard errors to achieve sound and robust results. 
	\par
	Our findings suggest that aggressive initial home strategies in the Serie A league positively influence home-away goal differences, 
	home win probability, and home resulting points, with a dynamic effect. Moreover, coaching strategies play a pivotal role in shaping team success, even when focusing only on teams playing at home. We confirm that home win probability, likelihood of home team winning and home resulting points are significantly affected by referee's actions, as well as highly-aggressive actions and counter-attack strategies. We also find that offensive actions that do not lead to shots, such as nuanced crosses and corners, may foster counter-attack, reducing the overall performance of the home team. Our advancements in inference and prediction enhance the methodological toolkit for sports researchers and shed light on the dynamics of sports outcomes and the interplay between coaching decisions, HA, and match performance, contributing to the field of sports economics. For sports management, the findings offer guidance for decision-making processes related to team selection, player recruitment, and match preparation. Teams can leverage these insights to optimize coaching strategies and in-game tactics, ultimately improving performance.
	\par
	The sports economics literature has explored two key aspects influencing team performance in soccer: the impact of home advantage (HA) and the role of coaches' strategies. The literature on HA highlights its effects on referee decisions, crowd support, and team strategies. Research has found evidence of referee home bias \citep[e.g.,][]{SCO08,HOL21} and crowd influence on HA \citep[e.g.,][]{PON18,FIS21}. Few studies have analyzed HA's impact on team tactics, such as \citet{DEW13}, who found that home teams often adopt a more offensive style. Recent pandemic-related studies \citep[e.g.,][]{fischer2022,BEN23} leveraged ghost games to explore the causal impact of HA, yielding mixed results.\footnote{\citet{STEF07}, by a casual review of the most sports literature on HA, revealed that a home team wins more than it loses and scores more points than the visiting opposition, and finds that soccer and rugby have a stronger HA with respect to other sports. \citet{dohmen2016} surveyed the empirical literature on the behavior of referees in professional football and other sports. \citet{goncalves2021} presented a literature review on the effects of HA in team sports on coaches', teams', and players' decisions.} Studies on coaching often examine tactical decisions and their effects on performance. For example, \citet{MUE18} assessed managerial ability in the Bundesliga but did not account for match-specific tactics. \citet{MES20} analyzed formation decisions influenced by past strategies, while \citet{goes2021} used tracking data to evaluate player movements and team strategies. Recent work, such as \citet{bauer2023}, employed machine learning to classify dynamic formations but focused only on distinct phases of play, overlooking continuous in-game adjustments. Other studies, like \citet{lopez2022} and \citet{lorenzo2022}, explored playing styles and substitutions, emphasizing their roles in shaping performance.\footnote{The tactical formation is widely recognized as an important factor in soccer. However, in-game changes in formation and the distinction between offensive and defensive setups have rarely been considered in previous studies on match performance \citep{forcher2023}. Interest in tactical analysis in soccer has grown in recent years, driven by the increasing availability of player tracking data. For instance, the compact organization of defending teams has been a frequent subject of study \citep{forcher2024}} Table \ref{tab:selected_lit} contains a synthesis on the main empirical studies on these two topics in comparison with our paper.
	\par	
	The paper is structured as follows. Section \ref{sec:3_MODEL} describes the statistical modeling framework. Section \ref{sec:Results} presents the results of the model fit, model diagnostics and checking. Section \ref{sec:concl} concludes the paper, including a discussion of the impact of the study and possible future investigations. Additional analyses and extended results are reported in the Supplementary Material (hereby refereed to as SM).
	
	\section{Data and variables description}\label{sec:3_MODEL} \label{sec:4_VARS}
	
	We collected non-lemmatized soccer textual commentary data from the sports section of the Virgilio website, dedicated to the Italian Serie A league.\footnote{\url{https://sport.virgilio.it/calcio/serie-a}. 15 matches had no comments, including the Cagliari-Roma match in the 2012/13 season, which was awarded to Roma by default, with a score of 0-3. Commentary information on the other 14 matches was collected from Serie A (\url{https://www.legaseriea.it/it}) and ESPN (\url{https://www.espn.com/soccer/league}).} For the seasons between 2011/12 and 2013/14, we obtained 157,985 events/observations. 
	Additional sources were used to integrate missing information that influenced the results, such as penalizations following the \emph{``Calcioscommesse"} scandal in 2011 and 2012 or to resolve ambiguous information like homonymy of players.\footnote{The dataset were cross-checked for possible errors in the commentary and corrected. The complete process of data management is reported in the SM, Section \ref{sec:datamanage}}
	All commentaries were transformed into 0-1 dummy variables.
	For each event, we matched player's name with their respective team. Consequently, each event is linked to a team, either home or away. Using information about the official role within the team for each season, we also infer the scheme chosen by the coach based on the players present at any given minute. 
	\par
	In the following we sketch the construction of the scheme variable. Building on the role of each player, we assume that a team’s formation is defined by the number of defenders, midfielders, and forwards among its outfield players. In doing so, we do not account for role changes that players may undergo within the same match or across the same season. Additionally, we assume that teams employing identical formations adhere to the same tactical scheme. To convert this categorical variable into a  numeric one, we employ an offensive index that evaluates the goal-scoring potential of 10 outfield players, excluding goalkeepers. Specifically, the index weights players based on their roles, linked to the average goals scored historically by players in those positions. The weighted sum of these points is assigned according to player positions: defenders receive a weight of 1, midfielders 2, and forwards 3. The resulting weighted sum ranges from 10 to 30—representing the theoretical extremes where all players are either defenders or forwards. One advantage of this transformation is that it enables an intuitive interpretation of the index. An increase of one point reflects an elevation in a team’s offensive setup and viceversa. Furthermore, this metric accounts for a team’s potential to create scoring opportunities, providing a quantifiable measure of offensive capability.\footnote{Additional player indices can be explored to construct an offensiveness index for coaches' strategies, as exemplified in \citet{CEF23} and \citet{CEF24}.}
	\par
	To create a balanced dataset, we aggregate multiple events occurring within a given minute, ensuring that every minute of a match contains at least one event. Conversely, some minutes may feature multiple events. This process involved distinguishing between variables that can be aggregated (e.g., the dummy variable for fouls, which can be summed if multiple fouls are called by the referee within the same minute) and those that cannot (e.g., a variable representing a team's formation, which is assumed to remain constant unless explicitly changed by the coach). Additionally, matches often have variable numbers of events after the 45th minute of the first half and after the 90th minute of the second half, due to differing ending times. To address this, we allocated all the extra-time events either to the 45th or the 90th minute. However, since the referee determines the amount of extra time, we include a variable for ending time assigned by the referee to control for potential implicit home/away bias in their decisions. Finally, to ensure a balanced panel structure, we added the missing minutes for each match, resulting in a dataset comprising 1,140 matches, each standardized to 90 minutes. For further details on the dataset, see the SM, Section \ref{sec:A_empirical}.
	\par
	To account for the variable number of events (intensity) during a match, we add a weighted version of the explanatory variables to the models. The weight is determined as follows: 
	\begin{equation*}
		\omega_{it} \equiv  \dfrac{n_{it}}{n_i}+1,
	\end{equation*}
	where $n_{it}$ denotes the number of events at minute $t$ of match $i = 1,\dots,N$, and $n_i$ is the total number of events of match $i$. 
	For time points $t\in \left\{ 45, 90\right\}$, the weight incorporates the number of events occurring during  extra time. The product of the number of events and the weights is normalized to $91$. From now on, for a generic variable $X$, its corresponding weighted version is denoted as $\tilde{X}$.
	\par
	Our sample contains information on $N=1,140$ matches. We denote by $m_{ij}$, the match $i$, $i=1,\dots, N$, of the $j$-th championship season, $j \in \{ 2011, 2012, 2013\}$. The dataset contains the following variables (all the vectors are row-vectors, unless otherwise indicated):
	\begin{itemize}[noitemsep, nosep]
		\item  Response variables (see Section~\ref{sec:dependent})
		\begin{itemize}
			\item $y_{i}^{(1)}$: Goal difference between home and away teams (integer valued); 
			\item $y_{i}^{(2)}$: Home team win. Binary variable equal to 1 if the home team wins the match, 0 otherwise;
			\item $y_{i}^{(3)}$: Points gained in the match by the home team. Takes the values 0 (loss), 1 (tie), 3 (win). 
		\end{itemize}
		\item Predictors  (see Sections~\ref{subsubsec:coach_schemes}--\ref{subsubsec:ref})
		\begin{itemize}[noitemsep, nosep]
			\item $\textbf{s}_i = (s_{i,1}, s_{i,2}, s_{i,3}, s_{i,1}^{*}, s_{i,2}^{*}, s_{i,3}^{*})$: coach scheme at the beginning(no asterisk)/ end(asterisk) of the match;
			\item $\bw_i = (w_{i,1}, \dots, w_{i,8})$: team actions; 
			\item $\bz_i = (z_{i,1}, \dots, z_{i,5})$: referees' actions. 
		\end{itemize}
		\item Control variables $\bc_i = (c_{i,1}, c_{i,2}, c_{i,3}, c_{i,4})$ (see Section~\ref{subsubsec:control})
		\begin{itemize}[noitemsep, nosep]
			\item $c_{i,1}$: home stadium filling index;
			\item $c_{i,2}$: extra time at the end of match;
			\item $c_{i,3}$: difference between the home and away teams ranking points before the match.
			\item $c_{i,4}$: normalized version of $c_{i,3}$ to lie in $[-1,1]$.
		\end{itemize}
		\item Fixed effects  $\mathbf{k}_i = (\mathbf{k}_{i}^{s}, \mathbf{k}_i^{d},\mathbf{k}_{i}^{e})$ (see Section~\ref{subsubsec:fixed})
		\begin{itemize}[noitemsep, nosep]
			\item $\mathbf{k}_i^{s} = (k_{i}^{(11)},k_{i}^{(12)},k_{i}^{(13)})$: season effect;
			\item $\mathbf{k}_i^{d} = (k_{i,1}^{d}, k_{i,2}^{d})$: league day effect;
			\item $\mathbf{k}_i^{e} = (k_{i,1}^{e},k_{i,2}^{e},k_{i,3}^{e})$: extreme event effect.
			\item $\mathbf{k}_i^{T}$ : team effect.
		\end{itemize}
	\end{itemize}
	In general, all the variables are expressed as differences between the home and the away team. When appropriate, we have included in the models measurements only for the home ($H$) or the away ($A$) team, rather than its difference, and denoted these with the corresponding superscript. For instance, $w_1$ is the total number of crosses and it results $w_1 = w_1^H - w_1^A$. Also, weighted versions of the variables are denoted with an added tilde. Table~\ref{tab:codebook} contains the concise variable codebook. For descriptive statistics, see Tables~\ref{tab:dependent}-\ref{tab:additional2}. For the visual representation of data please consult the SM, Section~\ref{sec:destat}. Further discussion on the effect of HA can be inferred from SM, Figures~\ref{fig:SM_F_02}--\ref{fig:SM_F_12}.
	
	\subsection{Response/dependent variables}\label{sec:dependent}
	
	We select three variables to capture the key aspects of match results and identify three different modeling strategies to account appropriately for their nature. 
	\par
	The first dependent variable $y_{i}^{(1)}$ is the goal difference between home and away teams \citep{heuer2009} for match $i$. In our sample, this variable ranges from -7 to 6, with mean and median close to 0 indicating symmetry, as also confirmed by skewness tests. The standard deviation (1.67) and coefficient of variation (4.35) suggest low variability relative to the mean. Although the distribution exhibits a slight leptokurtosis, it is close to a normal distribution. This suggests that the goal difference variable behaves approximately as a continuous variable with a symmetric distribution. Hence, we use a standard Ordinary Least Squares (OLS) regression for this first analysis.
	\par	
	The second dependent variable $y_{i}^{(2)}$ is dichotomous and takes the value $1$ when the home team wins match $i$, and $0$ otherwise (either tie or loss).\footnote{In literature such dependent variable has been already employed to study HA \citep{JAM10}, while, over the last $30$ years other variables have been considered \citep[][p. 448]{PIC17}.} 
	In the seasons spanning from 2011/12 to 2013/14, out of a total of 1,140 matches, the home team was victorious in 531 instances, accounting for 46.58\% of the matches. Out of 609 non-winning matches (53.42\%), the home team either faced a defeat (311 matches, 27.28\%) or settled for a draw (298 matches, 26.14\%). \footnote{For insights on the subtle HA in match outcomes, see the SM, Section~\ref{subsec:dep}, which discusses the distribution of wins, losses, and draws, indicating a slight home team edge.}
	This suggests a slight advantage for home teams and highlights the potential impact of HA in determining match outcomes. It seems that matches are competitively balanced with a discernible edge for the home team. Nevertheless, while HA plays a role, it is not a decisive factor in every match and our generalized linear models approach allows to uncover how this depends on the coaches' strategies, team playing home, and in-game factors.
	\par
	The third dependent variable $y_{i}^{(3)}$ represents the points gained in match $i$: it takes the value 3 if the home team wins the match, 1 if it ties, and 0 if it loses it. This response variable is ordinal and has consistent cardinal differences between the values. This makes it suitable for an ordered logit model, which assumes that the dependent variable reflects an underlying latent continuous variable.\footnote{Note that the ordered logit model assumes that the relationship between the independent variables and the dependent variable is the same across the different thresholds (e.g., between loss and tie, and between tie and win), i.e. \emph{parallel- or proportional-odds assumption.}}
	\par
	Overall, the three specifications assess different aspects of the match dynamics. On the one hand, logit and ordered logit models are appropriate for categorical outcomes (e.g., win, draw, or loss). On the other hand, multiple regression on the goal difference $y_{i}^{(1)}$ allows us to analyze the intensity of the match outcome as it accounts for the pseudo-continuous nature of the goal difference.
	
	\subsection{Explanatory, control, and fixed effects variables}\label{sec:variables}
	In this section, we present the explanatory variables associated to coaches, teams and referees decision-making, as well as some control and fixed-effects variables. For the complete set see Table~\ref{tab:codebook}.\footnote{See Figures~\ref{fig:SM_F_02}--\ref{fig:SM_F_12}, SM, for a visual representation by seasons and ranked home teams.}
	
	\subsubsection{Coach actions/schemes}\label{subsubsec:coach_schemes}   
	We devise three different versions of this variable: $(s_{i,1}, s_{i,2}$, $s_{i,3})$ but use only one of them at a time in each model. They are based upon the offensiveness index described in Appendix~\ref{App:OffInd}, which, in turn, depends on the team's formation, a good proxy of the coach's decisions since these are reflected in the players' positioning on the field. All three scheme variables are defined as the difference in the offensiveness index between the home and away coach schemes for match $i$. The three versions differ in the way such index is defined. Specifically, for $s_{i,1}$, the offensiveness index evaluates the performance of all 10 outfield players; for $s_{i,2}$, it evaluates the active players, excluding those who have been sent off; and for $s_{i,3}$, the offensiveness index normalizes the relative performance by dividing $s_{i,2}$ by its attainable maximum (the case of having all forward players). To capture the reaction of the coach to prior events we also define initial and final schemes, $\bs_{i}$ and $\bs_{i}^{*}$, respectively. While the initial scheme is based on the starting lineup and reflects the coach's forward-looking strategies, the final scheme is based on the formation at the end of the match and represents the coach's retrospective adjustments. In Tables~\ref{tab:schemesI} and \ref{tab:schemesF} we present some descriptive statistics of the coach actions variables. The initial and final schemes follow similar patterns, but the final schemes present higher variability, confirming the adaptive nature of coaching decisions to the real-time result of the match and hinting at dynamic effects. 
	
	\subsubsection{Team actions}\label{subsubsec:team}  
	The variables $\bw_i =  (w_{i,h})$, with $h=1,\dots,8$ measure the activities of the teams through 8 different actions. They are computed as:
	$
	w_{i,h} \equiv \sum_{t=1}^{90} w_{i,h,t}, 
	h=1,\dots,8$.
	Specifically, the actions are: total number of crosses ($h=1$), corners ($h=2$), crosses from corners ($h=3$), crosses not from corners ($h=4$), corners without crosses ($h=5$), shots ($h=6$), goal kicks ($h=7$) and offsides ($h=8$). Note that $w_{i,3}, w_{i,4}, w_{i,5}$ are derived from the other variables. In Table~\ref{tab:team} we present some descriptive statistics for these variables. They reveal insights into the tactical execution and strategic decisions made by teams in soccer matches. The differences between home and away team actions provide a quantitative measure of the HA in offensive output. The mean differences across most actions are positive, indicating that home teams, on average, undertake more offensive actions than away teams. This HA is most pronounced in crosses and shots, essential elements of attacking play. The distribution of these differences again shows variability in match strategies, with skewness and kurtosis values suggesting a range of outcomes from balanced to highly aggressive home performances. The differences in goal kicks and offsides between home and away performances are minimal, indicating that teams maintain consistent counter-offensive and defensive tactics regardless of playing at home or away. A slightly negative value for home-away difference in goal kicks indicate that, on average, more home offensive actions tend to generate away goal kicks rather than the contrary.
	
	\subsubsection{Referee actions}\label{subsubsec:ref}  
	The variables denoted as $\bz_{i} =  (z_{i,h})$, with ${h = 1,\dots,5 }$ represent referee decisions for match $i$. Specifically, they measure the number of yellow cards ($h=1$), red cards ($h=2$), free kicks ($h=3$), penalty kicks ($h=4$), and total fouls ($h=5$) leading to free and penalty kicks. These are calculated analogously to $\bw_{i}$. In Table~\ref{tab:referee} we report some descriptive statistics of the referee actions. The distribution of yellow and red cards shows a nuanced picture of match discipline. The statistics for fouls leading to free kicks and penalties offer further insight into the tactical and disciplinary dimensions. The differences in yellow and red card distributions between home and away teams are relatively small, indicating that, while there may be slight biases or strategic differences in play style, overall disciplinary actions are evenly distributed across venues. The statistics for fouls leading to free kicks and penalties provide further information on the tactical and disciplinary dimensions. Even though there is a slight indication of HA influencing disciplinary measures, the descriptive statistics suggest that the teams maintain consistent levels of discipline irrespective of venue.
	
	\subsubsection{Control variables}\label{subsubsec:control}  
	
	The control variables are denoted as $\bc_{i} = (c_{i,h})$, with ${h = 1,\dots,4 }$. In particular, the home stadium filling index, ($c_{i,1} \in [0,1]$) is a continuous control variable computed as the proportion of spectators present at the home stadium relative to its capacity. This captures the influence of home fans compared to away fans.\footnote{It is assumed that, on average, home spectators occupy at least half of the available seats as shown by the percentage of spectators who have season seats for home matches.} 
	The variable $c_{i,2}  \in [0,15]$ is the extra time at the end of the match, measured in added minutes.\footnote{For the weighted extra time variable, we identified the number of events in each half’s extra time ($\mathrm{N}_{i,1}+\mathrm{N}_{i,2}={\mathrm{N}_i}$), and then divided these counts by the total number of events in the $i$-th match ${\mathrm{N}_i}$. We added 1 to each of these ratios and multiplied by the respective extra-time minutes ($c_{i,2,1}$ and $c_{i,2,2}$). Summing the two weighted components gives: $\tilde{c}_{i,2} = c_{i,2,1} \times \bigl(1 + \tfrac{\mathrm{N}_{i,1}}{\mathrm{N}_i}\bigr) + c_{i,2,2} \times \bigl(1 + \tfrac{\mathrm{N}_{i,2}}{\mathrm{N}_i}\bigr).$ This increases extra-time minutes for halves that have more events relative to the overall match.}
	This variable, determined by referees to compensate for lost time during the match, introduces an element of randomness to its duration. Finally, to account for the team effect, we introduce the control variable $c_{i,3}$ and its normalized version $c_{i,4}$. $c_{i,3}$ is the difference of points between the home and away teams before the match. $c_{i,4} \in [-1,1]$ is the ratio of ranking points over the maximum attainable ranking points before the match. 
	\par
	The home stadium filling index, $c_{i,1}$, presents a median of 0.56 and mean of 0.59. Its standard deviation is 0.19 and its skewness is 0.15, suggesting a slightly right-skewed but nearly mesokurtic distribution. The variable extra time $c_{i,2}$ has a mean of 5.71 minutes, with a small standard deviation and relatively small coefficient of variation. Its distribution is slightly right-skewed and leptokurtic, suggesting some heavier tail activity. Its weighted version $\tilde{c}_{i,2}$ ranges from 0.00 to 15.51, with a mean of 5.91 and a standard deviation of 1.79, indicating moderate dispersion around the mean, slight skewness and moderate kurtosis. Finally, $c_{i,4}$ has a mean of -0.01 and a standard deviation of 0.27, reflecting some disparity in team strengths. Its distribution exhibits moderate skewness and heavier tails compared to the Gaussian distribution.
	
	\subsubsection{Additional fixed effects: season, league day, extreme events, team}\label{subsubsec:fixed}  
	This group includes additional fixed effects denoted as $\mathbf{k}_i = (\mathbf{k}_{i}^{s}, \mathbf{k}_{i}^{d},\mathbf{k}_{i}^{e})$. First, $\mathbf{k}^{s}_{i} = (k_{i}^{(11)}, k_{i}^{(12)},$ $k_{i}^{(13)})$ are three dummies that capture the season effects for 2011, 2012, and 2013. Second, $\bk_{i}^{d}$ represents the league day effect that accounts for the specific match day within the championship. Specifically, we assign a rank to each distinct date of a match, giving the same rank to identical dates and incrementing by one for the next distinct date, as to create consecutive ranks without gaps. $k_{i,1}^{d}$ uses the absolute ranks, whereas in $k_{i,2}^{d}$ the ranks are scaled from $0$ to $1$, corresponding to the earliest and latest ranked dates, respectively. These variables are needed because the actual match date often differs from the scheduled date for reasons such as security or adverse weather. By referencing the actual playing date instead of the originally scheduled date, the ranking and relative date calculations accurately reflect the true chronological order of matches and thus ensure consistency in determining each team's standings. Third, we use $\mathbf{k}_i^{e} = (k_{i,1}^{e},k_{i,2}^{e},k_{i,3}^{e})$ to control for anomalous results, i.e. matches that ended with a difference of at least five goals between the home and away teams. The first of these dummy variables, $k_{i,1}^{e}$ indicates extreme goal difference matches in absolute value, whereas $k_{i,2}^{e}$ and $k_{i,3}^{e}$ account for positive and negative goal differences for the home team, respectively. The variable $k_{i}^{d}$ has a median and mean of about 0.51, near-zero skewness, and slightly lower-than-normal kurtosis. The two dummy variables $k_{i,2}^{e}$ and $k_{i,3}^{e}$, (extreme positive and extreme negative goal differences for the home team), are rare as they occur only nine times overall. The dummy variables $\bk_{i}^{T}$ add a fixed effect for each of the 26 teams. This is done to control for persistent, unobserved heterogeneity across the 26 Serie A teams in our dataset, as each team has distinct characteristics that can systematically influence both strategic choices and match outcomes. Details for controls and fixed effects are presented in Tables~\ref{tab:additional} and \ref{tab:additional2}, while Figure~\ref{fig_extreme_matches} shows the scatterplot of relative team effects $c_4$ against goal difference $y^{(1)}$, highlighting extreme matches $k_{1}^{e}$. 
	
	
	\section{Statistical modeling}\label{sec:Results}
	
	This section is devoted to the modeling exercise. As mentioned, we assess the impact of decision-making at the levels of coach, team and referee, on the performance of the home team. We fit a multiple regression model to analyze the intensity of match outcomes (Model 1), a logit model for the probability of home team win (Model 2), and an ordered logit model to study the probabilities of home team win, draw, and loss (Model 3). These three aspects of match performance provide a comprehensive picture of the factors influencing the outcome of a soccer match. Section~\ref{subsec:Specification} details the specifications used, whereas Section~\ref{subsec:Baseline} discusses the results of model fitting. Section~\ref{subsec:Selection} refines the full specifications via AIC and BIC to retain only significant effects and applies tests on residuals to confirm model validity. Section~\ref{subsec:averaging} uses Akaike‐weight averaging over top models to yield robust shrinkage estimates.  
	
	\subsection{Model specifications}\label{subsec:Specification}
	In the following, for a generic match $i$, with $i=1,\dots,N$, and the response $y_{i}^{(j)}$, with $j=1,2,3$, we adopt the following notation where the variable $\bX_i$ includes all the predictors
	\begin{align*}
		\bX_i &= (\bs_i, \bw_i,  \bz_i, \bc_i, \bk_i).
	\end{align*}
	Recall that the vectors of explanatory variables $\bs_{i}$, $\bw_{i}$, and $\bz_{i}$ correspond to the three decisional levels, \emph{coach}, \emph{team}, and \emph{referee} actions, whereas $\bc_i$ and $\bk_i$ are the vector of control variables and fixed effects, respectively. Since the three specifications do not necessarily include the whole set of predictors  $\bX_i$, we define the set of predictors for the $j$-th model as  $\bX_i^{(j)} = \bX_i\bR^{(j)}$, with $j=1,2,3$ where we use the block diagonal selector matrix $\bR^{(j)}$. We denote the corresponding parameter (column) vector as $\btheta^{(j)}$ so that the linear predictor for match $i$ and model $j$ results $\bX_i^{(j)}\btheta^{(j)}$. We assume the parameter vectors to be column-vectors. 
	\begin{model}\label{mod:2}
		The response variable $y_{i}^{(1)}$ represents home-away goal differences so that we specify a multiple regression OLS model for it:
		\begin{align}
			\E\left[y_{i}^{(1)}\mid \bX_i^{(1)}\right] &=  \bX_i^{(1)}\btheta^{(1)}, \qquad i=1,\dots,N,\\
			y_{i}^{(1)} &= \alpha_0 + \bs_{i}^{(1)}\balpha_s +  \bw_{i}^{(1)} \balpha_w+  \bz_{i}^{(1)}\balpha_z
			+  \bc_i^{(1)}\bEta_c  +  \bk_i^{(1)}\bEta_k +\eps_i,
		\end{align}
	\end{model}
	with  $\eps_i \sim \mathcal{N}(0, \sigma^2)$. The parameter vector results $\btheta^{(1)} =(\alpha_0, \balpha_s^\top, \balpha_w^\top, \balpha_z^\top, \bEta_c^\top, \bEta_k^\top)^\top$. 
	
	\begin{model}\label{mod:1}
		Here $y_{i}^{(2)}$ is a binary variable indicating home-team-win so we model the probability of winning the $i$-th match for the home team: 
		\begin{align}
			\E \left[y^{(2)}_i \mid \bX_i^{(2)} \right] &=  \Pbb\left(y^{(2)}_{i} = 1\right) = g\left(\bX_i^{(2)}\btheta^{(2)}\right); 
			\quad g(x) = \dfrac{e^x}{1+e^x} \in [0, 1], \nonumber \\
			\bX_i^{(2)}\btheta^{(2)}  &= \beta_0 +  \bs_{i}^{(2)}\bBeta_s + \bw_{i}^{(2)}\bBeta_w  +  \bz_{i}^{(2)}\bBeta_z  +  \bc_i^{(2)}\bphi_c +  \bk_i^{(2)}\bphi_k, \label{eq:mod2_score}
		\end{align}
	\end{model}	
	where the function $g(\cdot)$ represents the standard logistic cumulative distribution function. The parameter vector results 
	$\btheta^{(2)} = (\beta_0, \bBeta_s^\top, \bBeta_w^\top, \bBeta_z^\top, \bphi_c^\top, \bphi_k^\top)^\top$. 
	
	\begin{model}\label{mod:3}
		We model the probability of match win, draw and loss for the home team using an ordered logit model for cross-sectional data:
		\begin{align}
			\nonumber 
			\Pbb\left(y_{i}^{(3)} \le h\right) &= g\left(\pi_h - \bX_i^{(3)}\btheta^{(3)}\right), 
			\quad h=0,1; \qquad g(x) = \dfrac{e^x}{1+e^x}; \\
			\Pbb\left(y_{i}^{(3)} = 3 \right)  &= 1 - \Pbb\left(y_{i}^{(3)} \le 1 \right), \label{eq:ologit_highcat}\\
			\nonumber \bX_i^{(3)} \btheta^{(3)} &= \bs_i^{(3)} \bgamma_s +  \bw_i^{(3)}\bgamma_w + \bz_i^{(3)}\bgamma_z
			+ \bc_i^{(3)}\bpsi_c
			+ \bk_i^{(3)}\bpsi_k. \label{eq:mod_ologit_linpred}
		\end{align}
	\end{model}
	\noindent
	Here the response variable $y_{i}^{(3)} \in\{0,1,3\}$ (\emph{loss}, \emph{draw}, or \emph{win}) is treated as an ordinal outcome with two cutpoints $\pi_0$ and $\pi_1$ (satisfying $\pi_0 \le \pi_1$). As above, $g(\cdot)$ is the standard logistic cumulative distribution function. The cumulative probabilities $\Pbb(y_{i}^{(3)} \le h)$ for $h=0,1$ are modeled via logistic curves.  
	In this specification, the \emph{proportional-odds} assumption is retained. One can visualize the shifts in the conditional distribution of $y_{i}^{(3)}$ as $\bX_i^{(3)}$ changes, where an increasing $\bX_i^{(3)}\btheta^{(3)}$ shifts probability mass towards higher-ordered categories. The parameter vector results $\btheta^{(3)}=	(\pi_0,\pi_1,\bgamma_s^\top,\bgamma_w^\top,\bgamma_z^\top, \bpsi_c^\top,\bpsi_k^\top)^\top.$
	To test for parallel regression assumption, we augment the traditional \citet{BRA90} proportional odds test by employing a nonparametric bootstrap for the logistic fits in the procedure proposed by \citet{SCH20}. Each cutpoint $\pi_h$ is treated separately, yielding distinct binary logistic regressions of the form  $\pi_h - \bX_i^{(3)}\btheta^{(3)}$. Estimates are then computed on bootstrapped sample, producing robust variance-covariance matrices.   These threshold-specific estimates are assembled into a single Wald-type statistic, using a contrast matrix to compare the unconstrained $\bBeta$’s against the proportional-odds requirement of parameters' equality. Under the null hypothesis of the parallel regression assumption, both the overall (omnibus) and variable-specific test statistics follow (asymptotically) a chi squared distribution with degrees of freedom tied to the number of parameters and cutpoints.
	
	\subsection{Baseline Models}\label{subsec:Baseline}
	We begin by estimating Model~\ref{mod:2} via a Gaussian generalized linear model (GLM), equivalent to ordinary least squares under the identity link. Maximum likelihood estimation yields the coefficient vector 
	together with heteroskedasticity-consistent (HC) variance estimates. Similarly, Model~\ref{mod:1} is also estimated by maximum likelihood with robust standard errors. For Model~\ref{mod:3}, we implement an ordinal logit specification that jointly estimates the regression coefficients and threshold terms, by maximizing the likelihood under a logistic link. Standard errors are derived via nonparametric bootstrap resampling to accommodate potential model misspecification. For each model we start by fitting six baseline specifications, corresponding to the 6 different choices for the scheme variable, i.e. $(s_1,s_2,s_3,\tilde{s}_1,\tilde{s}_2,\tilde{s}_3)$. Table~\ref{estmod:1} reports the results for the scheme $\tilde{s}_2$ for Model 1, 2, 3, since it consistently resulted the best fit among the 6 choices. The results for the remaining schemes are reported in the SM, Tables~\ref{tab:regression_summary}, \ref{tab:logit_regression_summary}, \ref{tab:baseline_ologit}.     
	To evaluate model performance, we compute metrics such as Akaike’s information criterion (AIC) \citep{AKA74}, Schwarz’s Bayesian information criterion (BIC) \citep{SCH78}, and goodness-of-fit measures including adjusted $R^2$, \( \bar{R}^2\) (for pseudo-continuous outcomes), McFadden's $R_{MF}^2$, Nagelkerke's $R_N^2$, deviance, and classification metrics (for discrete outcomes).
	Across all specifications, $s_2$ is consistently positive ($p$-values $<0.001$) and $\tilde{s}_2^{*}$ is significantly negative ($p$-values $<0.001$). For the team variables, $\tilde{w}_1$ is significantly negative ($p$-values $<0.005$), whereas $\tilde{w}_2$ is also negative but borderline significant. Both $\tilde{w}_6$ and $\tilde{w}_7$ are consistently positive, whereas $\tilde{w}_8$ is only borderline significant. Among referee effects, both $\tilde{z}_1$ and $\tilde{z}_2$ are significantly negative, $\tilde{z}_3$ is never significantly different from zero, and $\tilde{z}_4$ is positive and significant across the three models. As for the control variables, $c_1$ (and its powers) are not significant, $\tilde{c}_2$ has a modest negative effect, and $c_4$ is strongly positive ($p$-values $<0.001$). Fixed effects $k^e_{2}$ and $k^e_{3}$ yield significant positive/negative coefficients, whereas the remaining fixed effects resulted nonsignificant. 
	For the baseline ordinal–logit specification reported (season~2011 as the reference level), the estimated thresholds of $-1.54$ and $0.17$ yield, with all regressors set to zero, predicted probabilities of $0.177$ for a loss ($y^{(3)} = 0$), $0.366$ for a draw ($y^{(3)} = 1$), and $0.458$ for a win ($y^{(3)} = 3$); neither threshold differs significantly from zero.
	
	Overall, the three fits are comparable in terms of $R^2$, whereas accuracy and sensitivity vary between Models 2 and 3, with category-specific metrics on the latter detailed in the table. The Brant test ($p=0.76$) does not indicate major proportional-odds violations. The results confirm that baseline models are benchmarks for assessing the impact of decision-making at the coach, team, and referee levels on the home team’s performance. They capture key statistical relationships before incorporating more complex specifications. 
	\par Our findings are broadly consistent with existing literature, while introducing important new elements that strengthen our central thesis. We confirm earlier mixed evidence \citep{DEP12} by showing that aggressive opening tactics reliably improve performance, even after controlling for end-match attacking levels and team-fixed effects. We find known results, such as the negative “cross paradox” \citep{SAR18a} and the strong impact of penalties \citep{SAR18b}, while contributing novel insights into the strategic value of goal-kicks and the behavioral interpretation of offsides. 
	The table also confirm the existence of real-time influence of coaching tactics, demonstrating that dynamic decisions made by coaches can significantly shape match outcomes. This implies that timely and informed tactical choices function as managerial interventions that optimize scarce resources under uncertainty. Just as firms adjust strategies in response to market conditions, coaches adapt tactics—such as player positioning, possession management, and team momentum—in real time to maximize expected returns, measured in terms of goal difference or win probability.
	\begin{table}[!h]
		\centering
		\footnotesize
		\caption{Fit and performance of baseline regression models with robust standard error. Standard errors for Model~1 and 2 use the sandwich estimator HC3, those for Model~3 use a nonparametric bootstrap ($B=1000$). Baseline for Model~3 is $k^{(11)}=1$. Number of observations $N = 1 139$.}\label{estmod:1}
		\resizebox{0.85\textwidth}{!}{
			\begin{tabular}{l c rl rl rl}
				\toprule
				\textbf{Models}&        & \multicolumn{2}{c}{Model 1} & \multicolumn{2}{c}{Model 2} & \multicolumn{2}{c}{Model 3} \\
				\textbf{Description} &\textbf{Variable} & Coeff. & $p$-value & Coeff. & $p$-value & Coeff. & $p$-value \\
				\midrule
				\textbf{Scheme} &&&&&&& \\
				Initial Scheme 2 & $s_{2}$             &  0.30  & (0.000) & 0.54  & (0.000) & 0.53  & (0.000) \\
				Final Weighted Scheme 2 &$\tilde{s}_{2}^*$   & -0.25 & (0.000) & -0.50 & (0.000) & -0.49 & (0.000) \\
				\midrule
				\textbf{Team} &&&&&&& \\
				Weighted Crosses &$\tilde{w}_{1}$   & -0.01 & (0.000) & -0.03 & (0.000) & -0.02 & (0.001) \\
				Weighted Corners &$\tilde{w}_{2}$   & -0.02 & (0.048) & -0.02 & (0.331)   & -0.04 & (0.066) \\
				Weighted Shots &$\tilde{w}_{6}$   &  0.04  & (0.000) & 0.07  & (0.000) & 0.07  & (0.000) \\
				Weighted Goal Kicks &$\tilde{w}_{7}$   &  0.03  & (0.000)  & 0.05  & (0.004)  & 0.04  & (0.008) \\
				Weighted Offsides &$\tilde{w}_{8}$   &  0.03  & (0.070)   & 0.04  & (0.123)   & 0.04  & (0.069) \\
				\midrule
				\textbf{Referee} &&&&&&& \\
				Weighted Yellow Cards & $\tilde{z}_{1}$   & -0.06 & (0.013)   & -0.11 & (0.021)   & -0.10 & (0.014) \\
				Weighted Red Cards &$\tilde{z}_{2}$   & -1.06 & (0.000) & -2.06 & (0.000) & -1.99 & (0.000) \\
				Weighted Free Kicks & $\tilde{z}_{3}$   & -0.01 & (0.141) & 0.00  & (0.964)   & -0.00 & (0.727) \\
				Weighted Penalty Kicks &$\tilde{z}_{4}$   &  0.33 & (0.000) & 0.47  & (0.008)  & 0.48  & (0.002) \\
				\midrule
				\textbf{Controls} &&&&&&& \\
				Linear Stadium Filling &$c_1$     & 0.39  & (0.886) &  0.18 & (0.976)     &  1.05 & (0.852) \\
				Quadratic Stadium Filling &$c_1^2$   & -0.32 & (0.949) & -0.21 & (0.985)     & -1.70 & (0.868) \\
				Cubic Stadium Filling &$c_1^3$   & -0.29 & (0.920) & -0.40 & (0.949)     &   0.20 & (0.972) \\
				\addlinespace
				Weighted Extra Minutes &$\tilde{c}_2$   & -0.04 & (0.118) & -0.09 & (0.034)     & -0.03 & (0.429) \\
				Relative Team Effect &$c_4$     & 1.47  & (0.000) &  2.04 & (0.000)     & 2.27  & (0.000) \\
				\midrule
				\textbf{Fixed effects} &&&&&&& \\
				Season 2011/12 &$k^{(11)}$    & 0.57  & (0.239) & 0.18   & (0.868) & - & - \\
				Season 2012/13 &$k^{(12)}$    & 0.48  & (0.323) & 0.22   & (0.843) & -0.05 & (0.746) \\
				Season 2013/14 &$k^{(13)}$    & 0.51  & (0.304) & 0.30   & (0.784) & 0.03 & (0.859) \\
				\addlinespace
				Scaled Ranked League Day &$k_{2}^{d}$ & -0.04 & (0.764) & 0.04   & (0.868) & -0.09 & (0.698) \\
				\addlinespace
				Extreme Positive Goal Difference &$k_{2}^{e}$    & 3.97  & (0.000) & 14.40  & (0.000) & 13.13 & (0.000) \\
				Extreme Negative Goal Difference &$k_{3}^{e}$    & -4.83 & (0.000) & -13.90 & (0.000) & -14.81 & (0.000) \\
				\midrule
				\textbf{Thresholds} &&&&&&& \\
				0|1 & $\tau_0$ & -& -& -& -& -1.54 & (0.131) \\
				1|3 & $\tau_1$& -& -& -& -&  0.17 & (0.868) \\
			\end{tabular}
		}
		%
		\resizebox{1\textwidth}{!}{
			\begin{tabular}{lccccccccc}
				\toprule
				Model & AIC & BIC & Deviance & $\bar{\mathrm{R}}^2$ $|$ $\bar{\mathrm{R}}_\mathrm{N}^2$ & Accuracy & Sensitivity & Specificity & F1 & Brant\\
				\midrule
				1 & 3809 & 3924 & 1814 & 0.44 & - & - & - & - &-\\
				2 & 1159 & 1270 & 1115 & 0.44 & 0.75 & 0.77 & 0.73 & 0.77&-\\
				3 & 1890 & 2006 & 1844 & 0.45 &0.63 & \scriptsize{0.65, 0.23, 0.83} & \scriptsize{0.86, 0.87, 0.67} & \scriptsize{0.64, 0.29, 0.75}& 0.76\\
				\bottomrule
			\end{tabular}
		}
	\end{table}

	\subsection{Augmented Models}\label{subsec:Selection}
	We augment the baseline models of Section~\ref{subsec:Baseline} using a large pool of block-nested candidates and performing full subset search. The results are presented in Table~\ref{tab:unified}, where each of the dependent variables has two fits: one for the best model identified through the AIC and one for the BIC.\footnote{See Table~\ref{tab:mod_blocks_extended}, SM, for an overview of specification combinations.} 
	Non significant coefficients are omitted with some exceptions. Also, to we used the following notation to report the significance of groups of variables that are omitted due to space constraints: S~=~significant; NS~=~not significant; -~=not included. The $p$-values are reported in parenthesis next to parameter estimates. The AIC-selected specifications include the 26 team fixed-effects $\bk^T$ (we omitted the single $p$-values and only report if some of them are significant), while the BIC-selected specifications include the continuous team effect $\bc_4$ . Both of them increase the in-sample classification performance (even against the weighted baseline specification). For Model 1, the AIC preferred model showed that 10 out of 26 team fixed effects are significant (not shown). 
	The second version of the coaches' schemes $s_2$ is frequently selected as the optimal choice under both AIC and BIC criteria.
	The results consistently indicate that the initial scheme $s_2$ exerts a highly significant positive effect, while the final scheme $s_2^*$ exhibits a highly significant negative effect. These findings highlight the notable home-away contrast, wherein crosses $(w_1)$ tend to have a negative impact, while the effect of shots $(w_6)$ and goal-kicks $(w_7)$ is positive, \emph{ceteris paribus}. The coefficients maintain both the magnitude and significance observed in the baseline. In summary, team fixed effects are present in the top three AIC specifications for Model 1, 2, and 3, while the top three BIC favor parsimonious models including season effects (Model 1), weighted minutes extra time (Model 2), and continuous team effects (Model 3). 
	For Model 3, the AIC-best ordered-logit fit (Juventus baseline, extreme scorelines excluded) gives thresholds of $-6.33$ and $-4.49$ ($p\approx0$), implying \(\mathbb{P}(y^{(3)}=0,1,3)=0.002,0.010,0.989\), when all covariates equal zero. The BIC-best fit (baseline: non-extreme goal differences) yields thresholds of $-1.38$ and $-0.36$, giving \(\mathbb{P}(y^{(3)}=0,1,3)=0.20,0.39,0.41\).  In both cases a home win is the most likely outcome—especially pronounced for Juventus.
	\par
	The robustness of the conclusions drawn from the models is further corroborated by allowing and testing different kinds of interactions, such as coach scheme and referee actions. Overall, interactions such as coach-referee or team-referee are infrequently chosen in the top 1\% of AIC or BIC model sets, and even when they appear, they neither alter the direction nor the significance of the main effects associated with schemes, coaches, teams, or referees. Notably, for Models~1--3, coach-referee interactions appear in a handful of top 1\% AIC-ranked models but do not appear at all among the top 1\% BIC-ranked ones, mirroring the general pattern that interactions fail to affect the principal insights. 
	
	\begin{table}[!hb]
		\centering
		\footnotesize
		\captionsetup{size=small} 
		\caption{Rank-1 AIC- and BIC-selected models for Models~1--3 
			with unweighted in-game actions including block-interactions. Significance levels in parentheses. Model~1 uses robust HC3 SE. Below, $p$-values for SW (Shapiro--Wilk), KS (Kolmogorov--Smirnov), JB (Jarque--Bera), BP (Breusch--Pagan),  
			and RR (Ramsey RESET). Model~2 also has robust HC3 SE, with p-values for HL (Hosmer--Lemeshow). Model~3 employs nonparametric bootstrap ($B=1000$) for the SEs and $p$-values also on Brant (proportional-odds) test. S = significant, NS = not significant; ``--'' = excluded variable. Baseline is Juventus for Models~1 and ~2, if Intercept; for Model~3, if $\bk^{T}=$ S. For Models~2--3, $R_{MF}^2$ = McFadden's $R^2$; for Model~3, $R_{N}^2$ = Nagelkerke's $R^2$. For Model 3, standard $p$-values for LI = Lipsitz.
		} \label{tab:unified}
		\resizebox{0.725\textwidth}{!}{
			\begin{tabular}{lcrrrrrr}
				\toprule
				\textbf{Model} && \multicolumn{2}{c}{Model 1} & \multicolumn{2}{c}{Model 2} & \multicolumn{2}{c}{Model 3} \\
				\textbf{Description} &\textbf{Variable} & \multicolumn{1}{c}{AIC}&\multicolumn{1}{c}{BIC}  & \multicolumn{1}{c}{AIC}&\multicolumn{1}{c}{BIC}  & \multicolumn{1}{c}{AIC}&\multicolumn{1}{c}{BIC} \\
				\midrule
				&(Intercept) & - & 0.31 (0.000) & - & - & - & - \\
				\addlinespace
				\textbf{Coach} &&&&&&& \\
				Initial Scheme 2&$s_{2}$ & 0.28 (0.000) & 0.29 (0.000) & 0.55 (0.000) & 0.53 (0.000) & 0.53 (0.000) & 0.51 (0.000) \\
				Final Scheme 2 &$s_{2}^{*}$ & -0.25 (0.000) & -0.26 (0.000) & -0.53 (0.000) & -0.52 (0.000) & -0.52 (0.000) & -0.50 (0.000) \\
				\addlinespace
				\textbf{Team} &&&&&&& \\
				Crosses     &$w_1$ & -0.02 (0.000) & -0.02 (0.000) & -0.04 (0.000) & -0.03 (0.000) & -0.03 (0.000) & -0.03 (0.000) \\
				Corners     &$w_2$ & -0.01 (0.273) & -0.02 (0.048) & -0.02 (0.452) & -0.02 (0.299) & -0.03 (0.223) & -0.04 (0.049) \\
				Shots       &$w_6$ & 0.04 (0.000) & 0.04 (0.000) & 0.08 (0.000) & 0.07 (0.000) & 0.07 (0.000) & 0.06 (0.0000) \\
				Goal Kicks  &$w_7$ & 0.03 (0.000) & 0.03 (0.001) & 0.07 (0.001) & 0.06 (0.003) & 0.06 (0.001) & 0.04 (0.009) \\
				Offsides    &$w_8$ & 0.03 (0.023) & 0.02 (0.121) & 0.06 (0.057) & 0.04 (0.201) & 0.06 (0.034) & 0.04 (0.092) \\
				\addlinespace
				\textbf{Referee} &&&&&&& \\
				Yellow Cards &$z_1$ & -0.05 (0.031) & -0.06 (0.011) & - & -0.11 (0.013) & -0.09 (0.052) & -0.11 (0.010) \\
				Red Cards &$z_2$ & -1.11 (0.000) & -1.10 (0.000) & - & -2.14 (0.000) & -2.20 (0.000) & -2.06 (0.000) \\
				Free Kicks &$z_3$ & 0.00 (0.526) & 0.01 (0.172) & - & -0.00 (0.929) & -0.00 (0.920) & 0.00 (0.789) \\
				Penalty Kicks &$z_4$ & -0.31 (0.000) & -0.33 (0.000) & - & -0.45 (0.012) & -0.43 (0.010) & -0.46 (0.003) \\
				Home Yellow Cards&$z_1^H$ & - & - & -0.22 (0.001) & - & - &-\\ 
				Home Red Cards&$z_2^H$ & - & - & -2.23 (0.000) & - & -&-\\ 
				Home Fouls&$z_5^H$ & - & - & -0.02 (0.291) & - & -&-\\ 
				Away Yellow Cards&$z_1^A$ & - & - & -0.02 (0.719) & - & -&-\\ 
				Away Red Cards&$z_2^A$ & - & - & 2.48 (0.000) & - & -&-\\ 
				Away Fouls&$z_5^A$ & - & - & -0.00 (0.920) & - & -&-\\
				\addlinespace
				\textbf{Controls} &&&&&&& \\
				Linear Stadium Filling &$c_{1}$ & -1.26 (0.629) & - & -5.45 (0.394) & - & -5.56 (0.350) & - \\
				Quadratic Stadium Filling &$c_{1}^2$ & 0.99 (0.837) & - & 7.54 (0.520) & - & 6.79 (0.512) & - \\
				Cubic Stadium Filling &$c_{1}^3$ & -1.52 (0.591) & - & -5.57 (0.415) & - & -4.99 (0.387) & - \\
				\addlinespace
				Weighted Extra Minutes&$\tilde{c}_{2}$ & - & - & - & -0.06 (0.000) & - & - \\
				Relative Team Effect&$c_{4}$ & - & 0.03 (0.000) & - & 0.05 (0.000) & - & 0.06 (0.000) \\
				\addlinespace
				\textbf{Fixed effects} &&&&&&& \\
				Team FEs &$\bk^{T}$     & S & - & S & - & S & - \\
				Extreme Goal Difference &$k_{1}^{e}$ & - & - & - & 0.63 (0.336) & - & - \\
				Extreme Positive Goal Difference &$k_{2}^{e}$ & 4.20 (0.000) & 4.16 (0.000) & 14.22 (0.000) & - & 12.78 (0.000) & 12.92 (0.000) \\
				Extreme Negative Goal Difference &$k_{3}^{e}$ & -4.43 (0.000) & -5.18 (0.000) & -13.91 (0.002) & - & -14.73 (0.000) & -15.56 (0.000) \\
				\addlinespace
				\textbf{Thresholds} &&&&&&& \\
				0|1 & $\tau_0$& - & - & - & - & -6.33 (0.000) & -1.38 (0.000) \\
				1|3 & $\tau_1$& - & - & - & - & -4.49 (0.000) & 0.36 (0.000) \\
				\midrule
				\multicolumn{8}{l}{\textbf{Goodness-of-fit}}\\
				\midrule
				AIC && 3777 & 3781 & 1123 & 1131 & 1839 & 1850 \\
				BIC && 3994 & 3862 & 1345 & 1202 & 2055 & 1931 \\
				DEV && 1704 & 1793 & 1035 & 1103 & 1753 & 1818 \\
				\addlinespace
				$\bar{R}^2$ && 0.47 & 0.43 & - & - & - & - \\
				$R_{MF}^2$ && - & - & 0.34 & 0.30 & 0.27 & 0.25 \\
				$R_{N}^2$ && - & - & 0.50 & 0.45 & 0.50 & 0.47 \\
				\addlinespace
				Accuracy && - & - & 0.78 & 0.75 & 0.65 & 0.64 \\
				Sensitivity && - & - & 0.79 & 0.77 & 0.68, 0.28, 0.84 & 0.65, 0.26, 0.84 \\
				Specificity && - & - & 0.77 & 0.73 & 0.87, 0.85, 0.73 & 0.88, 0.86, 0.68 \\
				Precision && - & - & 0.80 & 0.77 & 0.67, 0.39, 0.73 & 0.66, 0.39, 0.70 \\
				F1 && - & - & 0.79 & 0.77 & 0.67, 0.33, 0.78 & 0.65, 0.31, 0.76 \\
				\midrule
				\multicolumn{8}{l}{\textbf{Residual Diagnostics Tests} ($p$-values)}\\
				\midrule
				SW && 0.12 & 0.14 & - & - & - & - \\
				BP && 0.11 & 0.28 & - & - & - & - \\
				RR && 0.29 & 0.19 & 0.19 
				& 0.50 
				& - & - \\
				HL && - & - & 0.67 & 0.59 & 0.47 & 0.45 \\
				LI && - & - & - & - & 0.18 & 0.28     \\
				Brant && - & - & - & - & 0.99 & 0.95 \\
				\bottomrule
			\end{tabular}
		}
	\end{table}
	
	
	Beyond the inclusion of classification metrics and information criteria, we evaluated our regression models using several diagnostic tests. For Model 1, we computed the Shapiro-Wilk (SW) test of Gaussianity of the residuals, the Breusch-Pagan (BP) test for heteroskedasticity and Ramsey's RESET (RR) test for model specification \citep{RAM69} (i.e. omitted variables or incorrect functional forms). For Model 2, we computed above RR test, and the Hosmer-Lemeshow (HL) test of goodness-of-fit \citep{HOS13}. As for Model 3, we computed the generalized HL test, the \citet{LIP96} (LI) goodness-of-fit tests and the Brant test \citep{BRA90} that verifies the null hypothesis of proportional-odds.  The results are reported in Table~\ref{tab:unified}, bottom panels. All the top AIC and BIC specifications do not reject the null hypotheses. 
	
	From an economic perspective, these results reinforce our view that coaching decisions function as managerial choices under uncertainty. The consistent positive effect of early offensive strategies (confirmed across AIC–BIC model selection criteria) suggests that proactive tactical commitment yields high marginal performance, especially early in the game when resource allocation (e.g., player position) is most flexible. The repeated negative impact of crosses illustrates a strategic inefficiency, highlighting the importance of data-driven tactical refinement. In contrast, goal-kicks and shots maintain a strong, positive relationship with outcomes, indicating the value of direct, transition-based play as a high-performing tactical move. Corners show context-dependent returns, diminishing in more parsimonious BIC models, which may reflect their varying efficiency under different match conditions. Red cards and penalties emerge as large exogenous shocks with strong effects on outcomes which coaches must anticipate and mitigate. The insignificance of free-kicks (partly due to unmodeled location) reflects a form of information loss that weakens their strategic utility. Team fixed effects and continuous team performance scores further support the notion that organizational factors play a persistent role in shaping returns to tactical choices. The model selection procedures (AIC, BIC) and robustness across different specifications (e.g., referee action and offensive interaction terms) reinforce the stability and generalizability of the strategic insights, while clean residuals and strong fits indicate well-specified models. 
	
	\subsection{Model averaging}\label{subsec:averaging}
	Selecting from a broad pool of candidate models can induce a bias, leading to sub-optimal or non robust solutions. Various procedures have been proposed to overcome the problem, \citep[c.f. ][]{LUK10,BER13,KUC20} and we adopt the model-averaging framework derived from \cite{LUK10}. We focus on the top 2.5\% AIC candidates for Model 1 ($L=393$), and the top 5\% candidates for Model 2 ($L=317$) and Model 3 ($L=107$), excluding interaction terms. Each candidate set is further partitioned into specifications featuring coach variables in home--away differences (\emph{Set~1}) versus those incorporating separate home and away variants (\emph{Set~2}). For a generic parameter $\theta$, we indicate with $\hat\theta_l$ its estimate obtained under candidate $l$, where $l=1,\dots,L$. We derive the shrinkage estimator \(\tilde{\theta}\) and an approximation for its variance as follows: 
	$
	\tilde{\theta}=\sum_{l=1}^L w_l\,\hat{\theta}_{l}
	\quad\text{and}\quad
	\widehat{\Var}\![\tilde{\theta}]  \;=\; \sum_{l=1}^R  w_l\,\bigl[\widehat{\Var}\left(\hat{\theta}_{l}\right)
	\;+\;
	\bigl(\hat{\theta}_{l} - \tilde{\theta}_{j}\bigr)^2
	\bigr],
	$ where $w_l$ denotes the Akaike weight for the $l$-th specification (often interpreted as the posterior probability that model~$l$ is the best Kullback--Leibler approximation).\footnote{BIC weights may also be employed \citep{SCHO20}, e.g., Table~\ref{tab:boot1} to construct confidence intervals and \ref{tab:boot2} (SM) to construct confidence intervals.} 
	If $\theta$ is not included in a given specification, then $\hat{\theta}_{l}=0$ and $\widehat\Var[\hat{\theta}_{l}]=0$. Inference proceeds via a Student's \emph{t}-distribution whose degrees of freedom depend upon the mean number of parameters across the models within each set. The results are shown in Table~\ref{tab:modelavg} and showcase the robustness of our modeling exercise since they match closely the estimates obtained previously. The importance of coaching strategies together with teams actions and referee decisions is confirmed. Among the remaining variables, measures reflecting extreme goal-differences are significant, and the same applies to team fixed effects in Model~3. Table~\ref{tab:boot1} (Appendix) and~\ref{tab:boot2} (SM) 
	present 95\% confidence intervals from classical, Bias Corrected and Accelerated (BCa) bootstrap, and model-averaged approaches, for unweighted and weighted variables respectively. 
	The marginal effect evaluated at the mean using the BCa estimate (Appendix~\ref{sec:bca}) suggests that an aggressive opening strategy by the home team increases the probability of winning by roughly $9.44$\% to $16.17$\%, after accounting for match dynamics and external factors.
	
	The model-averaged results reinforce the interpretation of coaching tactics and in-game events as managerial decisions under uncertainty in business contexts. From an economic standpoint, the consistently positive impact of separated home- and away-team offensiveness indices (despite more conservative standard errors) supports the idea that aggressive play yields a positive expected return, regardless of venue. The persistent but context-sensitive harm of crosses (only significant in one model) underscores the concept of tactical inefficiency or overuse of low-yield strategies. By contrast, the reliability of shots, goal-kicks, and even offsides in maintaining their directional effects suggests the existence of productive, high-leverage channels of attack that are robust across tactical contexts. Referee decisions, particularly red cards, and penalties, function as external shocks. The weakly negative influence of yellow cards, especially for home teams, may indicate reputational costs or risk aversion, where caution dampens aggressive play. The omission of free-kicks from selected models signals their low strategic salience, possibly due to incomplete contextual information. The repeated selection of extreme-score controls and the significance of team fixed effects relative to a strong benchmark (Juventus) highlights the importance of organizational heterogeneity in shaping the marginal returns of strategy. That no residual pathologies are detected adds credibility to the model structure, suggesting that informational efficiency and rational inference underlie the tactical decisions being modeled. Overall, the averaged evidence supports a view of soccer strategy as an economically coherent process, in which coaches, as managers, continuously adapt to maximize expected outcomes given noisy signals, asymmetric information, and competitive pressure.
	
	\begin{table}
		\centering
		\captionsetup{size=small} 
		\caption{Model averaging results for Model 1, 2, and 3 using Akaike weights for the top AIC-ranked models (Sets 1 and Set 2). Model 1: top 2.5\% (393 specifications); Model 2: top 5\% (317 specifications); Model 3: top 5\% (107 specifications). ``--'' = excluded variable; if Model 3 with $\bk^{T}$, baseline is Juventus. Coefficients with $p$-values $>0.15$ are omitted.}
		\label{tab:modelavg}
		\resizebox{1\textwidth}{!}{
			\begin{tabular}{l c rl rl rl | rl rl rl}
				\toprule
				\textbf{Sets} & & \multicolumn{6}{c}{Set 1} & \multicolumn{6}{c}{Set 2} \\
				\cmidrule(lr){3-8} \cmidrule(lr){9-14}
				\textbf{Models} & & \multicolumn{2}{c}{Model 1} & \multicolumn{2}{c}{Model 2} & \multicolumn{2}{c}{Model 3} 
				& \multicolumn{2}{c}{Model 1} & \multicolumn{2}{c}{Model 2} & \multicolumn{2}{c}{Model 3} \\
				\cmidrule(lr){3-4} \cmidrule(lr){5-6} \cmidrule(lr){7-8} \cmidrule(lr){9-10} \cmidrule(lr){11-12} \cmidrule(lr){13-14}
				\textbf{Description} &\textbf{Variable} & Est. & $(p$-val$)$ & Est. & $(p$-val$)$ & Est. & $(p$-val$)$ & Est. & $(p$-val$)$ & Est. & $(p$-val$)$ & Est. & $(p$-val$)$ \\
				\midrule
				Initial Scheme 2 &$s_{2}$       &  0.30 & (0.000) &  0.52 & (0.000) &  0.44 & (0.022) & --    & --      & --    & --      & --    & --   \\
				Final Scheme 2 &$s_{2}^{*}$   & -0.25 & (0.000) & -0.51 & (0.000) & -0.43 & (0.018) & --    & --      & --    & --      & --    & --   \\
				Home Initial Scheme 2 &$s_{2}^{H}$   & --    & --      & --    & --      & --    & --      &  0.34 & (0.000) &  0.56 & (0.000) &  0.52 & (0.000) \\
				Away Initial Scheme 2 &$s_{2}^{A}$   & --    & --      & --    & --      & --    & --      & -0.24 & (0.000) & -0.54 & (0.000) & -0.51 & (0.000) \\
				Home Final Scheme 2 &$s_{2}^{*H}$  & --    & --      & --    & --      & --    & --      & -0.28 & (0.000) & -0.51 & (0.000) & -0.49 & (0.000) \\
				Away Final Scheme 2 &$s_{2}^{*A}$  & --    & --      & --    & --      & --    & --      &  0.23 & (0.000) &  0.55 & (0.000) &  0.52 & (0.000) \\
				\addlinespace
				Crosses &$w_1$    & --    & --      & -0.03 & (0.118) & --    & --      & -0.01 & (0.146) & -0.04 & (0.028) & --    & --   \\
				Shots &$w_6$    &  0.04 & (0.002) &  0.08 & (0.000) &  0.06 & (0.002) &  0.04 & (0.000) &  0.08 & (0.000) &  0.07 & (0.000) \\
				Goal Kicks &$w_7$    &  0.03 & (0.034) &  0.06 & (0.002) &  0.05 & (0.014) &  0.03 & (0.005) &  0.06 & (0.002) &  0.05 & (0.005) \\
				Offsides&$w_8$    &  0.03 & (0.111) &  0.06 & (0.094) &  0.05 & (0.071) &  0.02 & (0.126) &  0.05 & (0.105) &  0.06 & (0.050) \\
				\addlinespace
				Yellow Cards&$z_1$    & -0.05 & (0.025) & --    & --      & -0.09 & (0.060) & -0.05 & (0.028) & --    & --      & -0.09 & (0.055) \\
				Home Yellow Cards&$z_1^H$  & --    & --      & -0.18 & (0.118) & --    & --      & --    & --      & -0.20 & (0.040) & --    & --   \\
				Red Cards&$z_2$    & -1.05 & (0.000) & --    & --      & -2.01 & (0.000) & -1.10 & (0.000) & --    & --      & -2.18 & (0.000) \\
				Home Red Cards&$z_2^H$  & --    & --      & -1.63 & (0.099) & --    & --      & --    & --      & -1.87 & (0.013) & --    & --   \\
				Away Yellow Cards&$z_1^A$  & --    & --      &  1.83 & (0.089) & --    & --      & --    & --      &  2.04 & (0.017) & --    & --   \\
				Penalty Kicks&$z_4$    & -0.32 & (0.000) & --    & --      & -0.42 & (0.016) & -0.32 & (0.000) & --    & --      & -0.43 & (0.011) \\
				\addlinespace
				Extreme Positive Goal Difference&$k_{2}^{e}$ &  3.96 & (0.000) & 10.79 & (0.069) & 12.37 & (0.000) &  4.11 & (0.000) & 13.46 & (0.000) & 11.88 & (0.000) \\
				Extreme Negative Goal Difference&$k_{3}^{e}$ & -4.93 & (0.000) & -10.64 & (0.114) & -14.05 & (0.000) & -4.89 & (0.000) & -13.25 & (0.000) & -14.19 & (0.000) \\
				\addlinespace
				Team FEs &$\bk^{T}$ & \multicolumn{2}{c}{--} & \multicolumn{2}{c}{--} & \multicolumn{2}{c}{S} & \multicolumn{2}{c}{--} & \multicolumn{2}{c}{--} & \multicolumn{2}{c}{S} \\
				\midrule
				$L$ (\# models) & & \multicolumn{2}{c}{210} & \multicolumn{2}{c}{290} & \multicolumn{2}{c}{86} 
				& \multicolumn{2}{c}{183} & \multicolumn{2}{c}{27} & \multicolumn{2}{c}{21} \\
				\bottomrule
			\end{tabular}
		}
	\end{table}
	
	
	\section{Conclusions}\label{sec:concl}
	
	This paper findings consistently demonstrate that coaching strategies exert a measurable and economically meaningful impact on match outcomes. Across model-averaging approaches and robustness checks, aggressive offensive tactics (particularly those deployed early in the match) are associated with improved performance, both at home and away. This supports the strategic value of proactive play. 
	The persistence of negative effects from crosses reinforces the idea that not all attacking actions are equally efficient, highlighting the need for tactical precision and data-informed decision-making. Meanwhile, goal-kicks and shots emerge as reliable sources of advantage, underscoring their role in effective transition play.
	Referee decisions (especially red cards and penalties) remain powerful external shocks, analogous to exogenous interventions in economic models. Their continued significance across models highlights the importance of risk management in coaching strategies. The relative insignificance of free-kicks and the selective significance of yellow cards point to the situational nature of tactical risks and rewards.
	Furthermore, the consistent appearance of extreme-score controls and team-fixed effects emphasizes the role of persistent team-level factors, such as organizational quality, institutional knowledge, or managerial competence, in shaping tactical effectiveness.
	Overall, our results depict coaching decisions as strategic, data-responsive actions aimed at maximizing outcomes under uncertainty. By integrating rich in-game data with modern statistical methods and economic reasoning, we offer a robust framework for understanding soccer as a domain of rational outcome-optimized decision-making, comparable to strategic management in competitive markets.
	
	Our results demonstrate that coaching strategies and tactical decisions have a persistent, significant impact on match outcomes, even when controlling for a wide range of contextual, referee, and team-level variables. Through exhaustive model averaging and subset selection, we recover a compact and stable set of strategic patterns, offering both statistical and economic insight.
	Home advantage (HA) remains robust across specifications: the BIC-optimal intercept in Model 1 is strongly positive, and Model 3 confirms the home advantage, even after controlling for tactics, referee behavior, and contextual factors. This underscores the structural relevance of HA as a latent, outcome-driving factor.
	Aggressive opening formations consistently improve team performance, regardless of whether models adjust for the team’s end-match tactics, separate home- and away-side offensiveness indices, or re-weight episodes by on-ball intensity. The stability of this coefficient across specifications strengthens the economic interpretation of these choices as early, high-return strategic investments under uncertainty.
	Technical actions follow consistent behavioral patterns: the "cross paradox" appears across all averaged models, while shots, goal-kicks, and offsides remain reliable predictors, reinforcing the internal consistency of the estimates. Referee decisions, especially red cards, and penalties, continue to operate as exogenous shocks with large, significant impacts. Meanwhile, yellow cards are weakly negative, and free-kicks are rarely selected, likely due to contextual ambiguity or informational gaps in historical data.
	Interestingly, stadium occupancy (cubic fill) fails to reach significance, suggesting that the quality or composition of fan support may be more influential than raw attendance, an aspect not directly captured in this dataset, and a promising area for future research.
	Finally, the study’s main methodological contribution lies in its use of time-match commentary to construct panel data of tactical episodes. This approach achieves substantial granularity at a lower computational cost than video tracking data, offering a scalable and efficient framework for analyzing dynamic in-game strategies. As such, the study not only clarifies the economics of coaching decisions but also opens up new avenues for cost-effective tactical analysis in sports economics.
	
	The control for the endogeneity of coach selection is an important point that may be considered in future investigations. Coaches are not randomly assigned to teams; instead, their recruiting is often influenced by factors such as team performance expectations, management strategies, and historical success. This endogenous relationship may bias the estimation of coaching effects if not carefully accounted for. Future studies could incorporate instrumental variables or panel data methods to address this issue. Indeed, coaching strategies and tactics are dynamic and subject to periodic change, driven by both internal factors (e.g., injuries, player availability) and external conditions (e.g., opponent strength, match context). These shifts may influence the interpretation of coaching effects over time. A dynamic approach, capturing these temporal variations, could provide deeper insights into how tactical adjustments evolve and affect match outcomes.
	
	Lastly, careful evaluation of potential confounders is essential to enhance the robustness of future studies. Addressing these limitations will provide a more comprehensive understanding of the dynamics influencing match outcomes.

	
	\section*{Financial disclosure}
	None reported.

	\section*{Conflict of interest}
	The authors declare no potential conflict of interests.
	
	\appendix
	\phantomsection
	\section*{Appendix}

	\refstepcounter{section}      
	\setcounter{subsection}{0}
	\renewcommand\thesubsection{\Alph{section}.\arabic{subsection}}
	
	\makeatletter
	\renewcommand\theHsubsection{\Alph{section}.\arabic{subsection}}
	\makeatother
	
	\subsection{Offensiveness index}\label{App:OffInd}
	The offensiveness index is separately computed for the home team, denoted as $s_{i,t,1}^{H}$, and the away team, $s_{i,t,1}^{A}$, at a given time $t$ in match $i$. The home-away differential in this index is represented as $s_{i,t,1}$. This measure remains invariant under red card occurrences and is only subject to modifications when there is a change in the team's formation. 
	A secondary index, $s_{i,t,2}$, also captures the offensiveness differential between teams. However, unlike $s_{i,t,1}$, it dynamically adjusts to account for red card incidents during match $i$.
	A third index, $s_{i,t,3}$, is derived as a transformation of the previous measure. Specifically, for the home team:
	\begin{equation}
		s_{i,t,3}^{H} \equiv \frac{s_{i,t,2}^{H}}{\text{Active players at } t} \times \frac{10}{30} \in \left[0, 1\right], \ i=\{1, \dots, N\}, \ t=\{1, \dots, 90\}.
	\end{equation}
	Here, the number of active players at time $t$ is typically 10 but may decrease due to red card penalties. The denominator, 30, represents the upper bound, which assumes all outfield players adopt an attacking role. An analogous formulation holds for the away team. 
	The home-away differential of this transformed offensiveness index, denoted as $s_{i,t,3} = s_{i,t,3}^H-s_{i,t,3}^A$, is computed at each minute $t$ within match $i$. 
	The key time points of interest are the initial formation at $t=0$ and the final configuration at $t=90$.
	
	\subsection{Measures of association}\label{sec:5_ASSOC} 
\begin{figure}[!ht]
		\captionsetup{size=small} 
		\centering
		\hspace*{0.5cm}
		\includegraphics[width=0.75\linewidth]{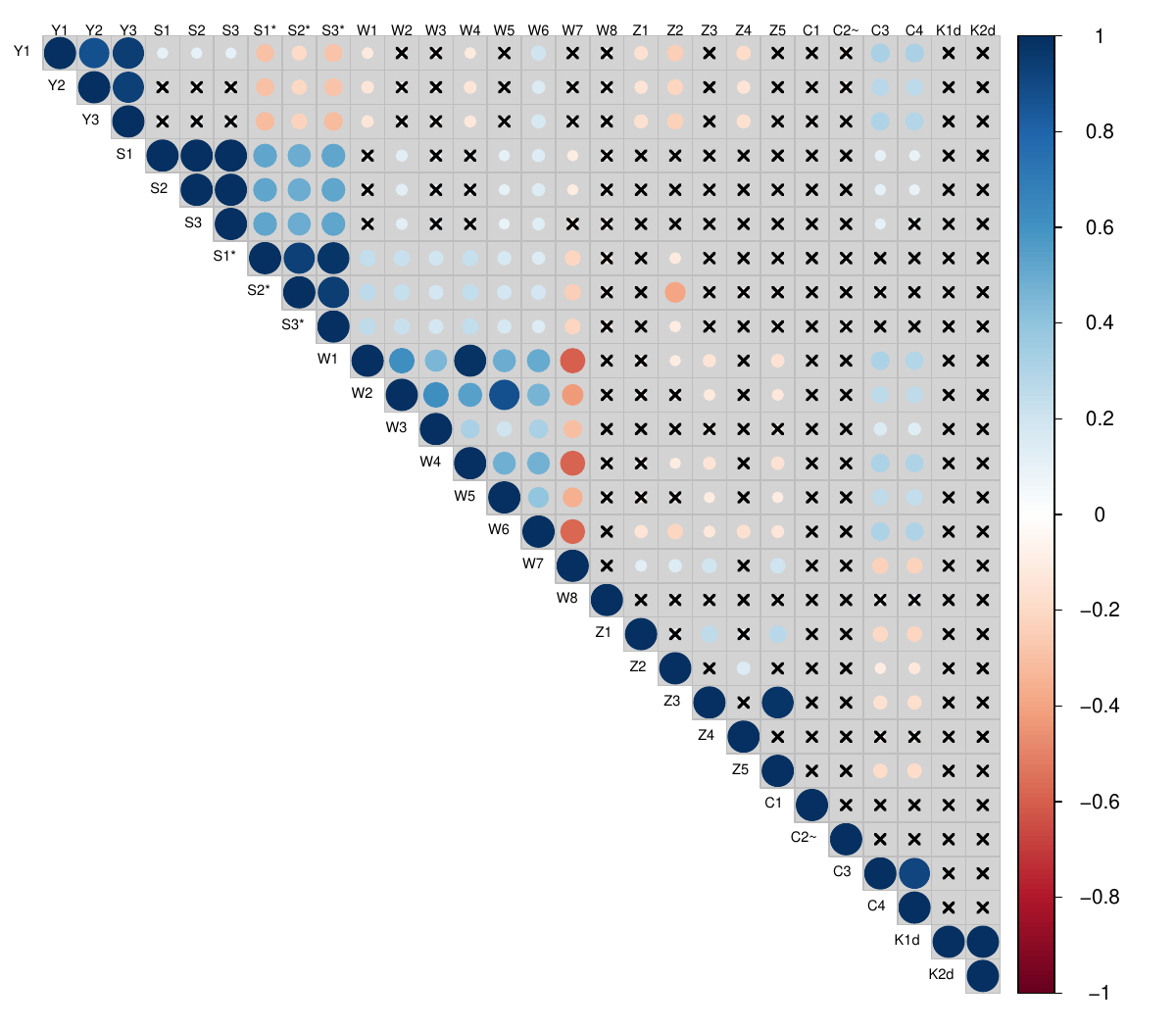}
		\caption{Spearman's rank correlation matrix, significance level at $0.1\%$ level. The vertical bar on the right side of the graph represents correlations ranging from $-1$ (deep red) to $+1$ (dark blue). The size of the circle reflects the strength of the correlation. For instance, darker circles are larger than their lighter counterparts, which are smaller in size. The $\pmb{\times}$ symbol denotes a correlation with a $p$-value greater than $0.1\%$.}
		\label{fig_cor_cs_spear}
	\end{figure}

	In Table~\ref{tab:codebook}, we analysed the dataset of $1,139$ observations and $146$ variables.\footnote{One observation was excluded due to missing values linked to a penalized match in the $2012/2013$ season involving Cagliari vs Roma. This match was canceled due to public order concerns. Subsequently, a score of $0$ to $3$ was awarded in favor of Roma, as stated by Article $17$ of Federcalcio.} 
	Figure~\ref{fig_cor_cs_spear} shows the Spearman's rank correlation coefficients and significance. The response variable \(y^{(1)}\) indicates a positive correlation (\(p\)-value \(< 0.1\%\)) with the initial schemes \(\bigl(s_1, s_2, s_3\bigr)\), the number of shots \(\bigl(w_6\bigr)\), and the discrete and continuous team effects \(\bigl(c_{i,3}, c_{i,4}\bigr)\). In contrast, \(y^{(1)}\) exhibits a negative relationship with the final schemes \(\bigl(s_1^{*}, s_2^{*}, s_3^{*}\bigr)\), the counts of crosses and crosses not originating from corners \(\bigl(w_1, w_4\bigr)\), and both yellow and red cards and penalty kicks \(\bigl(z_1, z_2, z_4\bigr)\). The remaining dependent variables, \(y^{(2)}\) and \(y^{(3)}\), generally align with these trends and levels of significance, except that the initial schemes do not reach significance at the \(0.1\%\) threshold.
	Pearson's correlation for \(y^{(1)}\) reveals significant negative relationships with final schemes, yellow and red cards, and penalty kicks. In contrast, \(y^{(1)}\) exhibits a positive correlation with the initial scheme \(\bigl(s_2\bigr)\), the number of shots, and the discrete and continuous team effects.
	The dependent variables \(y^{(2)}\) and \(y^{(3)}\) do not demonstrate significant correlation with the initial scheme. 
	However, in addition to mirroring the significant correlations observed for \(y^{(1)}\), \(y^{(2)}\) and \(y^{(3)}\) both show a negative correlation with the counts of crosses and crosses not derived from corners. Furthermore, \(y^{(3)}\) displays a slight positive correlation with the number of offsides.
	Using the \(\chi^2\) test on the binary response \(y^{(2)}\) and the considered covariates, the null hypothesis is rejected for final schemes, crosses, crosses not originating from corners, shots, offsides, yellow and red cards, and penalty shots.\footnote{For more details, see Table~\ref{tab:SM_cor}, SM.}

	\subsection{Bootstrap confidence intervals}\label{sec:bca}
	
	We extend our investigation of coaching strategies beyond the use of HC estimators by incorporating automated bootstrap standard errors for the parameters in linear, logistic, and ordered logistic regressions, as detailed in \citet{EFR20}. This approach generalizes the Bias-Corrected and Accelerated (BCa) bootstrap \citep{EFR87} by employing an innovative acceleration factor that simplifies computations.\footnote{Providing second-order adjustments to address both the non-normality of the estimator and its inherent bias.}
	
\begin{figure}[!h]
	\captionsetup{size=small} 
	\centering
	
	\begin{subfigure}[b]{0.32\linewidth}
		\includegraphics[width=\linewidth]{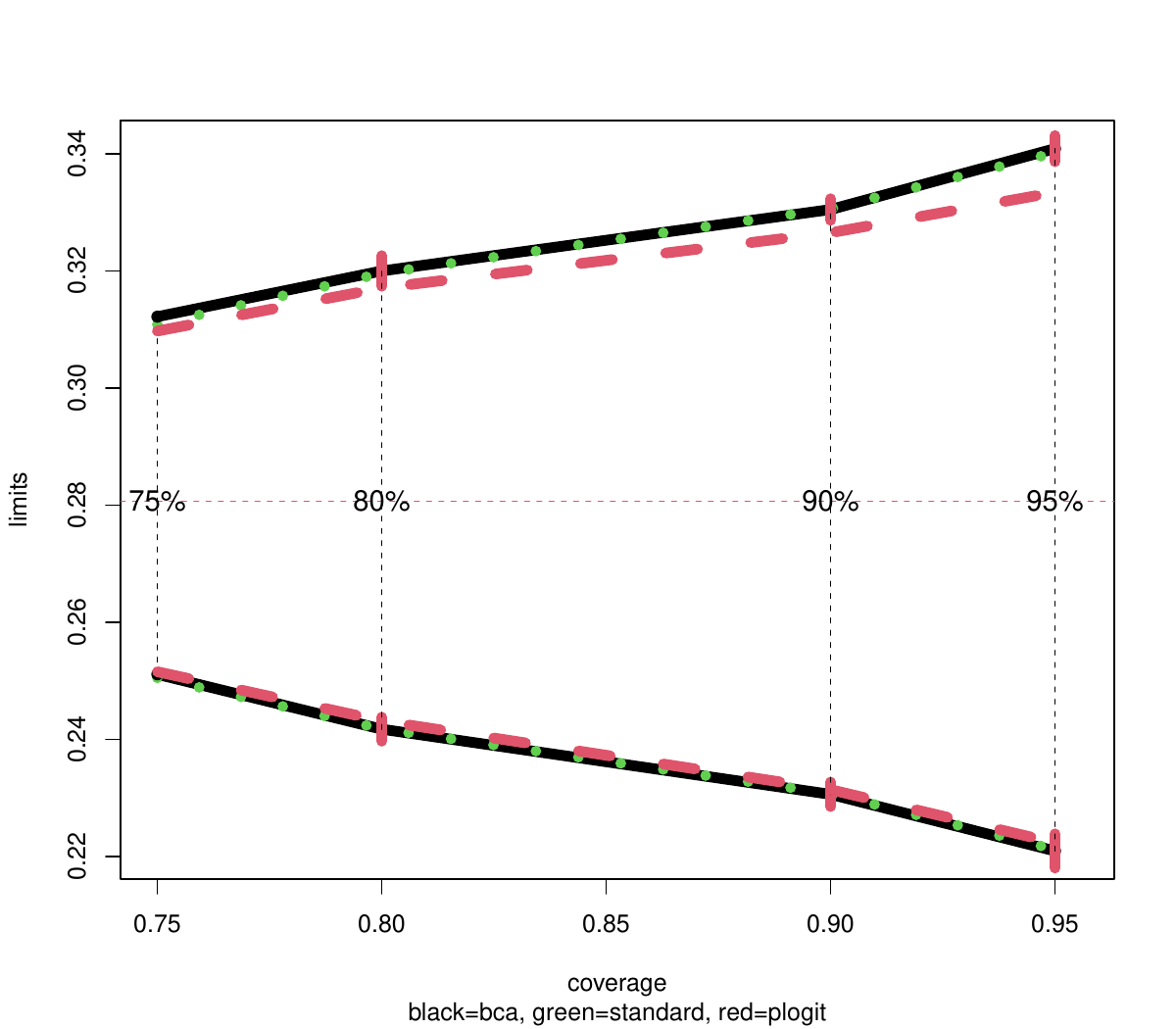}
		\caption{Model 1}
		\label{fig:compare_AIC1}
	\end{subfigure}
	\hfill
	\begin{subfigure}[b]{0.32\linewidth}
		\includegraphics[width=\linewidth]{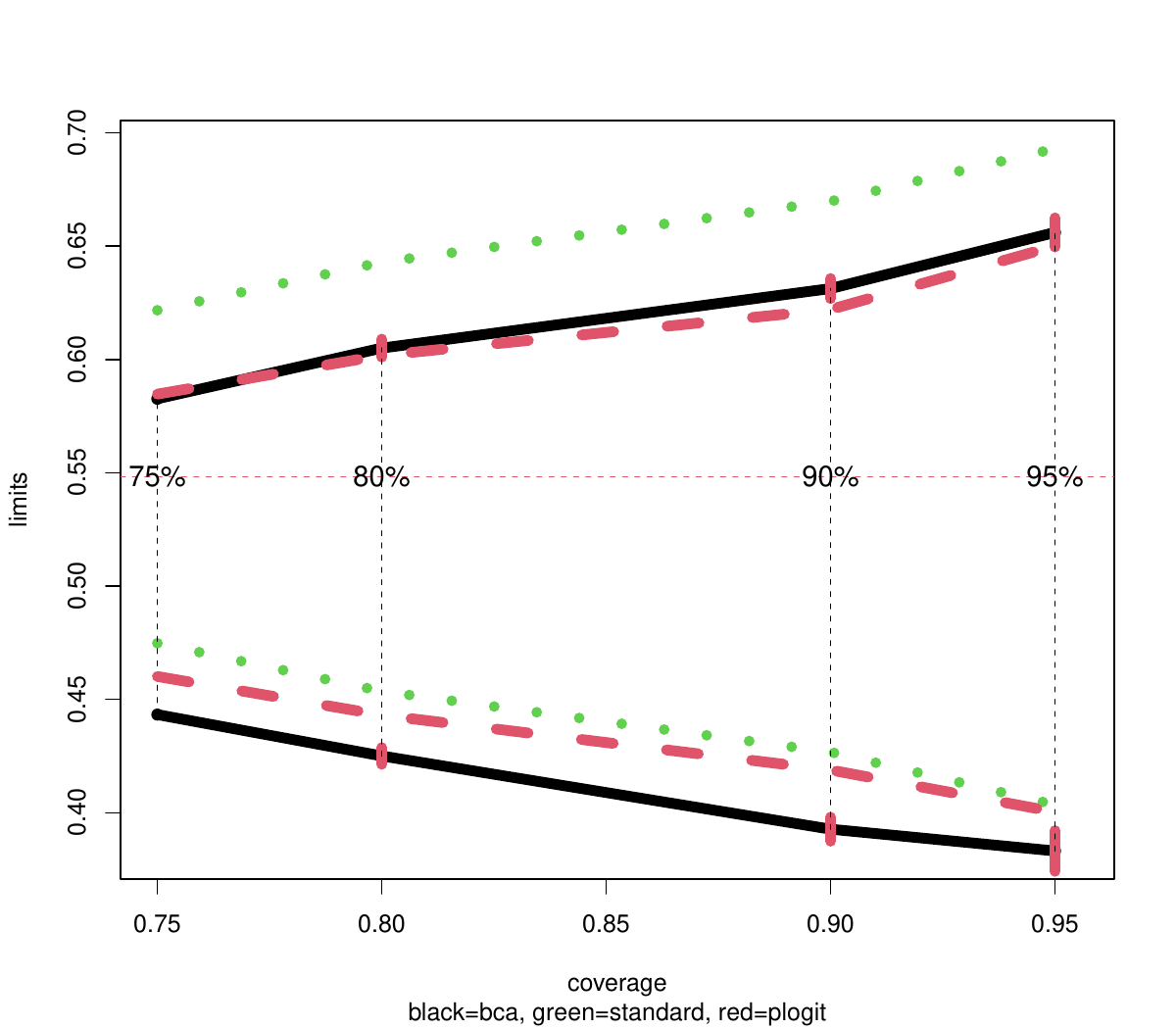}
		\caption{Model 2}
		\label{fig:compare_AIC2}
	\end{subfigure}
	\hfill
	\begin{subfigure}[b]{0.32\linewidth}
		\includegraphics[width=\linewidth]{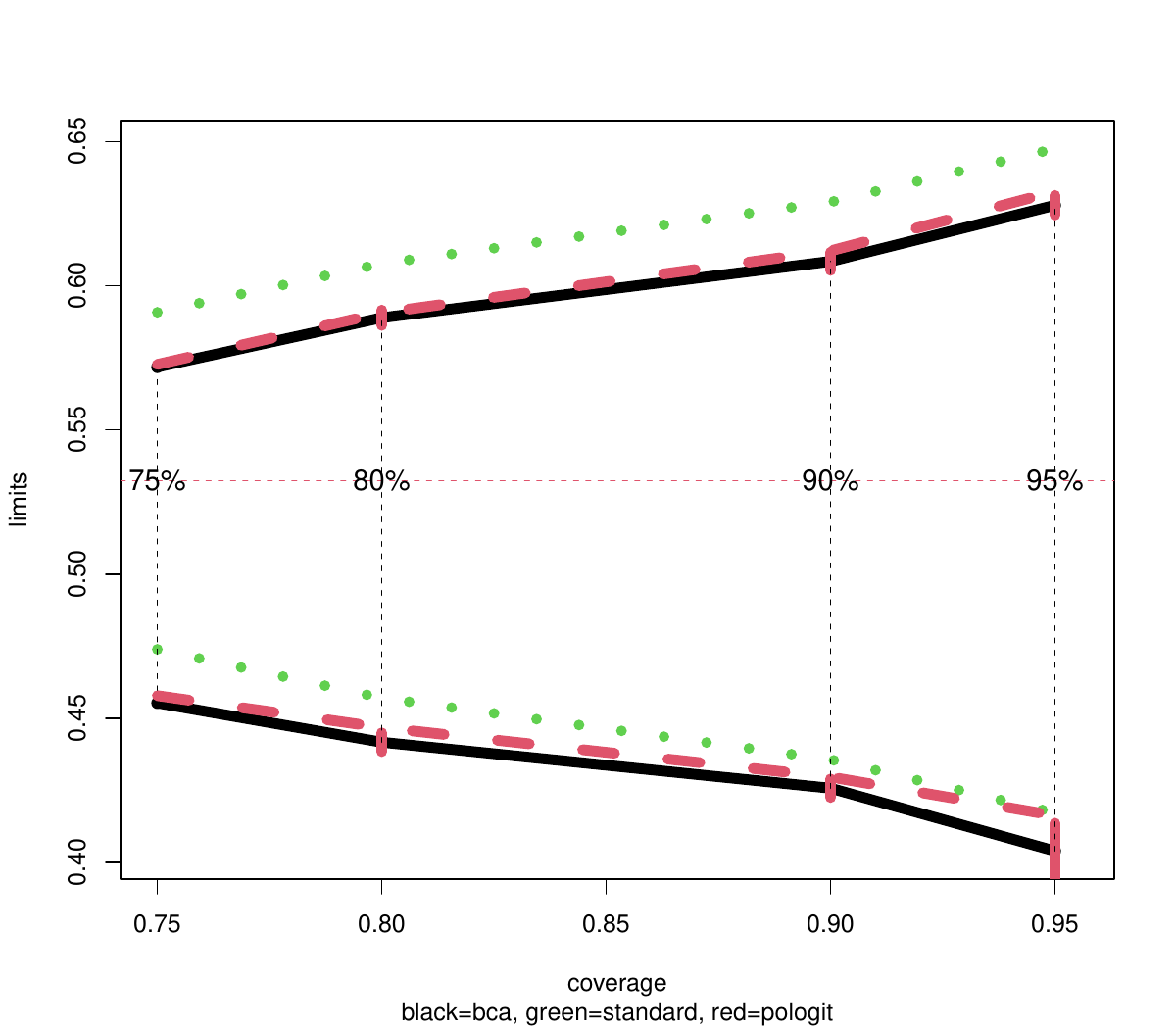}
		\caption{Model 3}
		\label{fig:compare_AIC3}
	\end{subfigure}
	\caption{Comparison of CIs for best-AIC ranked Models 1, 2, and 3 using asymptotic theory (green dots), BCa bootstrap (black line), and BCa for penalized regressions (red dashed), based on the best AIC-ranked specifications, for the initial scheme $s_2$ parameter.}
	\label{fig:compare_plog}
	\end{figure}

	\begin{table}[!h]
		\centering
		\resizebox{0.5\textwidth}{!}{
			\begin{tabular}{clcccc}
				\toprule
				&& \multicolumn{2}{c}{\textbf{AIC}} & \multicolumn{2}{c}{\textbf{BIC}} \\
				\cmidrule(lr){3-4} \cmidrule(lr){5-6}
				\textbf{Response} & \textbf{Estimator ($s_2$)} & $\mathrm{CI}_{2.5}$ & $\mathrm{CI}_{97.5}$ & $\mathrm{CI}_{2.5}$ & $\mathrm{CI}_{97.5}$\\
				\midrule
				\textbf{$y^{(1)}$} & Classical & 0.2213 & 0.3401 & 0.2367 & 0.3506 \\
				& BCa & 0.2210 & 0.3409 & 0.2327 & 0.3488 \\
				& BCa $\mathrm{L}_1/\mathrm{L}_2$ & 0.2224 & 0.3334 & 0.2169 & 0.3333 \\
				& Model Avg. & 0.2299 & 0.3692 & 0.2032 & 0.3575\\ 
				\midrule
				\textbf{$y^{(2)}$} & Classical & 0.4164 & 0.6847 & 0.4055 & 0.6493 \\
				& BCa & 0.3831 & 0.6561 & 0.4003 & 0.6436 \\
				& BCa $\mathrm{L}_1/\mathrm{L}_2$ & 0.4003 & 0.6501 & 0.4028 & 0.6441 \\
				& Model Avg. & 0.2981 & 0.7424 & 0.0917 & 0.7625 \\ 
				\midrule
				\textbf{$y^{(3)}$} & Classical & 0.4251 & 0.6422 & 0.4086 & 0.6100\\
				& BCa & 0.4040 & 0.6279 & 0.4102 & 0.6032 \\
				& BCa $\mathrm{L}_1/\mathrm{L}_2$ & 0.4163 & 0.6327 & 0.4116 & 0.6055 \\
				& Model Avg. & 0.0649 & 0.8116 & 0.1543 & 0.9110 \\ 
				\bottomrule
			\end{tabular}
		}
		\caption{Asymptotic, BCa bootstrap, and model averaging CIs for different estimators at the 95\% level for the initial scheme \(s_2\), based on the best AIC and BIC-ranked specifications. All bootstrap CIs are computed using \(B=2000\) resamples. $\mathrm{L}_1/\mathrm{L}_2$ = regularized/penalized estimation. Avg.=averaging. Model Avg. for Model 3 uses $B=1000$ repetitions. See Figure~\ref{fig:bca_dist_combined}, SM, for the distribution of BCa replicates.}
		\label{tab:boot1}
	\end{table}
	Table~\ref{tab:boot1} offers a subset of our findings, emphasizing the 95\% asymptotic and BCa confidence intervals (CIs) for the initial scheme, \(s_2\), across the top AIC and BIC specifications in each of the three models. All bootstrap CIs are based on \(B=2000\) resamples. The results align with those obtained via asymptotic theory and HC standard errors (in Model~1 and~2), as well as bootstrapped standard errors (in Model~3). In particular, for the best-AIC specification of Model 1, a one-unit increment in the disparity of initial schemes between home and away teams, \(s_2\), is linked to a marginal effect at the mean ranging from 9.44\% to 16.17\%, and an average marginal effect varying from 5.69\% to 9.75\%, on home win -- at the 95\% confidence level, conditional on other in-game actions.
	Figure~\ref{fig:compare_plog} provides an additional comparison by illustrating BCa confidence intervals derived from a classical asymptotic theory, BCa bootstrap, and BCa bootstrap using penalized Gaussian regression for Model~1, a penalized logistic regression for Model~2 (with regularization via \(k\)-fold cross-validation), and an elastic net regularization \citep{WUR21} for Model~3. The shrinkage procedure described by \citet{EFR20} is designed to reduce the width of the confidence interval, which remains tighter than its counterpart obtained via BCa under Gaussian and logistic link functions. In all cases, the estimated intervals are narrower relative to the classical and BCa methods, underscoring the possibility of bias, particularly in Models~2 and~3. For initial schemes \(s_2\), all parameter estimates indicate a positive effect; similarly, for the final scheme estimates suggest a negative effect. 
	There is no conflicting evidence among classical, BCa, regularized BCa, or model averaging methods. These results collectively suggest significant dynamic influences of schemes on home outcomes, measured by goal difference and winning probability.
	
	\begin{landscape}
		
		\subsection{Tables}
		
		\noindent\vspace*{2.5cm}\begin{table}[!ph]
			\centering
			\resizebox{1.5\textwidth}{!}{
			\begin{tabular}{@{}llllll@{}}
				\toprule
				\textbf{Reference} & \textbf{Main focus} & \textbf{Method} & \textbf{League} & \textbf{No. Matches} 
				& \textbf{Season(s)}\\ 
				\midrule
				\citet{MUE18} & Strategies/tactics & OLS with fixed effects & Bundesliga & 6,426 
				&1993/94-2013/14 \\
				\citet{MES20} & Strategies/tactics  & Bayesian multi-level regression & European leagues & 9,127
				& 2012/13-2016/17 \\
				\cite{goes2021mod}& Strategies/tactics  & 4 machine learning classifiers & Eredivisie & 118
				& n/a \\
				\citet{lopez2022}& Strategies/tactics  &  Pearson's correlation and PCA & LaLiga  & 760 
				&2017/18-2018/19\\
				\citet{lorenzo2022}& Strategies/tactics  & Linear mixed model  & Bundesliga & 234
				& 2016/17 \\
				\citet{bauer2023}& Strategies/tactics  & Convolutional neural network & Bundesliga  & 2,142 
				& 2013/14-2019/20  \\
				\citet{SCO08} & HA/HB effect & OLS regression  & Serie A  & 686  
				& 2003/04-2004/05 \\
				\citet{PET10} & HA/HB effect& Natural experiment & Serie A-B &  842  
				&  2006/07\\
				\citet{DEW13} & HA/HB effect& Conditional FE Negative Binomial Regressions & Bundesliga & 8,029  
				& 1982/83-2007/08 \\ 
				\citet{HLA17} & HA/HB effect& OLS and Fixed-Effects regressions  &English Football Leagues  & 34,692  
				& 2000/01-2013/14 \\
				\citet{PIC17} & HA/HB effect&  Random effects panel-data regressions  & LaLiga & 2,651  
				& 2002/03-2009/10 \\
				\citet{PON18} & HA/HB effect& Ordered Probit & Serie A & 7,398  
				& 1991/92-2012/13\\
				\citet{FIS21} & HA/HB effect& OLS regression & Bundesliga 1,2,3&2,976  
				& 2017/18-2019/20\\
				\citet{HOL21} & HA/HB effect& OLS regression & Bundesliga, Serie A & 2,744 
				& 2015/16-2018/19\\
				\citet{fischer2022} & HA/HB effect& Difference-in-differences & Bundesliga 1-2 &6,120  
				& 2011/12-2020/21\\
				\citet{WUN21} & HA/HB effect& Linear mixed regressions models & 10  European professional leagues &37,888  
				& 2010/11-2019/20\\
				\citet{PRI22} & HA/HB effect & Hierarchical causal model & Premier League & 380 
				& 2020/21 \\
				\citet{REA22} & HA/HB effect&  Poisson regression &  UEFA and European top leagues &51,333 
				& 2011/12-2020/21\\
				\citet{BEN23} & HA/HB effect& Poisson regression & 17 European leagues  &25,327 
				&2015/16-2020/21\\
				This paper & HA/HB effect + Strategies/tactics & OLS, Logit, Ordered logit regressions + Model Avg& Serie A  & 1,140 
				&2011/12-2013/14\\
				\bottomrule
			\end{tabular}
		}
			\caption{Main studies on coaching, strategy and HA effects in soccer teams' performances.} 
			\label{tab:selected_lit}
		\end{table}
	\end{landscape}	
	
	\begin{table}
		\footnotesize
		\centering
		\captionsetup{size=small} 
		\resizebox{0.85\textwidth}{!}{
			\begin{tabular}{l l p{8cm}}
				\toprule
				\multicolumn{2}{l}{\textbf{Variables 
				}} & \textbf{Description}\\
				\toprule
				Response&&\\
				\cmidrule(lr){1-2} \cmidrule(lr){3-3}
				$y_{i}^{(1)}$ &  & Goal difference (home - away)\\
				$y_{i}^{(2)}$ &  & Binary: 1 if home-team wins, 0 otherwise\\
				$y_{i}^{(3)}$ &  & Points gained by home-team: 3 (win), 1 (draw), 0 (lose)\\
				\cmidrule(lr){1-2} \cmidrule(lr){3-3}
				Coach actions/schemes &&\\
				\cmidrule(lr){1-2} \cmidrule(lr){3-3}
				$s_{i,1}$, $s_{i,1}^{*}$& & Offensiveness index, type (1), see Appendix~\ref{App:OffInd} \\
				$s_{i,2}$, $s_{i,2}^{*}$& & Offensiveness index, type (2), see Appendix~\ref{App:OffInd} \\
				$s_{i,3}$, $s_{i,3}^{*}$& & Offensiveness index, type (3), see Appendix~\ref{App:OffInd} \\
				\addlinespace[1pt]
				\multicolumn{3}{l}{The index is computed at the beginning (no superscript) and at the end (asterisk) of the match.} \\
				\cmidrule(lr){1-2} \cmidrule(lr){3-3}
				\cmidrule(lr){1-2} \cmidrule(lr){3-3}
				Team actions &&\\
				\cmidrule(lr){1-2} \cmidrule(lr){3-3}
				$w_{i,1}$ & & Total number of crosses \\
				$w_{i,2}$ & & Total number of corners \\
				$w_{i,3}$ & & Number of crosses from corners\\
				$w_{i,4}$ & & Number of crosses not from corners\\
				$w_{i,5}$ & & Number of corners without crosses\\
				$w_{i,6}$ & & Number of shots  \\
				$w_{i,7}$ & & Number of goal kicks \\
				$w_{i,8}$ & & Number of offsides \\
				\cmidrule(lr){1-2} \cmidrule(lr){3-3}
				Referee actions &&\\
				\cmidrule(lr){1-2} \cmidrule(lr){3-3}	
				$z_{i,1}$ & & Difference in yellow cards (home - away) \\
				$z_{i,2}$ & & Difference in red cards (home - away) \\
				$z_{i,3}$ & & Difference in free kicks (home - away) \\
				$z_{i,4}$ & & Difference in penalty kicks (home - away) \\
				$z_{i,5}$ & & Difference in total fouls (home - away) \\
				\cmidrule(lr){1-2} \cmidrule(lr){3-3}
				Control Variables &&\\
				\cmidrule(lr){1-2} \cmidrule(lr){3-3}
				$c_{i,1}$ & & Home stadium filling index (proportion of occupied seats) \\
				$c_{i,2}$ & & Extra time (min.) at the end of the match \\
				$c_{i,3}$ & & Difference of ranking points (home - away) before the match (absolute)\\
				$c_{i,4}$ & & As $c_{i,3}$ but normalized to lie in $[-1,1]$\\
				\cmidrule(lr){1-2} \cmidrule(lr){3-3}
				Fixed Effects &&\\
				\cmidrule(lr){1-2} \cmidrule(lr){3-3}
				\multicolumn{2}{l}{$k_{i}^{(11)}, k_{i}^{(12)}, k_{i}^{(13)}$} & Season dummies (2011, 2012, 2013)\\
				\multicolumn{2}{l}{$k_{i,1}^{d}$}  & League day effect (rank)\\
				\multicolumn{2}{l}{$k_{i,2}^{d}$}  & League day effect (continuous $[0,1]$)\\
				\multicolumn{2}{l}{$k_{i,1}^{e}$}  & Extreme-event dummy (absolute difference of more than 4 goals)\\
				\multicolumn{2}{l}{$k_{i,2}^{e}$}  & Extreme-event dummy (positive difference of more than 4 goals)\\
				\multicolumn{2}{l}{$k_{i,3}^{e}$}  & Extreme-event dummy (negative difference of more than 4 goals)\\
				\multicolumn{2}{l}{$\bk_{i}^{T}$}    & Team effect dummy \\
				\bottomrule
			\end{tabular}
		}
		\caption{Concise variables' codebook.}\label{tab:codebook}
	\end{table} 
	
	\begin{table}
		\centering
		\captionsetup{size=small} 
		\small 
		\setlength{\tabcolsep}{4pt} 
		
		\begin{minipage}{\textwidth}
			\centering
			\begin{tabular}{lcccccccccc}
				\toprule
				& Min. & Q1 & Median & Mean & Q3 & Max. & SD & CV & Skew. & Kurt. \\
				\midrule
				$y^{(1)}$ & -7.00 & -1.00 & 0.00 & 0.38 & 1.25 & 6.00 & 1.67 & 4.35 & -0.09 (0.19) & 3.25 (0.10) \\
				\bottomrule
			\end{tabular}
		\end{minipage}
		
		\vspace{0.5cm} 
		
		\begin{minipage}{0.45\textwidth}
			\centering
			\begin{tabular}{ccc}
				\toprule
				$y^{(2)}$ & $n_k$  & $f_k$ \\
				\midrule
				0 & 609 & 53.42 \\
				1 & 531 & 46.58 \\
				\bottomrule
			\end{tabular}
		\end{minipage}%
		\hfill
		\begin{minipage}{0.45\textwidth}
			\centering
			\begin{tabular}{ccc}
				\toprule
				$y^{(3)}$ & $n_k$ & $f_k$ \\
				\midrule
				0 & 311 & 27.28\\
				1 & 298 & 26.14\\
				3 & 531 & 46.58\\
				\bottomrule
			\end{tabular}
		\end{minipage}
		\caption{Summary statistics for the response variables. Upper table: $y^{(1)}$ goal differences. The skewness ($p=0.19$) and kurtosis ($p=0.10$) tests assess the null hypothesis of Gaussianity.  Lower-left table: $y^{(2)}$ home-team match outcome (win=1, loss/tie=0). Bottom-right table: $y^{(3)}$ points gained by the home team (3 = win, 1 = draw, 0 = loss)). $n_k, f_k$ indicate absolute frequencies and  percentages, respectively. Min. = minimum, Q1 = first quartile, Q3 = third quartile, Max. = maximum, SD = standard deviation, CV = coefficient of variation, Skew. = skewness test, Kurt. = kurtosis test.}\label{tab:dependent}
	\end{table}
	
	\begin{table}
		\centering
		\captionsetup{size=small} 
		\resizebox{0.75\textwidth}{!}{
			\begin{tabular}{lcccccccccccc}
				& \multicolumn{3}{c}{Home} & \multicolumn{3}{c}{Away} & \multicolumn{3}{c}{H-A} & \multicolumn{3}{c}{H-A, weighted}\\
				\cmidrule(lr){2-4} \cmidrule(lr){5-7} \cmidrule(lr){8-10} \cmidrule(lr){11-13}
				& $s_{1}^{H}$ & $s_{2}^{H}$ & $s_{3}^{H}$ 
				& $s_{1}^{A}$ & $s_{2}^{A}$ & $s_{3}^{A}$ 
				& $s_{1}$ & $s_{2}$ & $s_{3}$
				& $\tilde{s}_{1}$ & $\tilde{s}_{2}$ & $\tilde{s}_{3}$\\
				\cmidrule(lr){2-4} \cmidrule(lr){5-7} \cmidrule(lr){8-10} \cmidrule(lr){11-13}
				Min. & 14.00 & 14.00 & 0.47 & 15.00 & 15.00 & 0.50 & -5.00 & -5.00 & -0.17 & -5.03 & -5.03 & -0.17 \\
				Q1 & 17.00 & 17.00 & 0.57 & 17.00 & 17.00 & 0.57 & -1.00 & -1.00 & -0.03 & -1.01 & -1.01 & -0.03 \\
				Median & 18.00 & 18.00 & 0.60 & 18.00 & 18.00 & 0.60 & 0.00 & 0.00 & 0.00 & 0.00 & 0.00 & 0.00\\
				Mean & 18.06 & 18.06 & 0.60 & 18.05 & 18.04 & 0.60 & 0.02 & 0.01 & 0.00 & 0.02 & 0.01 & 0.00\\
				Q3 & 19.00 & 19.00 & 0.63 & 19.00 & 19.00 & 0.63 & 1.00 & 1.00 & 0.03 & 1.01 & 1.01 & 0.03\\
				Max. & 21.00 & 21.00 & 0.70 & 21.00 & 21.00 & 0.70 & 5.00 & 5.00 & 0.17 & 5.10 & 5.10 & 0.17 \\
				\addlinespace
				SD & 1.09 & 1.09 & 0.04 & 1.12 & 1.12 & 0.04 & 1.58 & 1.58 & 0.05 & 1.59 & 1.60 & 0.05 \\
				CV & 0.06 & 0.06 & 0.06 & 0.06 & 0.06 & 0.06 & 100 & 120 & 101 & 101 & 121 & 101 \\
				\addlinespace
				Skew. & -0.53 & -0.53 & -0.53 & -0.45 & -0.45 & -0.45 & -0.05 & -0.05 & -0.05 & -0.05 & -0.05 & -0.05 \\
				$p$-value & 0.00 & 0.00 & 0.00 & 0.00 & 0.00 & 0.00 & 0.52 & 0.52 & 0.52 & 0.52 & 0.51 & 0.52\\
				Kurt. & 2.36 & 2.35 & 2.36 & 2.14 & 2.13 & 2.14 & 2.65 & 2.64 & 2.66 & 2.66 & 2.65 & 2.66 \\
				$p$-value & 0.00 & 0.00 & 0.00 & 0.00 & 0.00 & 0.00 & 0.01 & 0.00 & 0.01 & 0.01 & 0.00 & 0.01\\
				\cmidrule(lr){2-4} \cmidrule(lr){5-7} \cmidrule(lr){8-10} \cmidrule(lr){11-13}
			\end{tabular}
		}
		\caption{Summary statistics for initial schemes ($\bs_i$). Home and away (columns 1--6), and home-away (H-A) difference  (columns 7--12). $p$-values for skewness and kurtosis tests (null hypothesis of Gaussian distribution) are indicated below the test statistic. See the caption of Table~\ref{tab:dependent} for the legend.}
		\label{tab:schemesI}
	\end{table}
	
	\begin{table}
		\centering
		\captionsetup{size=small} 
		\small
		\resizebox{0.75\textwidth}{!}{
			\begin{tabular}{lcccccccccccc}
				& \multicolumn{3}{c}{Home} & \multicolumn{3}{c}{Away} & \multicolumn{3}{c}{H-A} & \multicolumn{3}{c}{H-A, weighted}\\
				\cmidrule(lr){2-4} \cmidrule(lr){5-7} \cmidrule(lr){8-10} \cmidrule(lr){11-13}
				& $s_{1}^{*H}$ & $s_{2}^{*H}$ & $s_{3}^{*H}$ 
				& $s_{1}^{*A}$ & $s_{2}^{*A}$ & $s_{3}^{*A}$ 
				& $s_{1}^{*}$ & $s_{2}^{*}$ & $s_{3}^{*}$
				& $\tilde{s}_{1}^{*}$ & $\tilde{s}_{2}^{*}$ & $\tilde{s}_{3}^{*}$\\
				\cmidrule(lr){2-4} \cmidrule(lr){5-7} \cmidrule(lr){8-10} \cmidrule(lr){11-13}
				Min. & 13.00 & 12.00 & 0.43 & 13.00 & 9.00 & 0.43 & -7.00 & -8.00 & -0.32 & -7.32 & -8.41 & -0.34\\
				Q1 & 17.00 & 17.00 & 0.57 & 17.00 & 17.00 & 0.57 & -2.00 & -2.00 & -0.07 & -2.06 & -2.07 & -0.07\\
				Median & 18.00 & 18.00 & 0.60 & 18.00 & 18.00 & 0.60 & 0.00 & 0.00 & 0.00 & 0.00 & 0.00 & 0.00\\
				Mean & 18.15 & 17.94 & 0.61 & 18.24 & 17.99 & 0.61 & -0.08 & -0.05 & 0.00 & -0.09 & -0.05 & 0.00\\
				Q3 & 19.00 & 19.00 & 0.63 & 19.00 & 19.00 & 0.63 & 2.00 & 2.00 & 0.05 & 2.03 & 2.06 & 0.05\\
				Max. & 24.00 & 24.00 & 0.80 & 23.00 & 23.00 & 0.81 & 8.00 & 9.00 & 0.27 & 8.22 & 9.19 & 0.27\\
				\addlinespace
				SD & 1.71 & 1.82 & 0.06 & 1.67 & 1.84 & 0.06 & 2.54 & 2.74 & 0.09 & 2.64 & 2.86 & 0.09\\
				CV & 0.09 & 0.10 & 0.10 & 0.09 & 0.10 & 0.09 & 30.1 & 54.9 & 21.4 & 30.2 & 54.1 & 21.4\\
				\addlinespace
				Skew. & -0.15 & -0.22 & -0.08 & -0.09 & -0.39 & 0.00 & 0.04 & 0.07 & -0.03 & 0.04 & 0.06 & -0.03\\
				$p$-value & 0.04 & 0.00 & 0.28 & 0.23 & 0.00 & 0.97 & 0.59 & 0.33 & 0.73 & 0.62 & 0.38 & 0.69\\
				Kurt. & 3.13 & 3.17 & 3.08 & 3.03 & 3.66 & 3.13 & 3.15 & 3.15 & 3.26 & 3.15 & 3.14 & 3.26\\
				$p$-value & 0.35 & 0.24 & 0.53 & 0.73 & 0.00 & 0.33 & 0.27 & 0.29 & 0.08 & 0.28 & 0.31 & 0.08\\
				\cmidrule(lr){2-4} \cmidrule(lr){5-7} \cmidrule(lr){8-10} \cmidrule(lr){11-13}
			\end{tabular}
		}
		\caption{Summary statistics for final schemes ($\bs_i^{*}$). Home and away (columns 1--6), and home-away (H-A) difference (columns 7--12). $p$-values for skewness and kurtosis tests (null hypothesis of Gaussian distribution) are indicated below the test statistic. See the caption of Table~\ref{tab:dependent} for the legend.}
		\label{tab:schemesF} 
	\end{table}

	\begin{table}
		\centering
		
		\footnotesize
		
		\begin{tabular}{lcccccccc}
			& \multicolumn{8}{c}{Home Team Actions} \\
			\cmidrule(lr){2-9}
			& $w_{1}^{H}$ & $w_{2}^{H}$ & $w_{3}^{H}$ & $w_{4}^{H}$ & $w_{5}^{H}$ & $w_{6}^{H}$ & $w_{7}^{H}$ & $w_{8}^{H}$ \\
			\cmidrule(lr){2-9}
			Min.  & 0.00  & 0.00  & 0.00  & 0.00  & 0.00  & 0.00  & 0.00  & 0.00  \\
			Q1    & 15.00 & 4.00  & 1.00  & 13.00 & 2.00  & 10.00 & 6.00  & 1.00  \\
			Median & 21.00 & 5.00  & 1.00  & 19.00 & 4.00  & 13.00 & 8.00  & 2.00  \\
			Mean  & 22.74 & 5.77  & 1.72  & 20.98 & 4.05  & 13.41 & 8.40  & 2.57  \\
			Q3    & 29.00 & 8.00  & 3.00  & 27.00 & 6.00  & 17.00 & 11.00 & 4.00  \\
			Max.  & 82.00 & 21.00 & 13.00 & 79.00 & 14.00 & 41.00 & 20.00 & 11.00  \\
			SD    & 11.64 & 3.06  & 1.55  & 11.35 & 2.47  & 5.47  & 3.50  & 1.94  \\
			CV    & 0.51  & 0.53  & 0.90  & 0.54  & 0.61  & 0.41  & 0.42  & 0.75  \\
			Skew. & 1.18  & 0.71  & 1.36  & 1.22  & 0.75  & 0.67  & 0.35  & 1.04  \\
			$p$-value & 0.00 & 0.00 & 0.00 & 0.00 & 0.00 & 0.00 & 0.00 & 0.00 \\
			Kurtosis & 5.34  & 3.83  & 6.79  & 5.65  & 3.59  & 4.04  & 2.92  & 4.28  \\
			$p$-value & 0.00 & 0.00 & 0.00 & 0.00 & 0.00 & 0.00 & 0.65 & 0.00 \\
			\cmidrule(lr){2-9}
			\addlinespace
			&\multicolumn{8}{c}{Away Team Actions} \\
			\cmidrule(lr){2-9}
			& $w_{1}^{A}$ & $w_{2}^{A}$ & $w_{3}^{A}$ & $w_{4}^{A}$ & $w_{5}^{A}$ & $w_{6}^{A}$ & $w_{7}^{A}$ & $w_{8}^{A}$ \\
			\cmidrule(lr){2-9}
			Min.  & 0.00  & 0.00  & 0.00  & 0.00  & 0.00  & 0.00  & 0.00  & 0.00  \\
			Q1    & 11.00 & 3.00  & 0.00  & 10.75 & 2.00  & 8.00  & 7.00  & 1.00  \\
			Median & 17.00 & 4.00  & 1.00  & 15.00 & 3.00  & 11.00 & 9.00  & 2.00  \\
			Mean  & 18.52 & 4.46  & 1.34  & 17.16 & 3.12  & 11.03 & 9.56  & 2.28  \\
			Q3    & 24.00 & 6.00  & 2.00  & 22.00 & 4.00  & 14.00 & 12.00 & 3.00  \\
			Max.  & 82.00 & 17.00 & 8.00  & 79.00 & 14.00 & 29.00 & 25.00 & 11.00  \\
			SD    & 10.21 & 2.63  & 1.38  & 9.85  & 2.11  & 4.70  & 3.78  & 1.85  \\
			CV    & 0.55  & 0.59  & 1.03  & 0.57  & 0.68  & 0.43  & 0.39  & 0.81  \\
			Skew. & 1.27  & 0.85  & 1.13  & 1.35  & 0.92  & 0.44  & 0.41  & 1.01  \\
			$p$-value & 0.00 & 0.00 & 0.00 & 0.00 & 0.00 & 0.00 & 0.00 & 0.00 \\
			Kurtosis & 5.92  & 4.08  & 4.13  & 6.30  & 4.34  & 3.12  & 3.24  & 4.06  \\
			$p$-value & 0.00 & 0.00 & 0.00 & 0.00 & 0.00 & 0.38 & 0.11 & 0.00 \\
			\cmidrule(lr){2-9}
			\addlinespace
			& \multicolumn{8}{c}{H-A Differences in Team Actions} \\
			\cmidrule(lr){2-9}
			& $w_{1}$ & $w_{2}$ & $w_{3}$ & $w_{4}$ & $w_{5}$ & $w_{6}$ & $w_{7}$ & $w_{8}$ \\
			\cmidrule(lr){2-9}
			Min.  & -49.00 & -14.00 & -8.00  & -47.00 & -13.00 & -22.00 & -21.00 & -10.00 \\
			Q1    & -5.00  & -2.00  & -1.00  & -4.00  & -1.00  & -3.00  & -5.00  & -1.00  \\
			Median & 4.00   & 1.00   & 0.00   & 3.00   & 1.00   & 2.00   & -1.00  & 0.00   \\
			Mean  & 4.22   & 1.31   & 0.38   & 3.82   & 0.93   & 2.39   & -1.16  & 0.29   \\
			Q3    & 13.00  & 4.00   & 2.00   & 12.00  & 3.00   & 8.00   & 3.00   & 2.00   \\
			Max.  & 63.00  & 19.00  & 12.00  & 63.00  & 11.00  & 37.00  & 16.00  & 10.00  \\
			SD    & 13.84 & 4.50  & 2.18  & 12.96 & 3.52  & 8.44  & 5.69  & 2.75  \\
			CV    & 3.28  & 3.44  & 5.68  & 3.39  & 3.80  & 3.54  & 4.90  & 9.53  \\
			Skew. & 0.17  & 0.04  & 0.25  & 0.19  & -0.02 & 0.16  & -0.10 & 0.08  \\
			$p$-value & 0.02  & 0.55  & 0.00  & 0.01  & 0.76  & 0.02  & 0.15  & 0.28  \\
			Kurtosis & 3.90  & 3.23  & 4.43  & 4.14  & 3.18  & 3.33  & 3.00  & 3.38  \\
			$p$-value & 0.00  & 0.12  & 0.00  & 0.00  & 0.21  & 0.04  & 0.92  & 0.02  \\
			\cmidrule(lr){2-9}
		\end{tabular}
	
	\caption{Summary statistics of match actions for home (H), away (A), and home-away (H-A) differences. See the caption of Table~\ref{tab:dependent} for the legend.}
	\label{tab:team}
	\end{table}
	
	\begin{table}
	\centering
	\captionsetup{size=small} 
	\small 
	\setlength{\tabcolsep}{4pt} 
	
	\begin{tabular}{lcccccc|cccccc}
		& \multicolumn{6}{c}{Yellow \& Red Cards} 
		& \multicolumn{6}{c}{Free Kicks \& Penalties} \\ 
		\cmidrule(lr){2-7} \cmidrule(lr){8-13} 
		& $z_{1}^{H}$ & $z_{1}^{A}$ & $z_{2}^{H}$ & $z_{2}^{A}$ & $z_{1}$ &  $z_{2}$
		& $z_{3}^{H}$ & $z_{3}^{A}$ & $z_{4}^{H}$ & $z_{4}^{A}$ & $z_{3}$ &  $z_{4}$ \\
		\cmidrule(lr){2-7} \cmidrule(lr){8-13} 
		Min. & 0.00 & 0.00 & 0.00 & 0.00 & -7.00 & -3.00 & 0.00 & 0.00 & 0.00 & 0.00 & -18.00 & -2.00 \\
		Q1 & 1.00 & 1.00 & 0.00 & 0.00 & -1.00 & 0.00 & 12.00 & 11.75 & 0.00 & 0.00 & -4.00 & 0.00 \\
		Median & 2.00 & 2.00 & 0.00 & 0.00 & 0.00 & 0.00 & 14.50 & 15.00 & 0.00 & 0.00 & 0.00 & 0.00 \\
		Mean & 2.25 & 2.47 & 0.12 & 0.16 & -0.22 & -0.04 & 14.49 & 14.66 & 0.09 & 0.16 & -0.17 & -0.07 \\
		Q3 & 3.00 & 3.00 & 0.00 & 0.00 & 1.00 & 0.00 & 17.00 & 18.00 & 0.00 & 0.00 & 4.00 & 0.00 \\
		Max. & 8.00 & 7.00 & 2.00 & 3.00 & 5.00 & 2.00 & 27.00 & 30.00 & 2.00 & 2.00 & 17.00 & 2.00 \\
		\addlinespace
		SD & 1.33 & 1.34 & 0.34 & 0.41 & 1.75 & 0.52 & 4.56 & 4.67 & 0.29 & 0.39 & 5.71 & 0.47 \\
		CV & 0.59 & 0.54 & 2.72 & 2.51 & 8.06 & 13.28 & 0.31 & 0.32 & 3.35 & 2.49 & 32.72 & 6.81 \\
		\addlinespace
		Skew. & 0.39 & 0.42 & 2.48 & 2.62 & -0.05 & -0.49 & -0.12 & -0.05 & 3.25 & 2.38 & 0.03 & -0.57 \\
		$p$-value & 0.00 & 0.00 & 0.00 & 0.00 & 0.52 & 0.00 & 0.09 & 0.49 & 0.00 & 0.00 & 0.70 & 0.00 \\
		Kurtosis & 2.92 & 2.87 & 7.87 & 10.57 & 3.01 & 5.99 & 3.46 & 3.57 & 12.87 & 7.95 & 2.85 & 6.47 \\
		$p$-value & 0.64 & 0.38 & 0.00 & 0.00 & 0.88 & 0.00 & 0.01 & 0.00 & 0.00 & 0.00 & 0.30 & 0.00 \\
		\cmidrule(lr){2-7} \cmidrule(lr){8-13} 
	\end{tabular}
	
	\caption{Summary statistics of yellow and red cards (left) and fouls leading to free kicks and penalties (right). Columns correspond to disciplinary actions for home and away teams, as well as their home-away differences. See the caption of Table~\ref{tab:dependent} for the legend.}
	\label{tab:referee}
	\end{table}
	
	\begin{table}
	\centering
	\footnotesize
	\captionsetup{size=small} 
	\begin{tabular}[t]{lcccccc}
		&$c_{1}$& $\tilde{c}_{2}$& $c_{3}$& $c_{4}$& $k_{1}^{d}$& $k_{2}^{d}$ \\
		\cmidrule{2-7}
		Min & 0.00 & 0.00 & -64.00 & -1.00 & 1.00 & 0.00\\
		Q1 & 0.47 & 5.11 & -6.25 & -0.16 & 24.00 & 0.24\\
		Q2 & 0.56 & 6.15 & 0.00 & 0.00 & 49.00 & 0.51\\
		Mean & 0.59 & 5.91 & -0.29 & -0.01 & 49.18 & 0.51\\
		Q3 & 0.70 & 7.22 & 7.00 & 0.14 & 75.00 & 0.78\\
		Max & 1.00 & 15.51 & 60.00 & 1.00 & 100.00 & 1.00\\
		\addlinespace
		SD & 0.19 & 1.79 & 13.85 & 0.27 & 28.45 & 0.30\\
		CV & 0.33 & 0.30 & -48.27 & -27.91 & 0.58 & 0.59\\
		\addlinespace
		Skew. & 0.15 & 0.21 & -0.13 & -0.13 & 0.01 & 0.01\\
		$p$-value  & 0.04 & 0.00 & 0.06 & 0.08 & 0.84 & 0.90\\
		Kurt. & 2.99 & 4.57 & 5.61 & 4.70 & 1.76 & 1.75\\
		$p$-value & 0.95 & 0.00 & 0.00 & 0.00 & 0.00 & 0.00\\
		\cmidrule{2-7}
	\end{tabular}
	\caption{Summary statistics for selected control and fixed effects variables. $c_{i,1}$: control for home stadium filling index, $\tilde{c}_{i,2}$: weighted extra time, $c_{i,3}$: discrete team effect, $c_{i,4}$: continuous team effect, $k_{i,1}^{d}$: discrete league day (calendar date ranked), $k_{i,2}^{d}$: continuous league day effect (0 to 1). See the caption of Table~\ref{tab:dependent} for the legend.}
	\label{tab:additional}
	\end{table}     
	
	\begin{table}
	\centering
	\scriptsize
	\captionsetup{size=small} 
	\renewcommand{\arraystretch}{1.2}
	\begin{tabular}{lrrrrrr}
		\toprule
		Value & $k^{(11)}$ & $k^{(12)}$ & $k^{(13)}$ & $k_{1}^{e}$ & $k_{2}^{e}$ & $k_{3}^{e}$ \\
		\midrule
		0 & 760~(66.67\%)  & 760~(66.67\%)  & 760~(66.67\%)  & 1131~(99.21\%)  & 1137~(99.74\%)  & 1134~(99.47\%)  \\
		1 & 380~(33.33\%)  & 380~(33.33\%)  & 380~(33.33\%)  & 9~(0.79\%)  & 3~(0.26\%)  & 6~(0.53\%)  \\
		\bottomrule
	\end{tabular}
	\caption{Distribution of categorical variables. $k^{(11)}, k^{(12)}, k^{(13)}$: Binary season fixed effects, $k_{1}^{e}$: Binary extreme goal difference, $k_{2}^{e}$: Binary extreme positive goal difference, $k_{3}^{e}$: Binary extreme negative goal difference. Columns represent variable names.}
	\label{tab:additional2}
	\end{table}
	
	\clearpage
	
	\subsection{Figures}
	\medskip
	
	\noindent\begin{figure}[h!]
	\captionsetup{size=small}
	\centering
	\begin{minipage}{1\linewidth}
		\centering
		\includegraphics[width=0.5\linewidth]{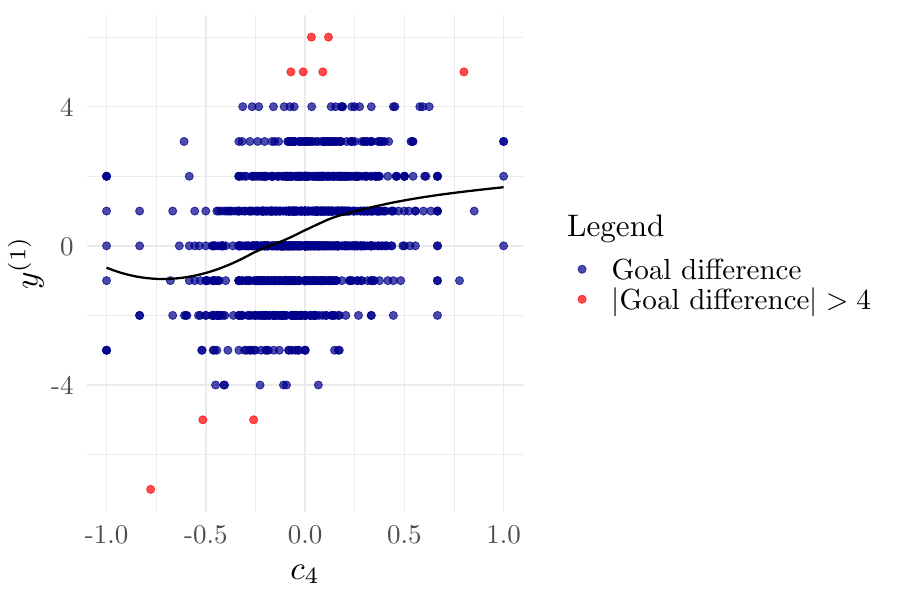}
		\caption{Scatterplot of the relative team effect, measured as the relative difference between ranking points up to the previous match ($c_4$), against the goal difference ($y^{(1)}$). Matches classified as extreme ($k_1^e$) are highlighted in red.}
		\label{fig_extreme_matches}
	\end{minipage}
	\end{figure}
	
	\bibliographystyle{abbrvnat} 
	\bibliography{references_v1_clean}
	
	\clearpage
	
	
	\phantomsection
	\section*{Supplementary Material}

	\setcounter{section}{1}
	\setcounter{subsection}{0}
	\renewcommand\thesection{\arabic{section}}
	\renewcommand\thesubsection{\arabic{subsection}}
	
	\makeatletter
	\renewcommand\theHsubsection{SM.\arabic{subsection}}
	\makeatother
	
	\subsection{Data management procedures for panel balancing}\label{sec:datamanage}
	
	Commentary was unavailable for fifteen matches in the Virgilio Serie~A database (\url{https://sport.virgilio.it/calcio/serie-a}). Commentary data for fourteen of these matches were reconstructed by consulting Lega~Serie~A (\url{https://www.legaseriea.it/it}) and ESPN play-by-play archives (\url{https://www.espn.com/soccer/league}). To preserve a coherent analytical frame, the 2012/13 Cagliari--Roma penalization (3--0) was excluded. Moment-by-moment remarks recorded at time~$\tau$, together with the initial line-ups, were subsequently integrated into the master panel. The qualitative text was lemmatized and translated into binary event indicators, whereas tactical configurations were inferred from ESPN positional diagrams (see Section~\ref{subsubsec:coach_schemes} and Appendix~\ref{App:OffInd}). Player substitutions triggered immediate updates of formation codes through a binary flag and the corresponding realignment of schemes; every change underwent a post-merge control.

	\subsubsection{Steps}
	
	\paragraph{Error management}\label{sec:err_man}
	The procedure summarized in Table~\ref{tab:DataManagement} (upper panel) opens with a scripted improvement of the raw match commentary. Conditional routines attach binary indicators for corner-originated crosses, $w_{3}^{H}$ and $w_{3}^{A}$, as well as for open-play crosses, $w_{5}^{H}$ and $w_{5}^{A}$. Additional dummy variables jointly register fouls, free-kicks, penalties, and handballs, the latter identified through handball-specific regular expressions applied to the lemmatized commentary. Two-way contingency matrices examine co-occurrence patterns—such as foul $\times$ free-kick—and guide the data cleaning step. Record-level revisions correct misclassified events—particularly fouls that lead to set pieces—after which all derived variables are recomputed to maintain data consistency.
	
	\paragraph{Data cleansing}
	The middle panel of Table~\ref{tab:DataManagement} documents a second data–cleaning step that removes whitespace from \texttt{cod\_partita} and \texttt{cod\_univoco}. The raw timestamp is decomposed into its \textit{year}, \textit{month}, and \textit{day} components, together with a consolidated \textit{date} field, whereas the legacy \textit{stagione} token is parsed into \textit{serie} and three seasonal indicators, $\{k^{(11)}, k^{(12)}, k^{(13)}\}$. Team identifiers are standardized (e.g., ``ACMilan'' $\rightarrow$ ``Milan'') and recoded as home–versus–away factors; league-day codes, half-time markers, and exact minute stamps are similarly normalized. Unique identifiers—match ID, event ID, and observation ID—index every game and discrete action. Supplementary variables capture coaches, referees, minute sequences, and intra-match quarters. The routine concludes by ordering the dataset by season, date, and match, removing redundant columns, and applying targeted manual corrections wherever diagnostic checks revealed outstanding irregularities.

	\paragraph{Panel balancing}
	Table~\ref{tab:DataManagement} (lower panel) provides an expanded account of the workflow for data balancing. 
	The raw record is partitioned into discrete match segments, metrics are aggregated, and novel variables are synthesized, producing a comprehensive panel dataset suitable for longitudinal analysis.
	The initial step involves categorizing the dataset into distinct temporal segments reflective of a soccer match's dynamics:
	\begin{itemize}[noitemsep, nosep]
		\item First half regular time (01-44 minutes): captures the majority of the first half excluding the added time.
		\item First half added time ($\geq 45$ minutes): reflects the stoppage time added at the end of the first half. Events in this period are aggregated to represent the added time as a singular observational unit.
		\item Second half regular time (46-89 minutes): similar to the first half, this covers the bulk of the second half excluding added time.
		\item Second half added time ($\geq 90$ minutes): aggregates events during the stoppage time at the match's conclusion.
	\end{itemize}
	Subsequent to data segmentation, new features are crafted: 
	\begin{itemize}[noitemsep, nosep]
		\item Cumulative metrics: variables like cumulative goals and team actions throughout the match provide a dynamic view of match evolution, highlighting momentum shifts and team performance over time.
		\item Differential metrics: differences in scores and specific actions between teams offer insights into competitive balance and match progression.
		\item Event occurrence: binary indicators for minute-specific events such as goals or fouls furnish a granular timeline of match proceedings.
		\item Quantitative insights: continuous variables capturing aspects like schemes or weighted actions. 
	\end{itemize}
	The final transformative phase ensures a balanced panel structure. 
	Each minute of match play is represented as a unique observation within each match, creating a consistent framework for longitudinal analysis. 
	The balancing procedure encompasses several critical actions:
	\begin{itemize}[noitemsep, nosep]
		\item Minute-by-minute observations: ensuring each match within the dataset has a uniform minute-by-minute representation, addressing gaps in the temporal sequence.
		\item Synthetic observation generation: for minutes lacking data, synthetic observations are generated, leveraging preceding non-missing data to maintain continuity.
		\item Variable aggregation: relevant metrics are aggregated or interpolated as necessary, aligning with the synthetic observations.
	\end{itemize}
	
	Multi–dimensional season, date, and event blocks are flattened with \texttt{melt()}, producing long tables whose value columns are relabelled (e.g.\ \texttt{minute}, \texttt{match\_id}, \texttt{season}).  From \texttt{minute} we derive \texttt{half} and \texttt{quarter}.  All blocks are inner–joined on (\texttt{match\_id}, \texttt{minute}) to form a single event ledger, sorted by the same keys and re-columned.  Team, coach, and referee data are merged, after which temporary variables are removed.

	
	\subsection{Statistical aspects of data transformation}\label{sec:A_empirical}
	The dataset, after completing missing match data, error management and cleaning process,
	is composed by $n=159{,}551$ observations of $K = 81$ variables ($K=65$ relevant), for $N=1{,}140$ matches, in $3$ seasons $2011/2012-2013/2014$, where (for match $i$ at minute $t$):
	\begin{itemize}[nosep, noitemsep]
		
		\item $\mathcal{C}_{i,t,\tau}$ is the random vector containing the lemmatized comment of the match describing the specific action at instant $\tau$; 
		
		\item $\boldsymbol{\mathcal{I}}_{i,t,\tau}=\left\{ \mathcal{I}_{i,t,\tau, 1}, \mathcal{I}_{i,t,\tau, 2}  \right\}$ is the vector containing the match and observation (of the event) identifiers, at instant $\tau$; 
		
		\item $\boldsymbol{\mathcal{T}}_{i,t,\tau} = \left\{ \mathcal{T}_{i,t,\tau, 1}, \dots , \mathcal{T}_{i,t,\tau, 6}   \right\}$ is the vector containing the time identifier indexes referred to the full date (YYYY/MM/DD), season, league day, half time, minute, and number of event inside the minute, at instant $\tau$;
		
		\item $\boldsymbol{\mathbf{\mathcal{N}}}_{i,t,\tau} = \left\{ \mathcal{N}_{i,t,\tau, 1}^H, \mathcal{N}_{i,t,\tau, 2}^H, 
		\mathcal{N}_{i,t,\tau, 1}^A, \mathcal{N}_{i,t,\tau, 2}^A,
		\mathcal{N}_{i,t,\tau, 3} \right\}$ is the vector containing the names of the home and away teams, their respective coaches for the match considered, and the referee arbitrating the match, at instant $\tau$.
		Note that in the following, the home team will be considered the reference level to which all differences will be computed (unless stated otherwise);
		
		\item $\mathbf{Y}_{i,t,\tau}=\left\{ Y_{i,t,\tau, 1}^H, Y_{i,t,\tau, 2}^H , Y_{i,t,\tau, 1}^A, Y_{i,t,\tau, 2}^A   \right\}$ are the random variables referred to the candidate dependent variables referred to goals scores by both the home and away team respectively and their respective auto goals, at instant $\tau$;
		
		\item $\mathbf{S}_{i,t,\tau, 0}= \left\{ S_{i,t,\tau, 0}^{H}, S_{i,t,\tau, 0}^{A} \right\}$ is the random vector containing the Bernoulli random variables referred to coaches' actions (substitutions), at instant $\tau$;
		
		\item $\mathbf{S}_{i,t,\tau}= \left\{ S_{i,t,\tau, 1}^{H}, S_{i,t,\tau, 1}^{A}, S_{i,t,\tau, 2}^{H}, S_{i,t,\tau, 2}^{A} \right\}$ is the random vector containing the discrete random variables referred to two types of coaches' offensiveness indexes, at instant $\tau$. 
		Note that $S_{i,t,\tau, 1}$ is the original variable, while $S_{i,t,\tau, 2}$ is the corrected version (with correction for substitutions);
		
		\item $\mathbf{S}^{'}_{i,t,\tau}= \left\{ S_{i,t,\tau, 3}^{H}, S_{i,t,\tau, 3}^{A} \right\}$ is the random vector containing the pseudo-continuous random variables referred to the coaches' offensiveness index derived from $S_{i,t,\tau, 2}$, at instant $\tau$;
		
		\item $\mathbf{W}_{i,t,\tau} = \left\{ W_{i,t,\tau, 1}^H, \dots, W_{i,t,\tau, 8}^H,  
		W_{i,t,\tau, 1}^A, \dots, W_{i,t,\tau, 8}^A
		\right\}$ is the random vector containing the Bernoulli random variables referred to the teams' actions, at instant $\tau$;
		
		\item $\mathbf{Z}_{i,t,\tau} = \left\{ Z_{i,t,\tau, 1}^H, Z_{i,t,\tau, 2}^H, 
		Z_{i,t,\tau, 3}^H, Z_{i,t,\tau, 4}^H, Z_{i,t,\tau, 1}^A, Z_{i,t,\tau, 2}^A, 
		Z_{i,t,\tau, 3}^A, Z_{i,t,\tau, 4}^A  \right\}$ is the random vector containing the Bernoulli random variables related to referee's actions on calling penalty, foul, yellow and red card for home and away teams, at instant $\tau$;
		
		\item The remaining $13$ variables are intermediate checks variables referred to the total number of: substitutions, crosses, shots, goal kicks, offsides, corner, free and penalty kicks, fouls, yellow and red cards, goals and auto-goals; adding the values for home and away.
	\end{itemize}
	
	The set of relevant variables and constants that require our focus can be summarized as the set:
	$	\left\{ 
	\cC_{i,t,\tau}, \bcI_{i,t,\tau}, \bcT_{i,t,\tau}, \bcN_{i,t,\tau},
	\bY_{i,t,\tau}, 
	\bS_{i,t,\tau, 0},
	\bS_{i,t,\tau},
	\bS^{'}_{i,t,\tau},
	\bW_{i,t,\tau},  
	\bZ_{i,t,\tau}
	\right\}.$
	See Tables~\ref{tab:combined_codebook_description} for a summary. In the following analysis, $\cC_{i,t,\tau}$ was used only as a check, reducing the set to
	$\left\{ 
	\bcI_{i,t,\tau}, \bcT_{i,t,\tau}, \bcN_{i,t,\tau},
	\bY_{i,t,\tau}, 
	\bS_{i,t,\tau, 0},
	\bS_{i,t,\tau},
	\bS^{'}_{i,t,\tau},
	\bW_{i,t,\tau},  
	\bZ_{i,t,\tau}
	\right\}.$
	One important consideration during the analysis is the presence of overlapping information within the variables, which can lead to potential multicollinearity. To address this issue, a strategy to avoid multicollinearity is to disaggregate the variable, as described in Section~\ref{sec:err_man}. 

	\subsection{Time aggregation}\label{sec:unbal_bal_s1}
	The initial event log comprises $n = 159{,}551$ observations distributed over $K = 81$ variables. Three forms of temporal irregularity hinder direct analysis: \emph{(i)} certain minutes contain no recorded action, \emph{(ii)} multiple events may occur within a single minute, and \emph{(iii)} the duration of stoppage time after minutes~45 and~90 differs across matches. To impose chronological balance, we generate a standardized minute panel for every match. First, a complete time grid is created (minutes~1--45, 46--90, plus stoppage). Second, concurrent events within each minute are collapsed to counts or most-recent-valued indicators. Third, the grid is left-joined to the raw log, with unobserved minutes padded by zeros. Finally, the match clock is propagated forward to the last observed minute. The resulting harmonized dataset provides one row per minute for every match and constitutes the foundation for the second-stage balancing:
	\begin{itemize}[noitemsep, nosep]
		\item[I.] For each match, and for each minute $t=\left\{1, \dots, 44\right\}$ in the first half, and for each minute $t=\left\{46, \dots, 89\right\}$ in the second half, aggregate all the events inside each minute $t$. Depending on the type of variables, different aggregations were performed:
		\begin{itemize}[noitemsep, nosep]
			\item For variables in $\bcI_{i,t,\tau}$, $\bcT_{i,t,\tau}$, $\bcN_{i,t,\tau}$, $\bS_{i,t,\tau}$ and $\bS^{'}_{i,t,\tau}$, the last value inside each minute is taken, with the exception of $\cT_{\tau, 6}$ (minute), which will be equal to the minute $t=1, \dots, 90$, under consideration, delivering respectively $\bcI_{i,t}$, $\bcT_{i,t}$, $\bcN_{i,t}$, $\bS_{i,t}$ and $\bS^{'}_{i,t}$. In particular:
			$\mathbf{S}_{i,t}= \left\{ S_{i,t, 1}^{H}, S_{i,t, 1}^{A}, S_{i,t, 2}^{H}, S_{i,t, 2}^{A} \right\}$, 
			$\mathbf{S}^{'}_{i,t}= \left\{ S_{i,t, 3}^{H}, S_{i,t, 3}^{A} \right\}$.
			\item For variables in $\bY_{i,t,\tau}$, $\bS_{i,t,\tau, 0}$, $\bW_{i,t,\tau}$, and $\bZ_{i,t,\tau}$, their sum is taken, i.e. 
			$\bY_{i,t} = \sum_{\tau=1}^{T_{i,t}} \bY_{i,t,\tau}$, 
			$\bS_{i,t,0} = \sum_{\tau=1}^{T_{i,t}} \bS_{i,t,\tau, 0}$, 
			$\bW_{i,t} = \sum_{\tau=1}^{T_{i,t}} \bW_{i,t,\tau}$, 
			$\bZ_{i,t} = \sum_{\tau=1}^{T_{i,t}} \bZ_{i,t,\tau}$,
			where $T_{t}$ is the total number of events inside minute $t=\left\{ 1,\dots, 44\right\}$. 
			Since each element of these vector variables is Bernoulli, the resulting $\bY_{i,t}$, $\bS_{i,t, 0}$, $\bW_{i,t}$, and $\bZ_{i,t}$ are the sums of Bernoulli random vector variables, i.e. each element is the count of the number of events observed at minute $t$.
			For instance, if in match $i$ at minute $t=30$ there were $3$ events for the variable home shots, then we write:
			$W_{i,30, 4}^H = \sum_{\tau=1}^{3} W_{i, 30, \tau, 4}^H = 3$.
			%
			It may also happen that inside the same minute different events occurred. For example, if at minute $25$ of match $i$ there were $4$ events for the variables 
			home substitutions, 
			away corners, 
			and home goal kicks. 
			In particular: one home substitution, two offsides for the away team, and one goal kick for the home team. Then we write:
			$S_{i,25, 0}^{H} = \sum_{\tau=1}^{4} S_{i,25,\tau, 0}^{H} = 1$, 
			$W_{i,25, 6}^A = \sum_{\tau=1}^{4} W_{i,25,\tau, 6}^A = 2$, 
			$W_{i,25, 5}^H = \sum_{\tau=1}^{4} W_{i,25,\tau, 5}^H = 1$.
		\end{itemize}
		\item[II.] For each match, and for each minute $t\geq 45$ in the first half, and for each minute $t\geq 90$ in the second half, aggregate all the events inside each minute $t$. The type of aggregation follows in the same manner as in the previous point.
	\end{itemize}
	The resulting balanced dataset is now composed of $n=102600$ observations for $K=59$ relevant observed variables. From this dataset, we proceeded to create cumulative, differenced, and further synthetic variables.\\
	
	\paragraph{Cumulative, differenced, ranking and cross-sectional variables}\label{subsec:cumdiffvar}
	For each match, identified by its unique match key, we compute cumulative counts of events through minute $T^{*}\!\in\!\{1,\dots,90\}$—goals, shots, corners, and related actions—separately for the home and visiting sides, and record their differences (home minus away). Similar procedures generate derived indicators, such as ``corners not followed by crosses’’. Seasonal rankings originate from this enriched ledger: the data are first partitioned by season, and points accrued in home and away matches determine the standings within each championship. Temporal continuity is maintained by forward-filling absent rows, while monotonicity tests confirm that ordinal positions never decline as fresh information accrues. The standings are then interpolated to the minute scale (minutes are referenced to match-day time, not to the exact kick-off timestamp), tracing the evolving balance between home-away as play unfolds. Minute-level ranks are merged back by match identifier and team name, followed by season-specific renaming of variables, yielding a panel dataset ready for subsequent empirical analysis.
	For instance, consider the dependent variable home goals for minute $t$ in match $i$, $Y_{i,t,1}^H$. Its cumulative value up to time $T^*$ is defined as:
	$Y_{i, 1:T^*, 1}^H = \sum_{t=1}^{T^*} Y_{i,t,1}^H$,
	where indicates the number of home goals scored up to minute $T^* \in [1, \dots, T]$, for match $i \in \{1, \dots, 1140\}$, and we write $Y_{i, T^*, 1}^H$ for simplicity. 
	Similarly, we define:
	$Y_{i,T^*,2}^H = \sum_{t=1}^{T^*} Y_{i,t,2}^H$, 
	$Y_{i,T^*,1}^A = \sum_{t=1}^{T^*} Y_{i,t,1}^A$, 
	$Y_{i,T^*,2}^A = \sum_{t=1}^{T^*} Y_{i,t,2}^A$,
	the cumulative action up to minute T* of home auto goals, away goals, and away auto goals, respectively. The remaining coach substitutions, and team and referee actions follow in similar manner. Then we define the difference between home and away cumulative variables. First, take teams' crosses: 
	$W_{i, T^*, 1} = W_{i, T^*, 1}^H - W_{i, T^*, 1}^A$,
	for match $i \in \{1, \dots, 1140\}$, up to minute $T^*$. The rest of the cumulative teams' actions follow in the same fashion:
	$W_{i, T^*, h}^H = \sum_{t=1}^{T^*} W_{i,t,h}^H$, 
	$W_{i, T^*, h}^A = \sum_{t=1}^{T^*} W_{i,t,h}^A$, 
	$W_{i, T^*, h} = W_{i, T^*, h}^H - W_{i, T^*, h}^A$, 
	$h=2, \dots, 8$,
	as well as those for referees' cumulative actions:
	$Z_{i, T^*, j}^H = \sum_{t=1}^{T^*} Z_{i,t,j}^H$, 
	$Z_{i, T^*, j}^A = \sum_{t=1}^{T^*} Z_{i,t,j}^A$, 
	$Z_{i, T^*, j} = Z_{i, T^*, j}^H - Z_{i, T^*, j}^A$, 
	$j=1, \dots, 4$.
	Setting $T^*=90$ delivers the variables referred to match $i$, where we avoid the term $T^*$ and write:
	\begin{align*}
	y^{(1)H}_i  &= Y_{i, 1}^H + Y_{i, 2}^A, &y^{(1)A}_i  &= Y_{i, 1}^A + Y_{i, 2}^H,    &y^{(1)}_i &= y^{(1)H}_i - y^{(1)A}_i, &\\
	w^{H}_{i,h} &= W_{i,h}^H,               &w^{A}_{i,h} &= W_{i,h}^A,                  & w_{i,h} &= w^{H}_{i,h} - w^{A}_{i,h}, &\text{for} \ h = 1, \dots, 8, \\
	z^{H}_{i,j} &= Z_{i,j}^H,               &z^{A}_{i,j} &= Z_{i,j}^A,                  & z_{i,j} &= z^{H}_{i,j} - z^{A}_{i,j}, &\text{for} \ j = 1, \dots, 4. 
	\end{align*}
	For the offensiveness index, it suffices to take the first and last value in  $\mathbf{S}_{i,t}$ and $\mathbf{S}^{'}_{i,t}$:
	\begin{align*}
	s_{i,1}^H & = S_{i,1,1}^H, &      s_{i,1}^A &=S_{i,1,1}^A,&            
	s_{i,2}^H & = S_{i,2,1}^H, &      s_{i,2}^A &=S_{i,2,1}^A,&          
	s_{i,3}^{H} & = S_{i,3,1}^{H}, &  s_{i,3}^{A} &=S_{i,3,1}^{A},\\
	s_{i,1}^{*H} & = S_{i,1,90}^{H}, &      s_{i,1}^{*A} &=S_{i,1,90}^{A},&            
	s_{i,2}^{*H} & = S_{i,2,90}^{H}, &      s_{i,2}^{*A} &=S_{i,2,90}^{A},&          
	s_{i,3}^{*H} & = S_{i,3,90}^{*H}, &     s_{i,3}^{*A}  &=S_{i,3,90}^{A},
	\end{align*}
	which delivers our offensiveness index for match $i$:
	\begin{align*}
	s_{i,1}    &=s_{i,1}^H - s_{i,1}^A &
	s_{i,2}    &=s_{i,2}^H - s_{i,2}^A &
	s_{i,3}    &=s_{i,3}^H - s_{i,3}^A, \\
	s_{i,1}^{*}    &=s_{i,1}^{*H} - s_{i,1}^{*A} &
	s_{i,2}^{*}    &=s_{i,2}^{*H} - s_{i,2}^{*A} &
	s_{i,3}^{*}    &=s_{i,3}^{*H} - s_{i,3}^{*A} 
	\end{align*}
	The resulting balanced panel is a $(n \times K)$ data frame, with $n=102600$ observations (rows), $K=247$ variables (columns), for $T=90$ minutes observed in $N=1140$ matches. Table~\ref{tab:codebook_variables} shows the codebook of the balanced panel.
	
	\paragraph{Weighted actions}
	Minute-specific pace is absorbed through the factor
	$\omega_{i,t}= \frac{n_{i,t}}{n_i}+1$,
	where $n_{i,t}$ counts events in minute $t$ of match $i$ and $n_i$ is the match total, so action-dense minutes exert greater pull.  Applying this weight yields
	$\tilde{\bS}_{i,t}= \bS_{i,t}\omega_{i,t}$, \ 
	$\tilde{\bS}'_{i,t}= \bS'_{i,t}\omega_{i,t}$, \ 
	$\tilde{\bW}_{i,t}= \bW_{i,t}\omega_{i,t}$, \ 
	$\tilde{\bZ}_{i,t}= \bZ_{i,t}\omega_{i,t}$,
	the tempo-adjusted team and referee covariates.

	\subsection{From panel data to cross-sectional data}\label{sec:cs}
	Data from three seasons are concatenated into the master table \texttt{dt}, with missing entries imputed. Additional attributes are appended: outcome flags (win, draw, loss), match-level cumulative event counts, and tactical indicators weighted by each event’s relative share of activity. Season-specific partitions (\texttt{dt\_s1}–\texttt{dt\_s3}) preserve aggregates—goals, shots, formations, and referee decisions—for modeling. Vertically stacking these blocks yields the cross-sectional file \texttt{dt\_cs}, onto which rankings and performance identifiers are merged to support interseason comparisons. A complete variable catalogue is provided in Table~\ref{tab:coded_var}.
	
	
	
	\subsubsection{Descriptive statistics}\label{sec:destat}
\begin{figure}[h]
	\centering
	\begin{minipage}[b]{0.3\textwidth}
		\centering
		\includegraphics[width=\linewidth]{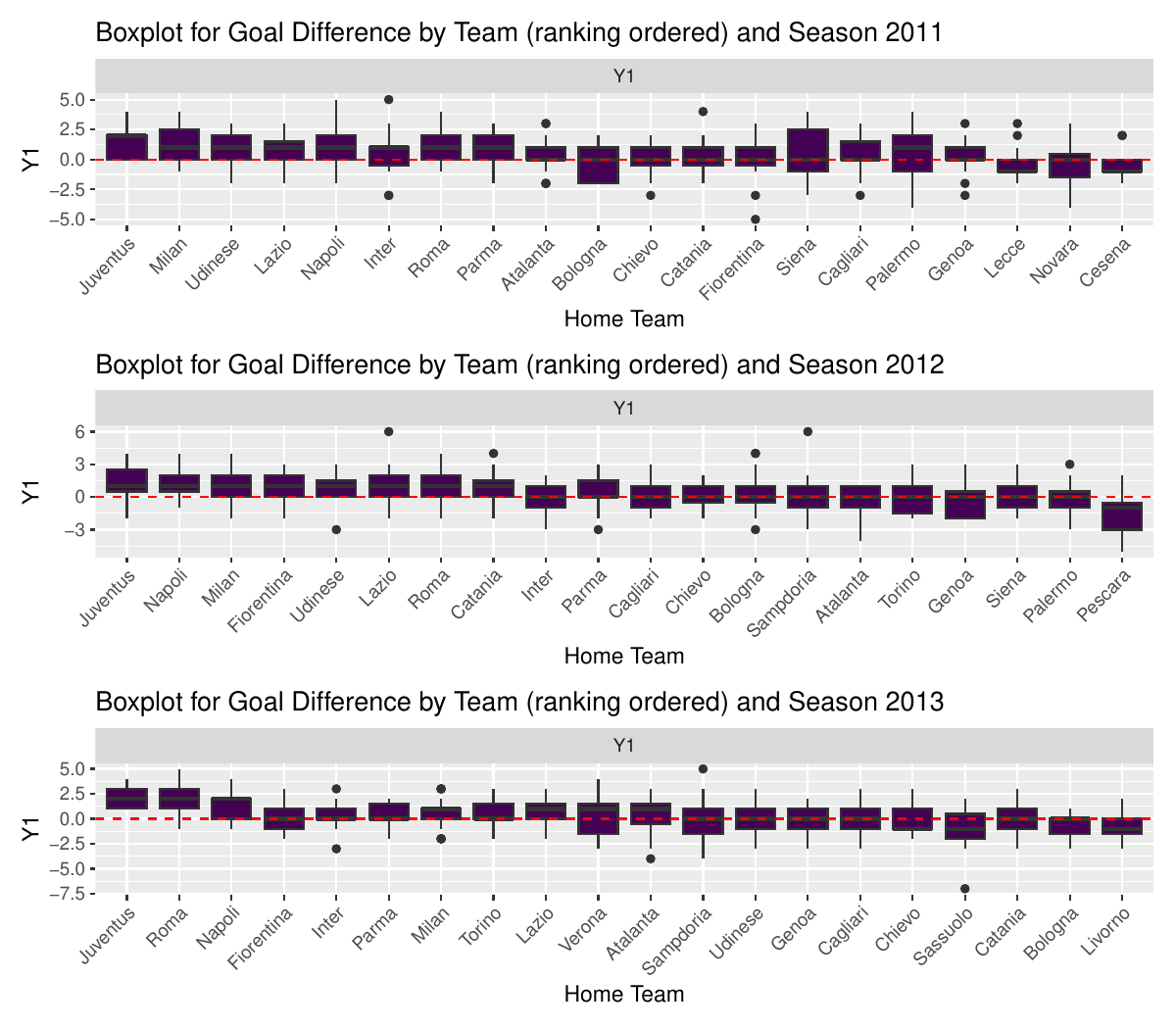}
		\caption{Boxplot of goal difference per season by home team ordered by seasonal ranking.}
		\label{fig:SM_F_02}
	\end{minipage}
	\hfill
	\begin{minipage}[b]{0.3\textwidth}
		\centering
		\includegraphics[width=\linewidth]{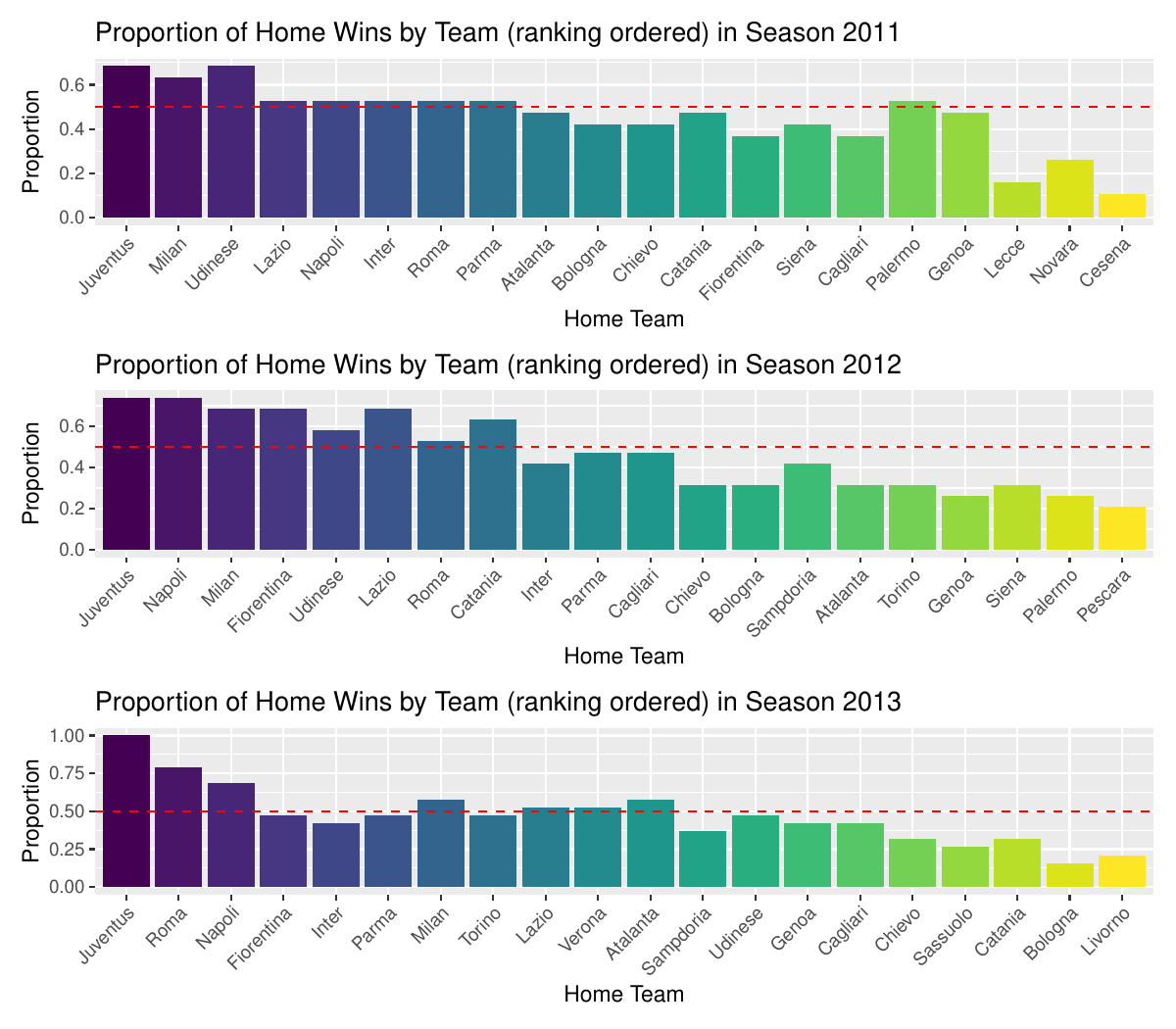}
		\caption{Proportion of wins per season by home team ordered by seasonal ranking.}
		\label{fig:SM_F_01}
	\end{minipage}
	\hfill
	\begin{minipage}[b]{0.3\textwidth}
		\centering
		\includegraphics[width=\linewidth]{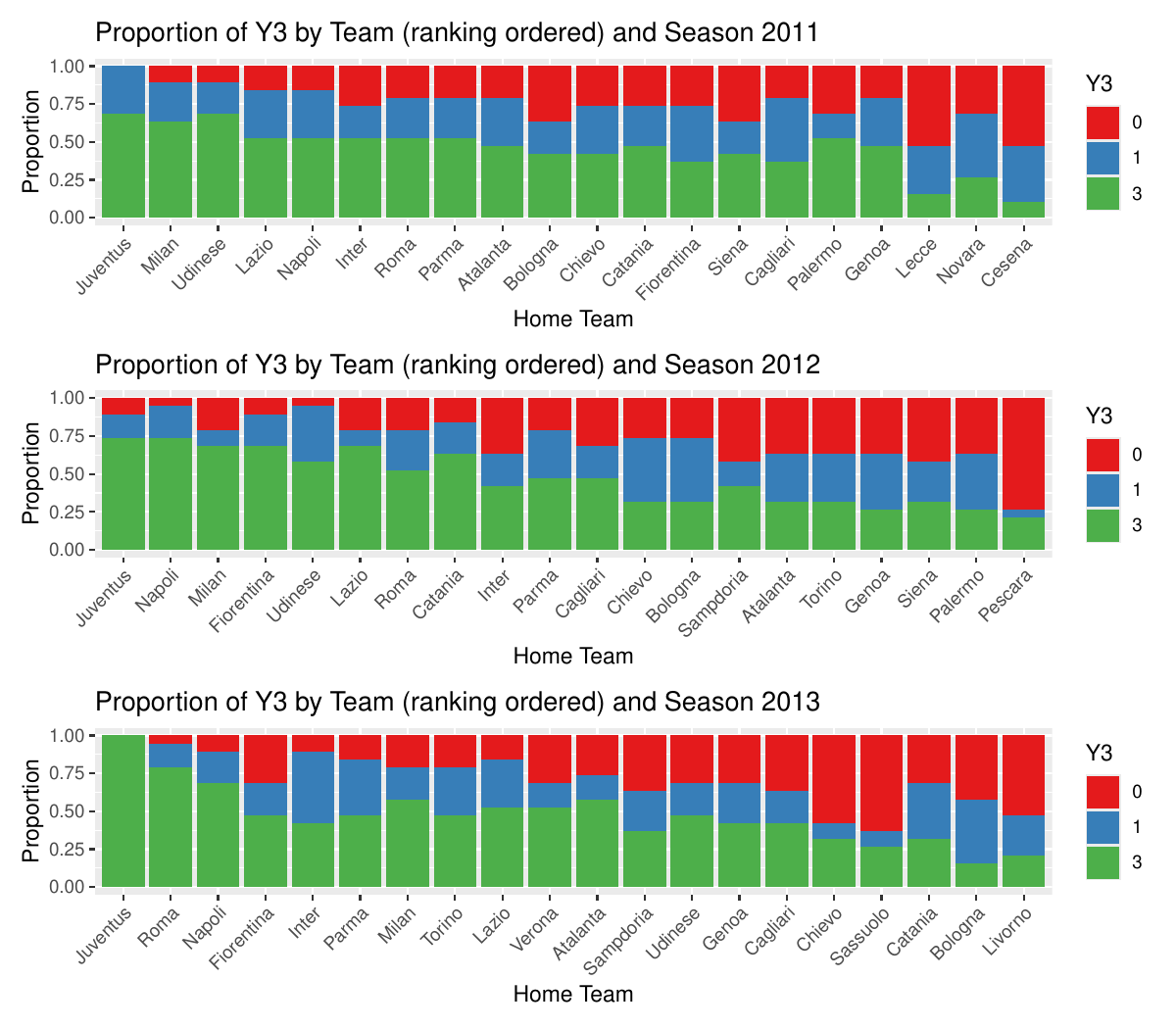}
		\caption{Proportion of losses, ties and wins per season by home team ordered by seasonal ranking.}
		\label{fig:SM_F_03}
	\end{minipage}
	\end{figure}

\begin{figure}[h]
	\centering
	\includegraphics[width=0.66	\linewidth]{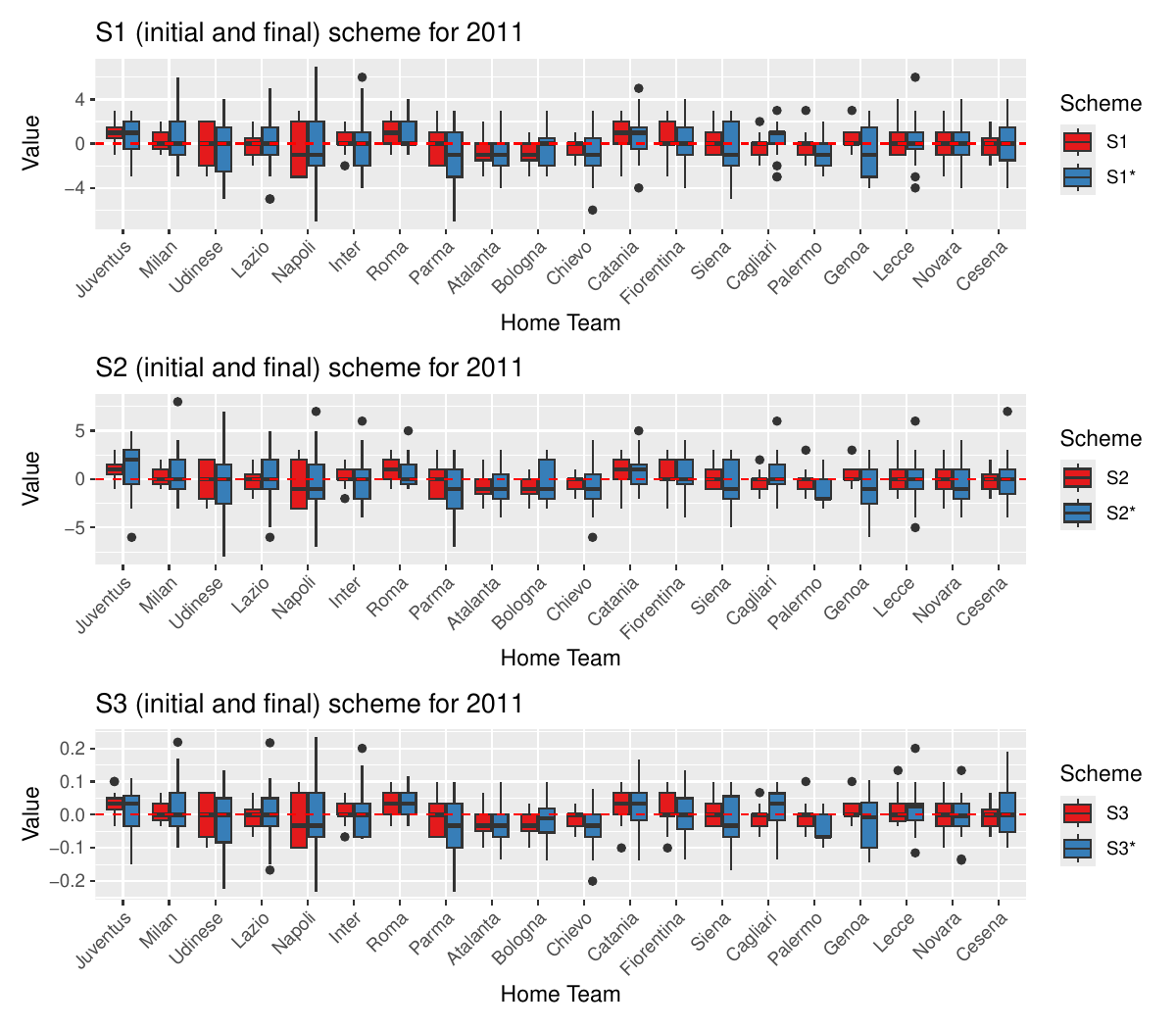}
	\caption{Boxplots for home-away differences in initial and final 
		schemes 1,2,3 by home team ordered by seasonal ranking for season 2011.}
	\label{fig:SM_F_04}
	\end{figure}

\begin{figure}[h]
	\centering
	\includegraphics[width=0.66	\linewidth]{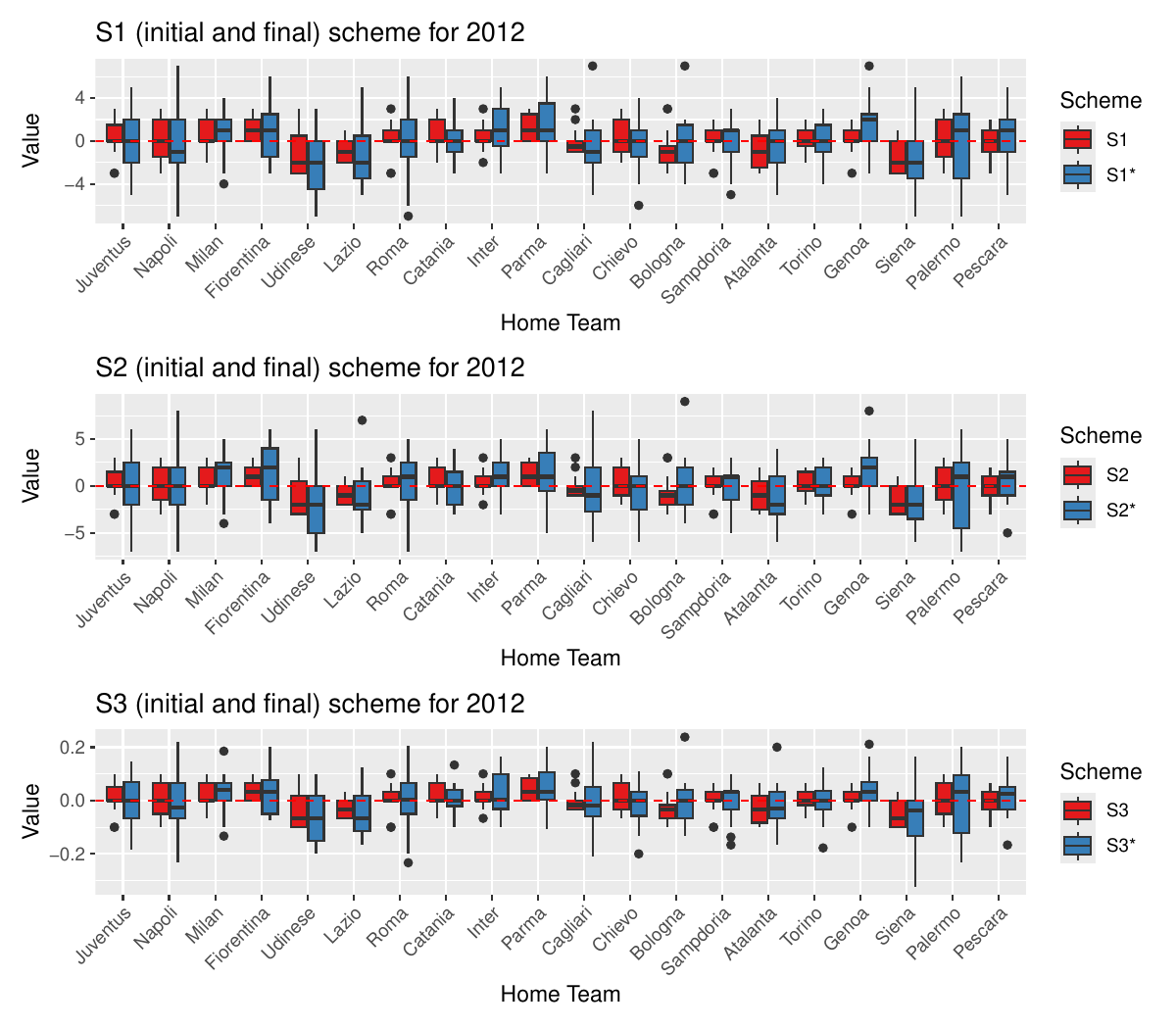}
	\caption{Boxplots for home-away differences in initial and final 
		schemes 1,2,3 by home team ordered by seasonal ranking for season 2012.}
	\label{fig:SM_F_05}
	\end{figure}

\begin{figure}[hp]
	\centering
	\includegraphics[width=0.66	\linewidth]{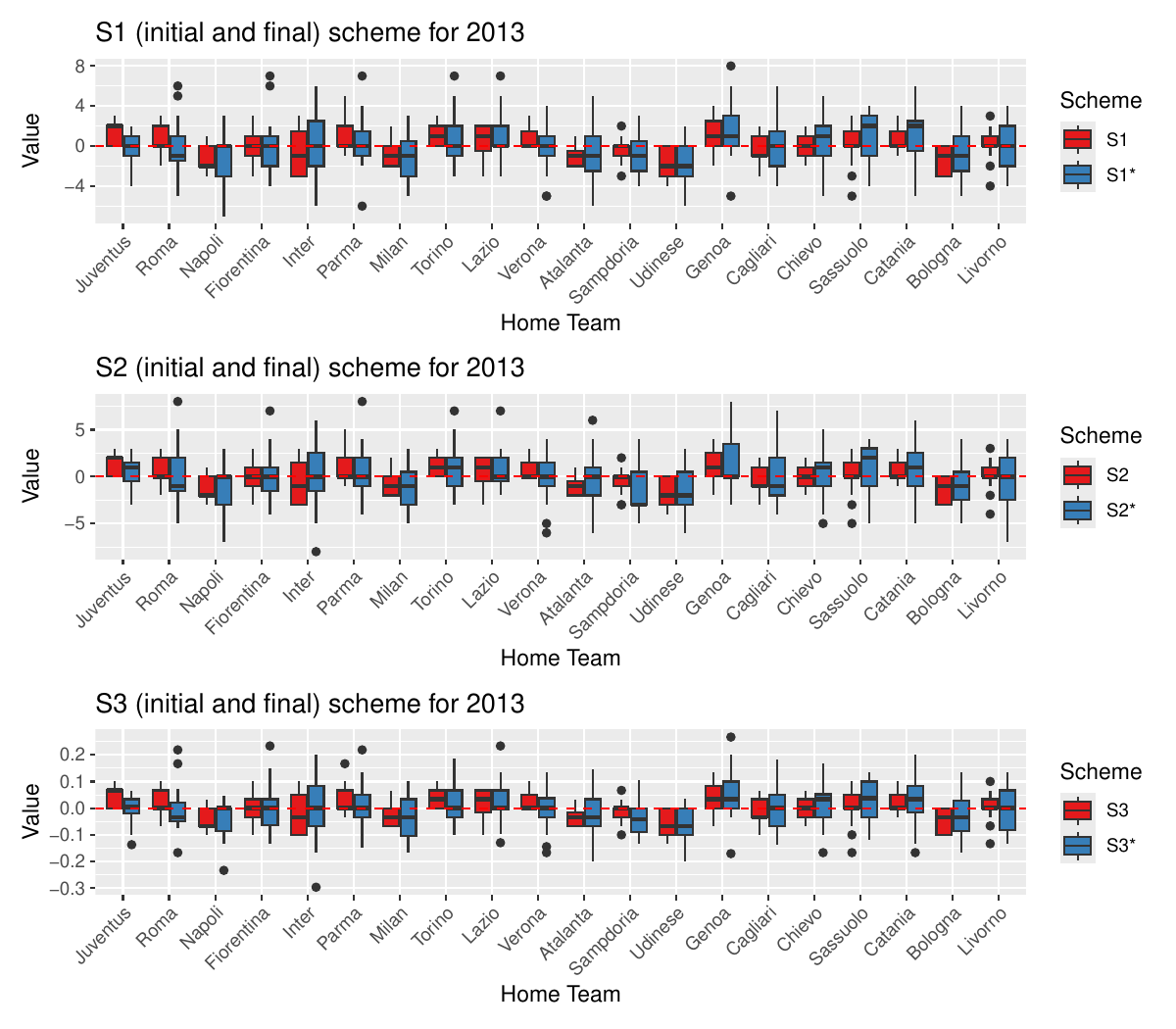}
	\caption{Boxplots for home-away differences in initial and final 
		schemes 1,2,3 by home team ordered by seasonal ranking for season 2013.}
	\label{fig:SM_F_06}
	\end{figure}

\begin{figure}[hp]
	\centering
	\includegraphics[width=1	\linewidth]{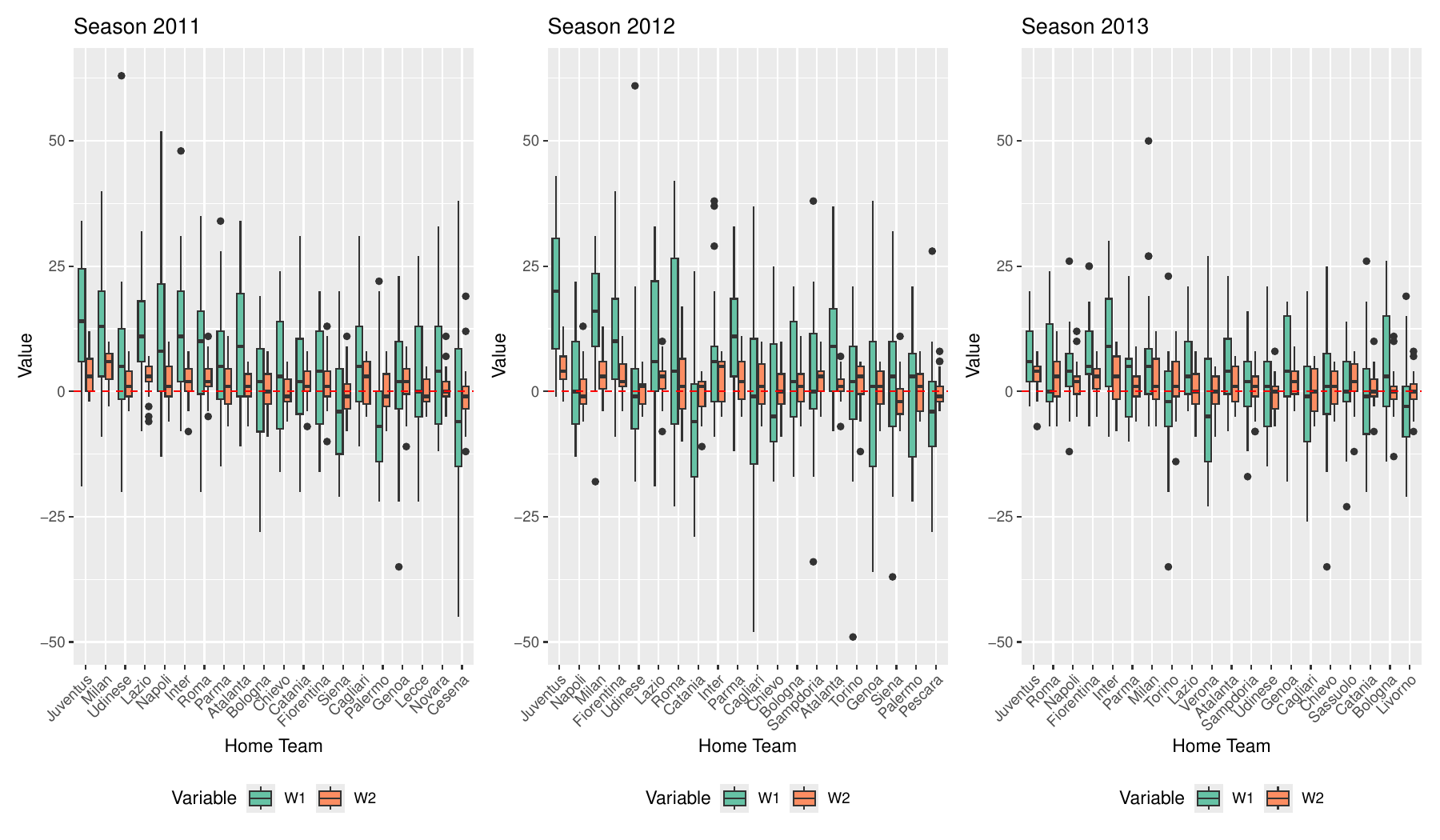}
	\caption{Boxplots for home-away differences in crosses and corners 
		by home team ordered by seasonal ranking and by season.}
	\label{fig:SM_F_07}
	\end{figure}

\begin{figure}
	\centering
	\includegraphics[width=1	\linewidth]{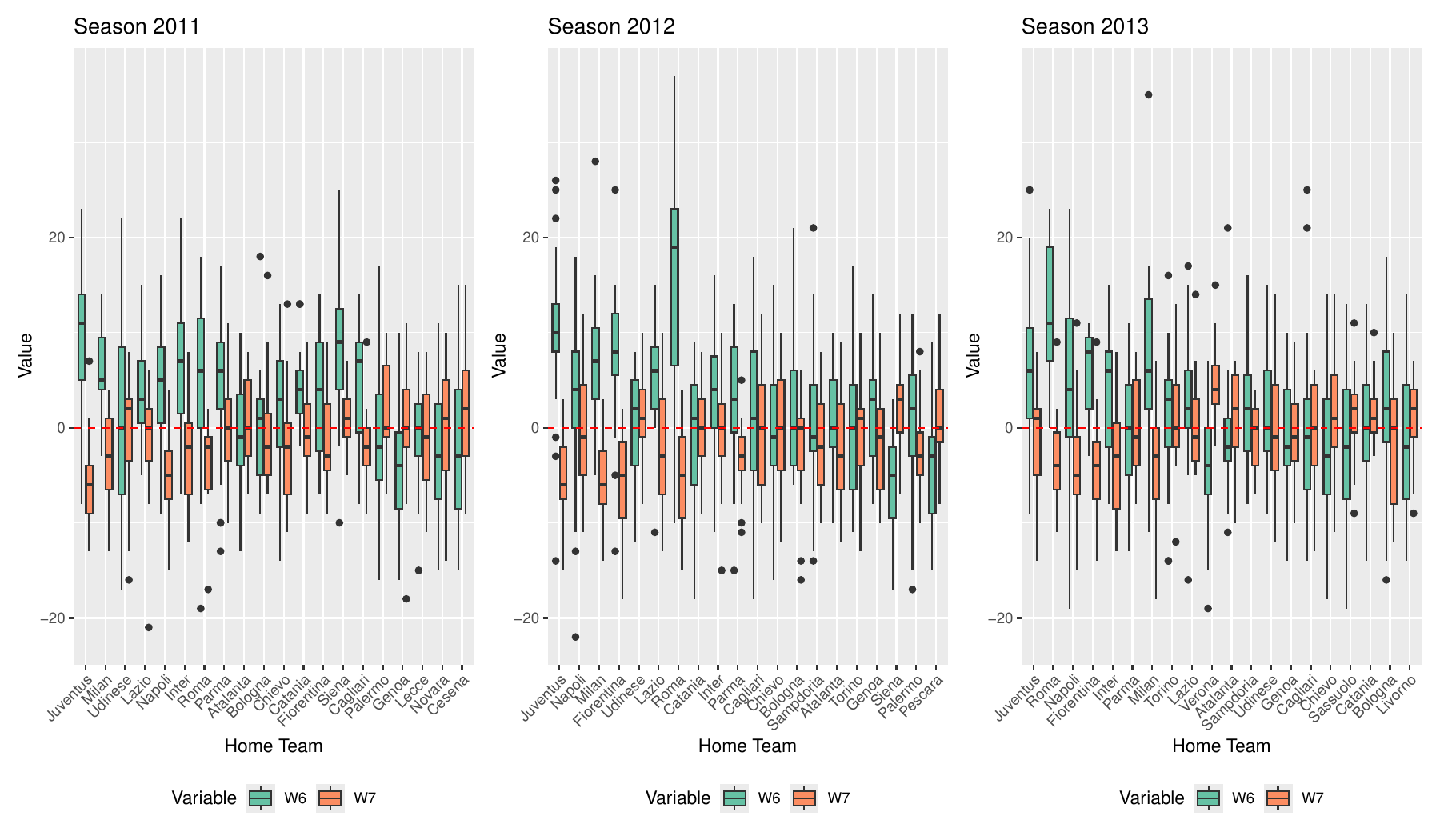}
	\caption{Boxplots for home-away differences in shots and goal kicks, by home team ordered by seasonal ranking and by season.}
	\label{fig:SM_F_08}
	\end{figure}

\begin{figure}
	\centering
	\includegraphics[width=1	\linewidth]{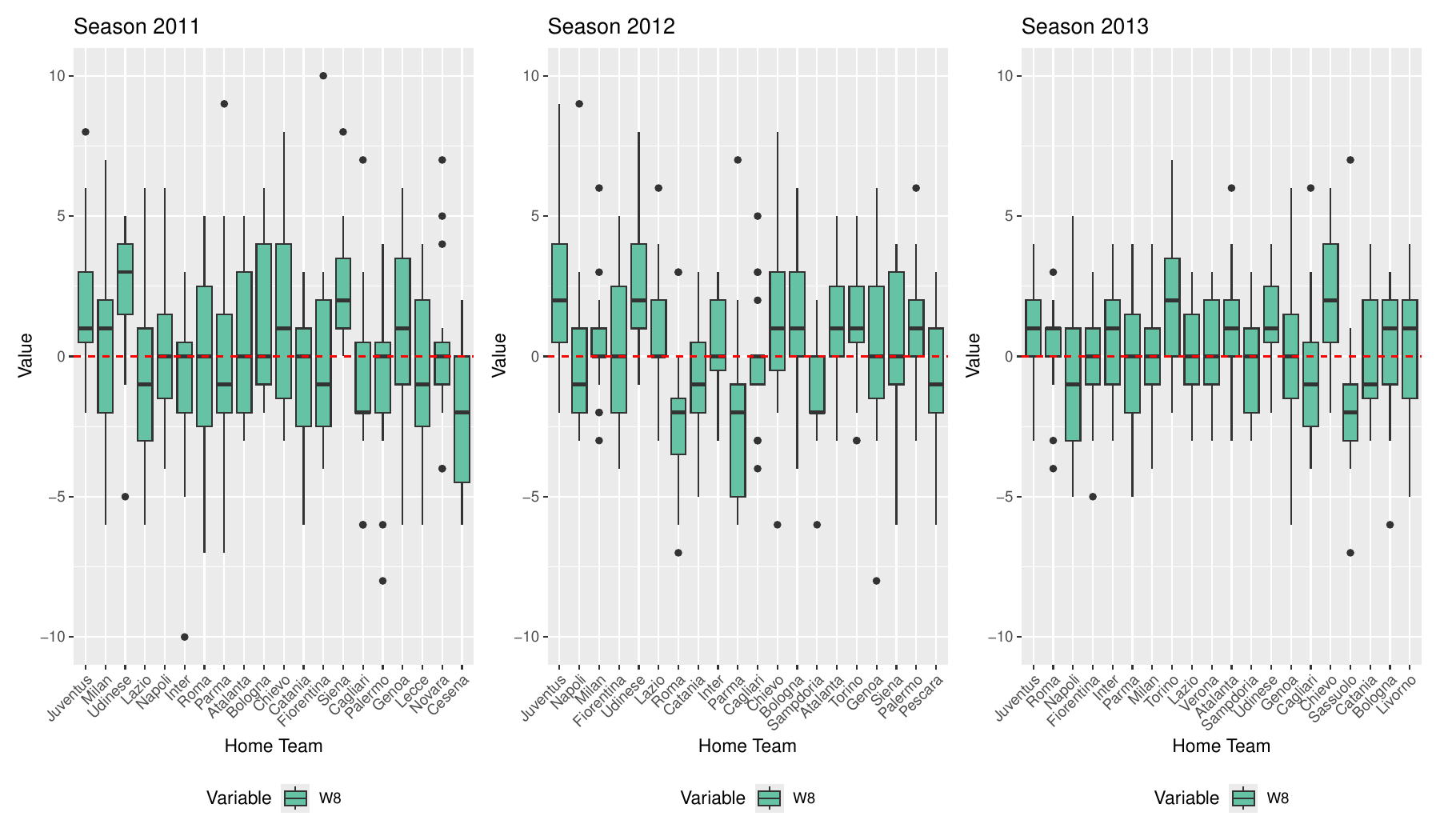}
	\caption{Boxplots for home-away differences in offsides by home team ordered by seasonal ranking and by season.}
	\label{fig:SM_F_09}
	\end{figure}

\begin{figure}
	\centering
	\includegraphics[width=1	\linewidth]{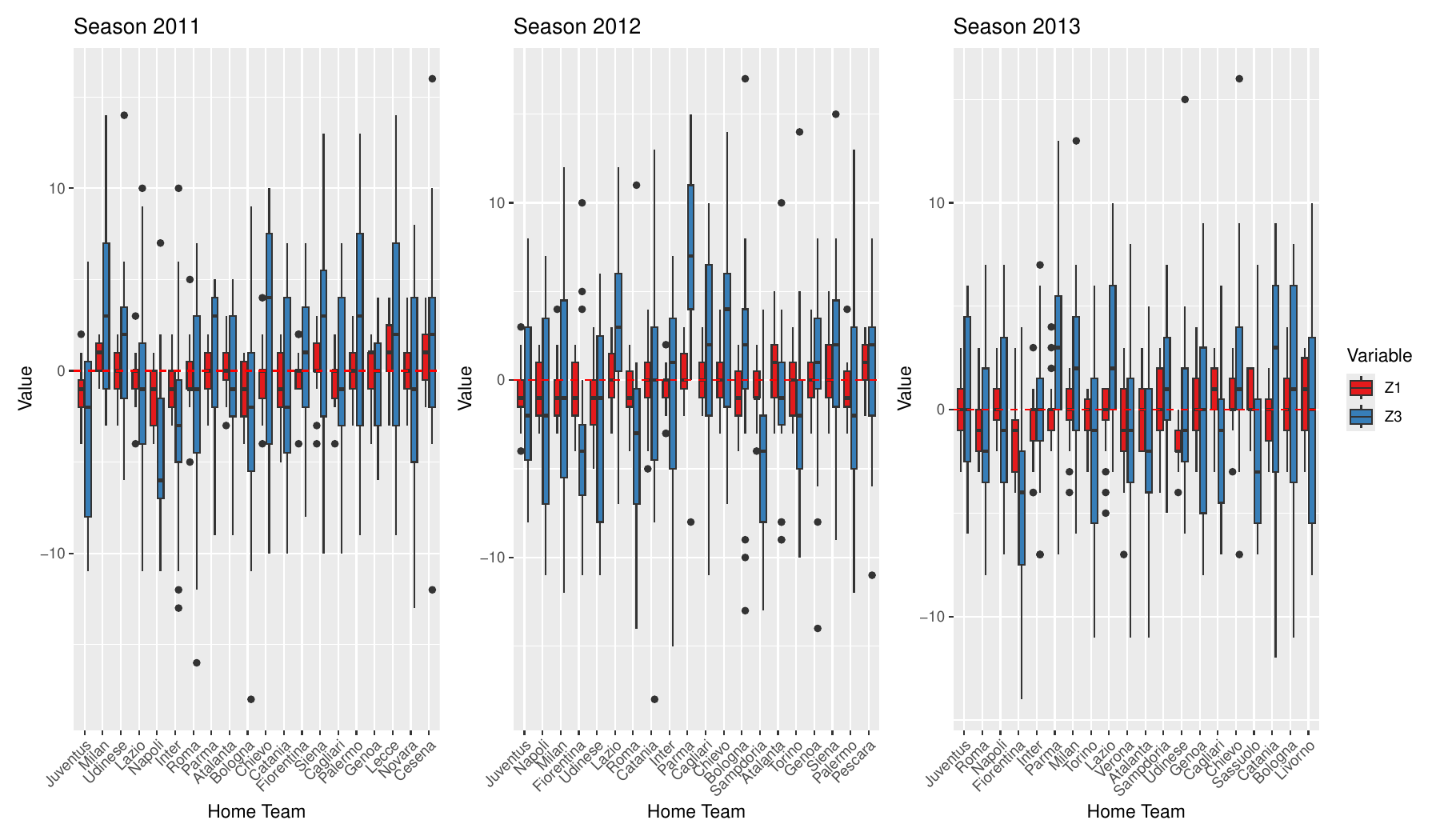}
	\caption{Boxplots for home-away differences in yellow cards and free-kicks by home team ordered by seasonal ranking and by season.}
	\label{fig:SM_F_15}
	\end{figure}

\begin{figure}
	\centering
	\includegraphics[width=1	\linewidth]{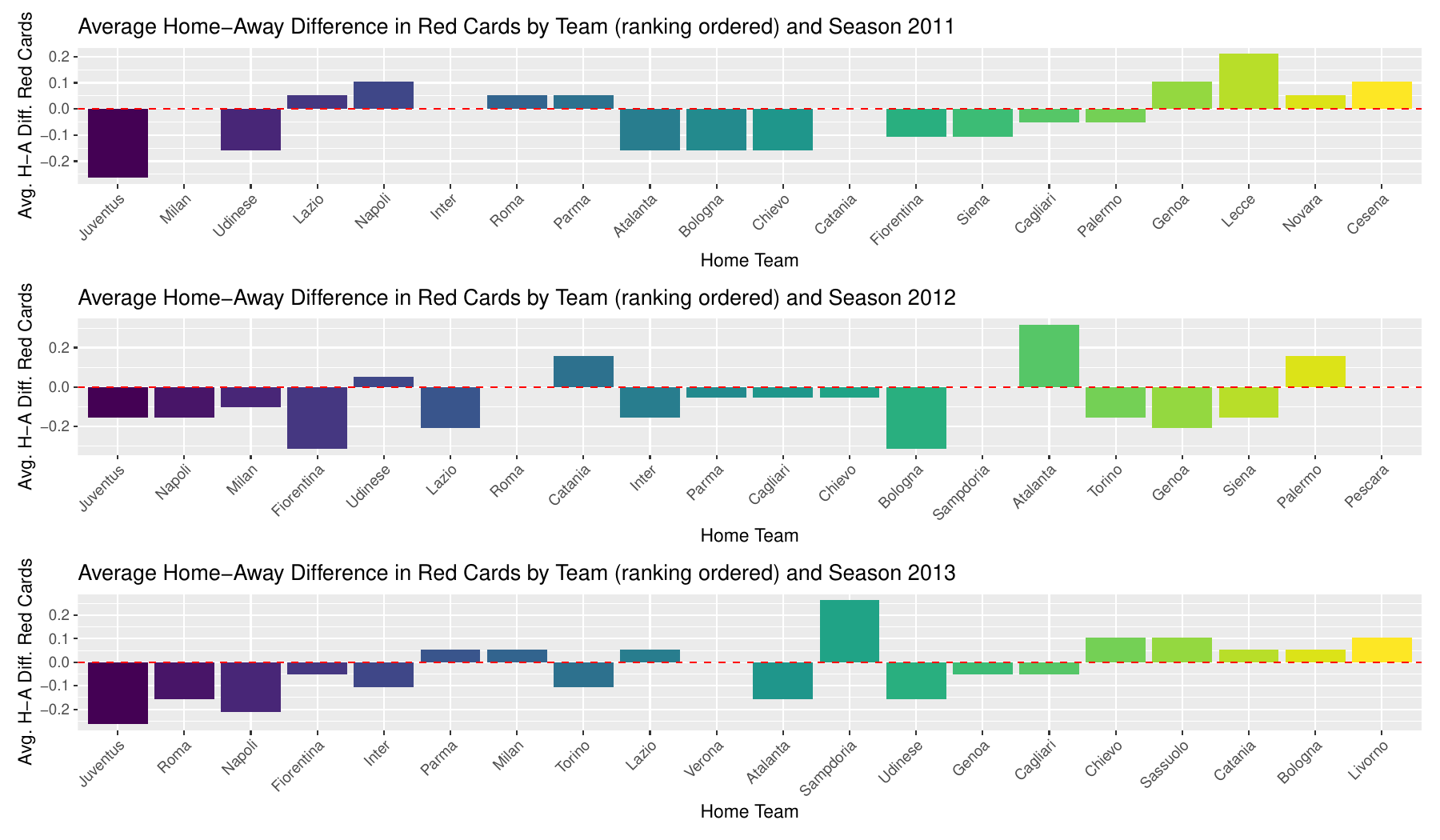}
	\caption{Average home-away differences in red cards by home team ordered by seasonal ranking and by season. Lack of a rectangle means the average difference is null.}
	\label{fig:SM_F_16}
	\end{figure}

\begin{figure}
	\centering
	\includegraphics[width=1	\linewidth]{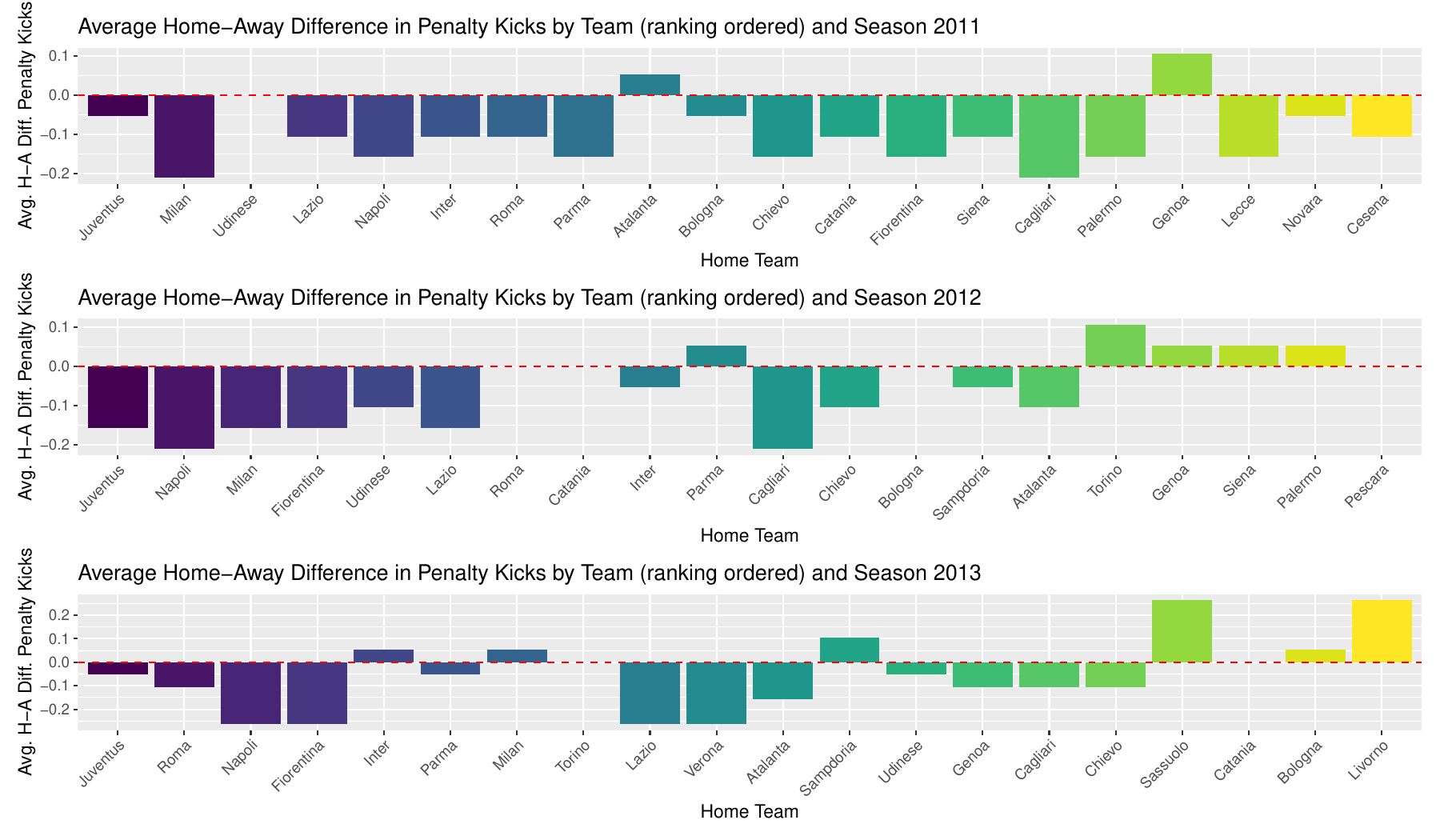}
	\caption{Average home-away differences in penalty kicks by home team ordered by seasonal ranking and by season. Lack of a rectangle means the average difference is null.}
	\label{fig:SM_F_17}
	\end{figure}

\begin{figure}
	\centering
	\includegraphics[width=1	\linewidth]{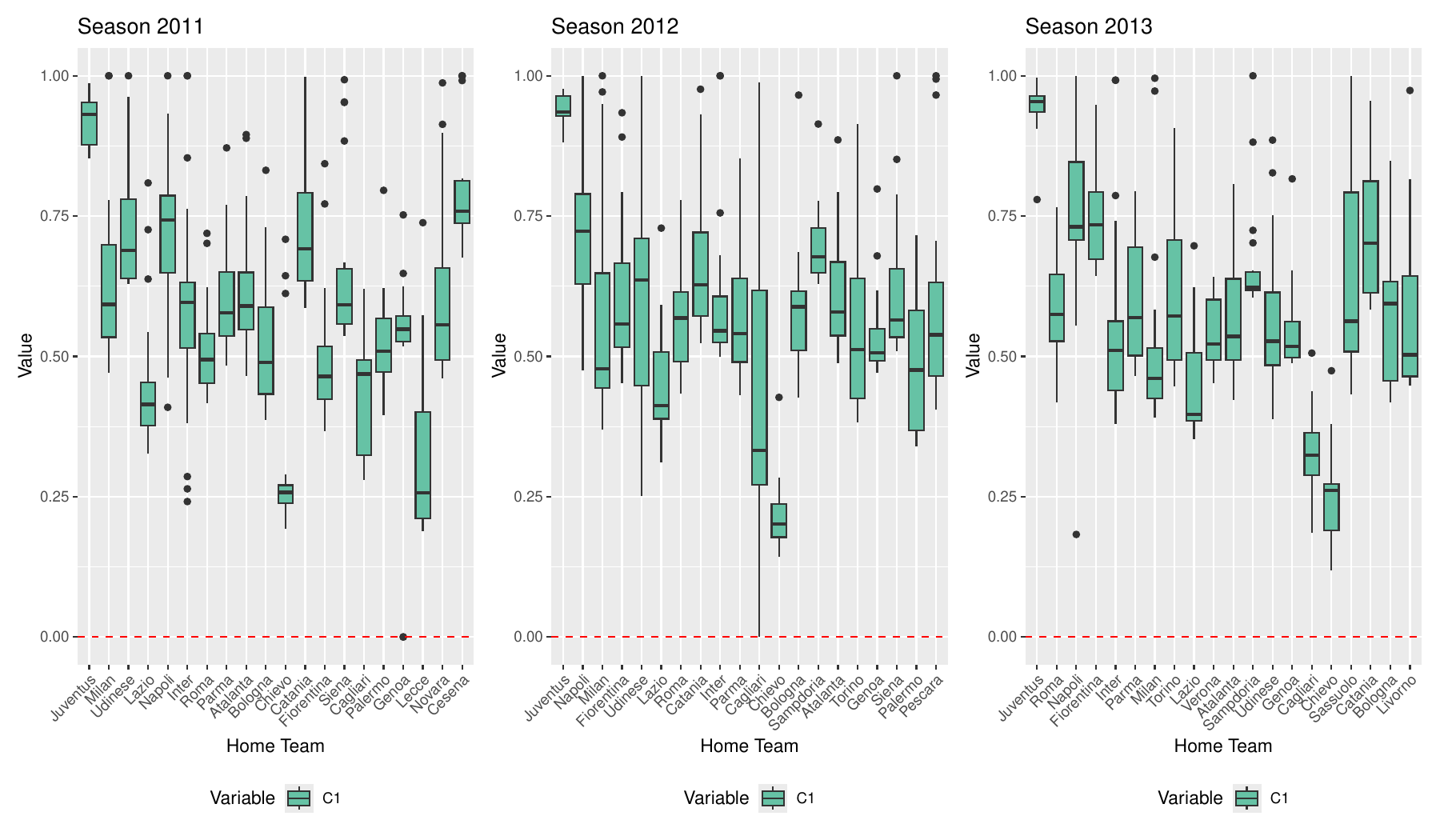}
	\caption{Boxplots for home stadium filling index by home team ordered by seasonal ranking and by season.}
	\label{fig:SM_F_14}
	\end{figure}

\begin{figure}
	\centering
	\includegraphics[width=1	\linewidth]{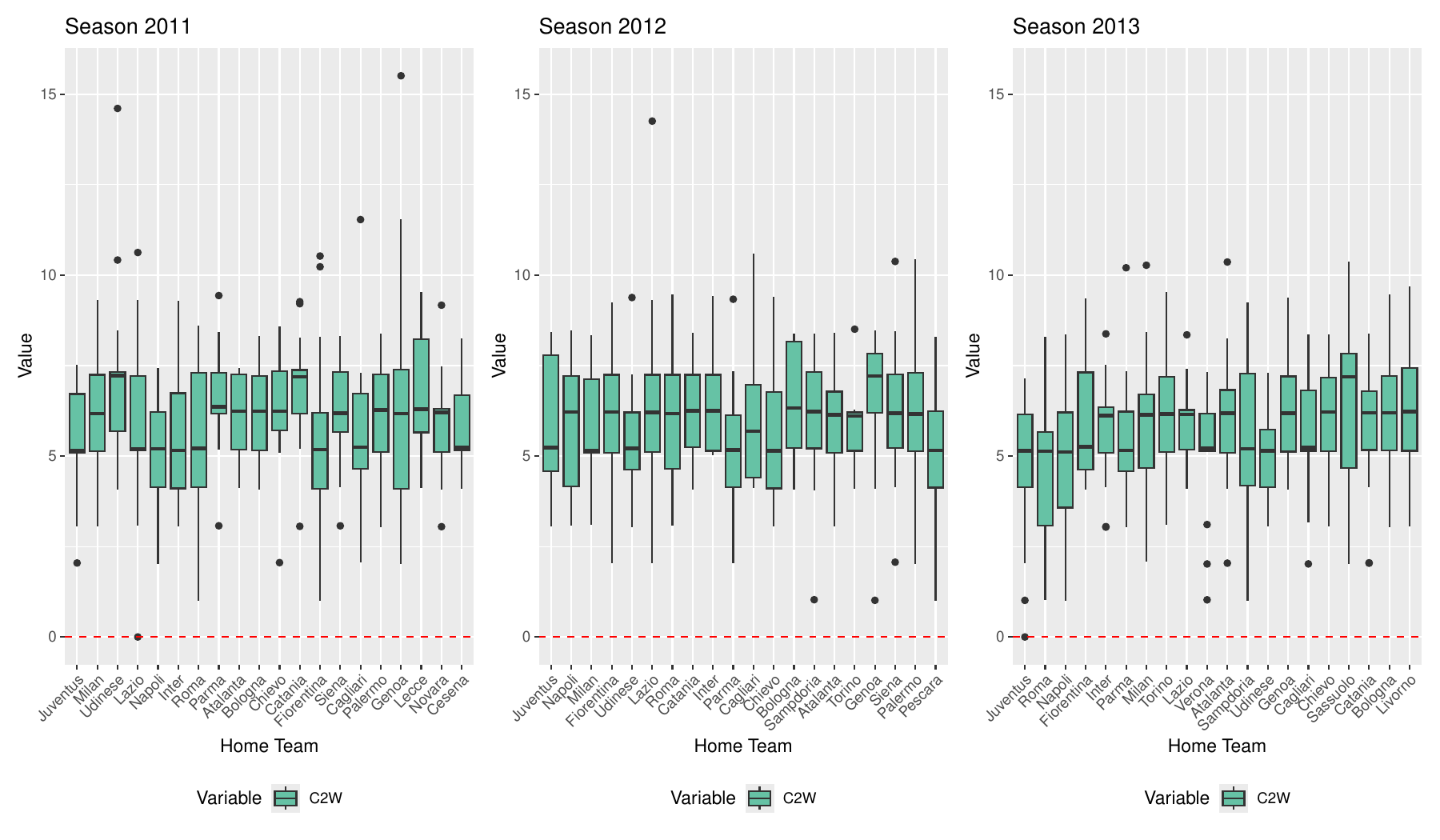}
	\caption{Boxplots for minutes extra time (weighted) by home team ordered by seasonal ranking and by season.}
	\label{fig:SM_F_10}
	\end{figure}

\begin{figure}
	\centering
	\includegraphics[width=1	\linewidth]{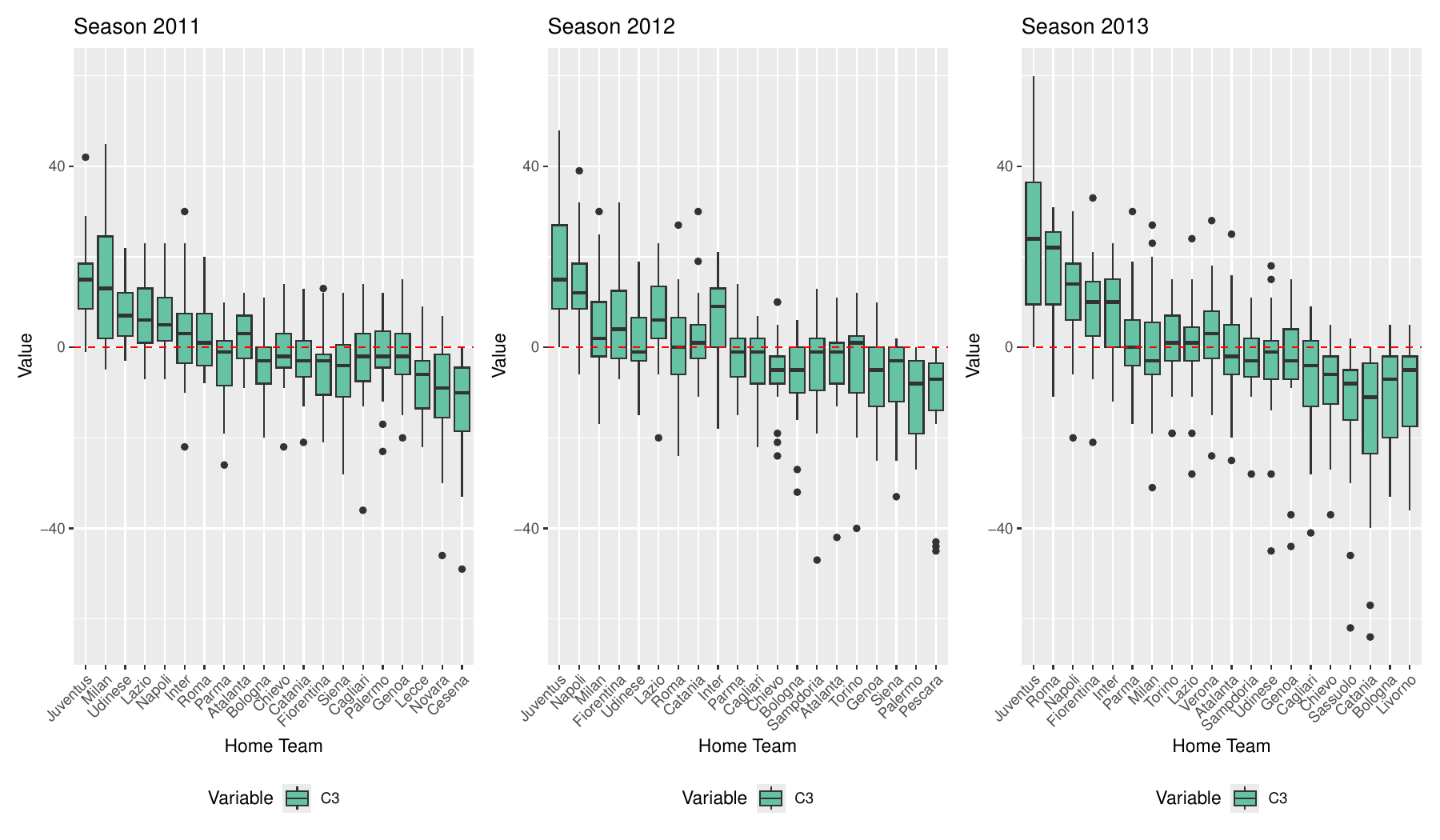}
	\caption{Boxplots for home-away differences in ranking points by home team ordered by seasonal ranking and by season.}
	\label{fig:SM_F_11}
	\end{figure}

\begin{figure}
	\centering
	\includegraphics[width=1	\linewidth]{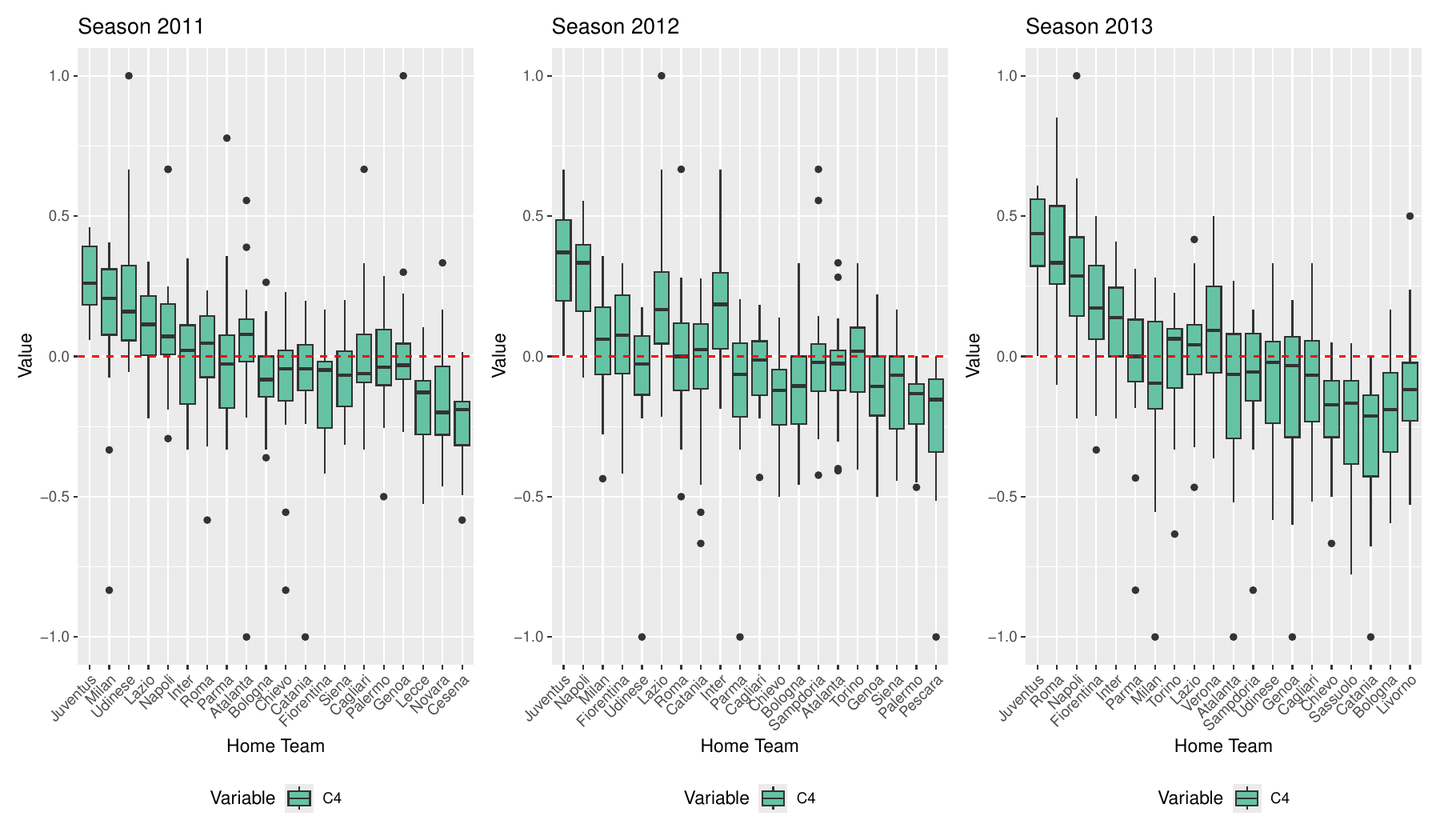}
	\caption{Boxplots for home-away relative differences in ranking points by home team ordered by seasonal ranking and by season.}
	\label{fig:SM_F_12}
	\end{figure}

	We analyze the response variables together with coaching strategies, team tactics, and refereeing choices. A systematic comparison of home and away performances,  tactical alignments, differences in teams' actions , and disciplinary measures display regularities that indicate a competitive balance,  adaptation of strategy, and evidence of field impact. Figures~\ref{fig:SM_F_02}--\ref{fig:SM_F_12} (SM) depict each construct, whereas Tables~\ref{tab:dependent}--\ref{tab:additional2} (Appendix) report descriptive statistics: measures of location (minimum, first quartile~(Q1), median, mean, third quartile~(Q3), maximum), dispersion (standard deviation~(SD), coefficient of variation~(CV)), distributional shape (skewness, kurtosis), and normality diagnostics based on skewness and kurtosis tests, each accompanied by its $p$-value. For both diagnostics the null hypotheses assert Gaussian skewness and Gaussian kurtosis, respectively.\\

	\paragraph{Dependent variables: paths to study HA}\label{subsec:dep}
	Table~\ref{tab:dependent} condenses overall match outcomes, while season- and team-specific (ranked) box-plots and proportions  are visualized in Figures~\ref{fig:SM_F_02}--\ref{fig:SM_F_03}. The goal differential $y^{(1)}$ spans $[-7,6]$, is centered at zero (median $0$, mean $0.38$), and displays a standard deviation of $1.67$ with a coefficient of variation of $4.35$. Skewness of $0.19$ and excess kurtosis of $0.10$ suggest an almost Gaussian distribution,  with mildly leptokurtic tails favoring the home side. For the binary outcome $y^{(2)}$, home teams win $531$ of $1{,}140$ contests ($46.58\%$), leaving $609$ non-victories ($53.42\%$). The trichotomous variable $y^{(3)}$ allocates $311$ results to losses ($27.28\%$), $298$ to draws ($26.14\%$), and again $531$ to wins. Thus, home advantage is present but not overwhelming; its magnitude depends on team strength and within-match dynamics, as confirmed by our multivariate analyses.\\

	\paragraph{Coach variables: strategic and tactical aspects}
	Tables~\ref{tab:schemesI} and~\ref{tab:schemesF} display the distribution of tactical formations at kick-off ($s$) and at full time ($s^{*}$). Initial dispersion is modest: the primary offensive metric for home teams, $s^{H}_{1}$, falls between 14 and 21 (SD~=~1.09), mirroring its away analogue (SD~=~1.12), while $s^{H}_{3}$ ranges from 0.47 to 0.70. Means, medians, and coefficients of variation align across home–away dyads—for example, the average for home-team $s^{H}_{1}$ is $18.06$, while for away is $18.05$, CV~=~0.06—accompanied by slight negative skew ($\text{skew}\approx-0.1$) and thin, near-Gaussian tails (excess kurtosis 2.13 and 2.36). By full time, $s^{*}$ exhibits wider support and greater volatility: $s^{*H}_{1}$ reaches 24 (80\% of the attainable maximum), and both SD and CV rise, indicating score-contingent tactical revisions. Higher-order moments shift minimally, implying that variance—rather than distributional form—distinguishes $s^{*}$ from $s$. Figures~\ref{fig:SM_F_04}–\ref{fig:SM_F_06} corroborate these patterns, revealing heterogeneous home–away differentials: positive (negative) values denote attacking (defensive) orientations, underscoring the idiosyncratic and adaptive nature of formation choice.\\

	\paragraph{Team variables: offensive, counter-offensive, and defensive actions}
	Table~\ref{tab:team} displays event frequencies for crosses, corners, their constituent actions, shots, goal kicks and offsides ($w_{1}$--$w_{8}$); season‐rank decompositions of the corresponding home–away gaps are illustrated in Figures~\ref{fig:SM_F_07}--\ref{fig:SM_F_09}. Offensive production is markedly higher for hosts: home crosses range from $0$ to $82$ (mean $22.74$) and shots from $0$ to $41$ (mean $13.41$), with both distributions exhibiting positive skew and leptokurtic right tails that signal occasional surges in wing play and finishing. Visiting teams display a comparable shape at lower magnitude (mean $w_{1}^{A}=18.52$, mean $w_{6}^{A}=11.03$). As a result, all home–away differentials are positive—particularly for crosses and shots—indicating a substantial yet variable home edge. By contrast, counterattacking and positional metrics are venue neutral: the average number of home goal kicks, $w_{7}^{H}$, is $8.40$ versus $9.56$ for the away side, while offsides average $2.57$ for hosts and $2.28$ for visitors, yielding near-zero gaps and a slight negative mean for goal kicks. In sum, offensive volume expands at home, whereas restart and defensive behaviors remain invariant with respect to venue.\\

	\paragraph{Referee variables: disciplinary actions}
	Table~\ref{tab:referee} shows disciplinary incidents—yellow and red cards—as well as foul-initiated restarts. Season‐specific and rank‐dependent variation is illustrated in Figures~\ref{fig:SM_F_15}--\ref{fig:SM_F_17}. Visiting teams absorb slightly more cautions (mean $2.47$) than hosts (mean $2.25$), and expulsions remain rare (mean $0.16$ away versus $0.12$ at home). Both series exhibit pronounced right skew and leptokurtic tails, signalling occasional high-temper encounters. Fouls leading to free‐kicks are nearly balanced (home mean $14.49$, away mean $14.66$). Penalty infractions are exceptional (home mean $0.09$, away mean $0.16$) and highly skewed, allowing isolated calls to influence outcomes disproportionately. Consequently, venue effects on disciplinary outcomes are modest: officiating is largely venue-neutral, with only a modest away excess in cautions and penalties (with significant impact on match's outcome).\\

	
	\paragraph{Control variables and fixed effects}
	\noindent Table~\ref{tab:additional}—with supporting graphics in Figures~\ref{fig:SM_F_14}–\ref{fig:SM_F_12}— summarizes the control covariates $c_{1}$–$c_{4}$. The stadium-occupancy ratio $c_{1}\,(0 \le c_{1} \le 1)$ clusters around half capacity (mean $0.59$, median $0.56$), showing moderate relative spread (CV $0.33$), small skew ($0.15$), and virtually mesokurtic tails (kurtosis $2.99$). Weighted extra minutes $\tilde{c}_2$ has a mean slightly above 5 minutes (mean $5.91$, median $6.15$), but a single extreme value of $15.51$ raises excess kurtosis to $4.57$; skew remains modest ($0.21$). The absolute difference in ranking points $c_3$ spans from $–64$ to $60$, balances at zero (mean $–0.29$, median $0$), yet displays pronounced leptokurtosis ($5.61$), indicating largely that most of the matches are evenly disputed with occasional rare events. Its relative analogue $c_{4}\,(-1 \le c_{4} \le 1)$ reproduces the same pattern on a unit scale (mean $-0.01$, SD $0.27$, kurtosis $4.70$), confirming that large home-away differences,  though infrequent, emerge from an otherwise sample of near balanced teams.

	
	\subsection{Additional material}\label{sec:further}
	Table~\ref{tab:SM_cor} presents the bivariate association between match–level predictors and the three response variables. For the continuous score \(y^{(1)}\) we report Pearson’s product--moment correlation \(\rho_{\mathrm{PM}}\) and its accompanying \(p\)-value. The dichotomous outcome \(y^{(2)}\) is evaluated through the point–biserial coefficient \(\rho_{\mathrm{PB}}\) and the \(\chi^{2}\) statistic, each with its significance level. The ordinal three–category response \(y^{(3)}\) is analysed via the Kruskal--Wallis statistic and its \(p\)-value. All covariates relate to coaching style, team behaviour, or refereeing decisions. %
	Table~\ref{tab:mod_blocks_extended} arranges the regressors into modular families~(A--J) and records the alternative parametrisations available within each set. Blocks~A--C capture coaching formations, team actions, and officiating calls; Blocks~D--J introduce indicators for extreme scores, home advantage, seasonality, calendar timing, stadium occupancy, extra time, and team fixed effects. Interaction blocks~K--M add every pairwise product among the coaching, team, and referee variables. The same hierarchy is retained for the weighted counterparts of each covariate. %
	The impact of these weighting schemes is illustrated in the fourth and fifth columns of the model summaries for the linear specification (Model~1), the logistic regression (Model~2), and the ordered--logit regression (Model~3), reported in Tables~\ref{tab:regression_summary}--\ref{tab:baseline_ologit}. Table~\ref{tab:regression_summary} displays Model~1 fitted with HC3 heteroskedasticity–consistent standard errors; Table~\ref{tab:logit_regression_summary} shows Model~2 estimated with the same sandwich estimator; Table~\ref{tab:baseline_ologit} reports the baseline ordered--logit coefficients with bootstrap standard errors drawn from 1{,}000 resamples. Each table lists the coaching metrics \(s_{1}\), \(s_{2}\), \(s_{3}\), \(\tilde{s}_{1}\), and \(\tilde{s}_{3}\), computed with raw counts in the first three columns and with weighted counts in the last two.%
	Building on the weighted framework, Table~\ref{tab:unified_combined} records the exhaustive model–selection exercise incorporating interaction terms. Table~\ref{tab:modelavg2} then summarises the model–averaged coefficients extracted from the top AIC–ranked specifications in Table~\ref{tab:unified_combined}. Finally, Table~\ref{tab:boot2} provides 95\% confidence intervals for \(s_{2}\)—asymptotic, BCa with \(B=2{,}000\) bootstrap draws, and model–averaged—computed from the leading AIC/BIC weighted models.

	\clearpage
	
	\subsection{Tables}\label{SM:apptab}
	\noindent \begin{table}[ph!]
	\centering 
	\scriptsize
	\resizebox{0.75\textwidth}{!}{%
		\begin{tabular}{@{}p{0.25\linewidth}p{0.7\linewidth}@{}} 
			\toprule
			\textbf{Phase: Error management} & \textbf{Description} \\
			\midrule
			Variable creation & Creation of new variables for crosses from corners, crosses not from corners, and combinations of fouls, free kicks, penalties, and handballs. \\
			\midrule
			Event analysis & Analysis of event relationships using two-way tables and regex for textual data, focusing on fouls, free kicks, and penalties. \\
			\midrule
			Data correction & Correction of dataset inaccuracies, including variable value adjustments and rectification of inconsistencies, particularly for event-related data. \\
			\midrule
			Enhancement & Refinement with additional variables for detailed tracking and analysis of game dynamics through interactions of fouls, free kicks, penalties, and handballs. \\
			\midrule
			Consistency update & Updating dependencies of newly created or modified variables to ensure consistent application across the dataset. \\
			\midrule
			Data integration & Combining original dataset with additional data to extend scope and fill data gaps, finalizing dataset. \\
			\bottomrule
			\addlinespace
			\toprule
			\textbf{Phase: Data cleansing} &  \\
			\midrule
			Data Cleaning & Removal of unnecessary spaces in identifiers. \\
			\midrule
			Date\&Time \newline Processing & Extraction and generation of date and time components (year, month, day, full date) from the \texttt{data} column; structuring of \texttt{stagione} into serie and season variables. \\
			\midrule
			Team\&Match \newline Refinement & Adjustment of team names for consistency; generation of factors for home and away teams; extraction and structuring of league day and match times. \\
			\midrule
			Identifier Generation & Creation of unique identifiers for matches and events. \\
			\midrule
			Additional Variables & Introduction or correction of variables for coaching information, match referees, and detailed event sequences (minute, quarter). \\
			\midrule
			Organizational Cleanup & Ordering of the dataset by season, date, and match; removal of redundant variables; manual data corrections for error rectification. \\
			\bottomrule
			\addlinespace
			\toprule
			\textbf{Phase: Data balancing} & \\
			\midrule
			Data Segmentation & Categorization of the dataset into distinct temporal segments reflecting soccer match dynamics: first half regular time, first half added time, second half regular time, and second half added time. \\
			\midrule
			Feature Crafting & Creation of cumulative and differential metrics, binary indicators for events, and continuous variables. \\
			\midrule
			Panel Structure \newline Balancing & Standardization of the dataset for longitudinal analysis, including minute-by-minute observations, synthetic observation generation for data gaps, and variable aggregation to ensure dataset integrity. \\
			\midrule
			Data Transformation and Merging & Arrays melted to long format for manageability, merging data by match identifier and minute, organizing chronologically, and enriching dataset for analysis readiness. Cleanup process ensures data relevance. \\
			\bottomrule
		\end{tabular}
	}
	\caption{Overview of data management procedure.}\label{tab:DataManagement}
	\end{table}
	
	\begin{table}[h!]
	\centering
	\footnotesize
	\resizebox{0.725\textwidth}{!}{%
		\begin{tabular}{@{}p{0.475\linewidth}p{0.5\linewidth}@{}}
			\toprule
			\textbf{Variable} & \textbf{Description} \\
			\midrule
			$\mathcal{C}_{i,t,\tau}$ & Lemmatized comment describing the specific action at instant $\tau$ (used only as a check) \\
			$\boldsymbol{\mathcal{I}}_{i,t,\tau} = \left\{ \mathcal{I}_{i,t,\tau,1}, \mathcal{I}_{i,t,\tau,2} \right\}$ & Match and observation identifiers at instant $\tau$ \\
			$\boldsymbol{\mathcal{T}}_{i,t,\tau} = \left\{ \mathcal{T}_{i,t,\tau,1}, \dots, \mathcal{T}_{i,t,\tau,6} \right\}$ & Time identifiers: full date, season, league day, half time, minute, and sequence in the minute \\
			$\boldsymbol{\mathcal{N}}_{i,t,\tau} = \left\{ \mathcal{N}_{i,t,\tau,1}^H, \mathcal{N}_{i,t,\tau,2}^H, \mathcal{N}_{i,t,\tau,1}^A, \mathcal{N}_{i,t,\tau,2}^A, \mathcal{N}_{i,t,\tau,3} \right\}$ & Names of home/away teams, coaches, and referee at instant $\tau$ \\
			$\mathbf{Y}_{i,t,\tau} = \left\{ Y_{i,t,\tau,1}^H, Y_{i,t,\tau,2}^H, Y_{i,t,\tau,1}^A, Y_{i,t,\tau,2}^A \right\}$ & Goals and autogoals by home and away teams at instant $\tau$ \\
			$\mathbf{S}_{i,t,\tau,0} = \left\{ S_{i,t,\tau,0}^H, S_{i,t,\tau,0}^A \right\}$ & Bernoulli variables for coaches' substitution actions at instant $\tau$ \\
			$\mathbf{S}_{i,t,\tau} = \left\{ S_{i,t,\tau,1}^H, S_{i,t,\tau,1}^A, S_{i,t,\tau,2}^H, S_{i,t,\tau,2}^A \right\}$ & Discrete variables for coaches' offensiveness indexes at instant $\tau$\\ 
			$\mathbf{S}^{'}_{i,t,\tau} = \left\{ S_{i,t,\tau,3}^H, S_{i,t,\tau,3}^A \right\}$ & Pseudo-continuous variables for additional offensiveness index  at instant $\tau$\\
			$\mathbf{W}_{i,t,\tau} = \left\{ W_{i,t,\tau,1}^H, \dots, W_{i,t,\tau,8}^H, W_{i,t,\tau,1}^A, \dots, W_{i,t,\tau,8}^A \right\}$ & Bernoulli variables for teams' actions at instant $\tau$ \\
			$\mathbf{Z}_{i,t,\tau} = \left\{ Z_{i,t,\tau,1}^H, \dots, Z_{i,t,\tau,4}^H, Z_{i,t,\tau,1}^A, \dots, Z_{i,t,\tau,4}^A \right\}$ & Bernoulli variables for referee’s actions: penalties, fouls, yellow/red cards \\
			\bottomrule
		\end{tabular}
	}
	\caption{Codebook and variable descriptions from the initial raw dataset.}
	\label{tab:combined_codebook_description}
	\end{table}
	
	
	\begin{table}[h!]
	\hspace*{-1.75cm}
	\centering
	\setlength\tabcolsep{5pt}
	\resizebox{0.8\textwidth}{!}{%
		\begin{minipage}[t]{0.475\textwidth}
			\centering\footnotesize
			\begin{longtable*}{@{}lp{0.15\linewidth}p{0.8\linewidth}@{}}
				\toprule
				\textbf{Variable} & \textbf{Col.} & \textbf{Description} \\
				\midrule
				\endfirsthead
				\multicolumn{3}{@{}l}{\dots continued}\\\midrule
				\textbf{Variable} & \textbf{Col.} & \textbf{Description} \\
				\midrule
				\endhead
				\midrule\multicolumn{3}{r}{\dots continued}\\
				\endfoot
				\bottomrule
				\endlastfoot
				
				\multicolumn{3}{@{}l}{\textit{Match identifiers \& timing}}\\\midrule
				MATCH\_ID      & [1]   & Unique match identifier. \\
				MINUTE         & [2]   & Minute of play. \\
				HOME\_TEAM     & [3]   & Home team name. \\
				AWAY\_TEAM     & [4]   & Away team name. \\
				HOME\_COACH    & [5]   & Name of home coach. \\
				AWAY\_COACH    & [6]   & Name of away coach. \\
				REFEREE        & [7]   & Match referee. \\
				SEASON         & [8]   & Season. \\
				HALFT          & [9]   & Half of match (1 or 2). \\
				QUARTER        & [10]  & Quarter. \\
				YEAR           & [11]  & Calendar year of match. \\
				MONTH          & [12]  & Calendar month of match. \\
				DAY            & [13]  & Day of month of match. \\
				LEAGUE\_DAY    & [14]  & League day. \\
				MATCHES        & [15]  & Cumulative matches played. \\
				L\_DAY         & [16]  & Discrete count of league days. \\
				
				\midrule
				\multicolumn{3}{@{}l}{\textit{Events, outcomes, and points}}\\\midrule
				NEVENTS        & [17]  & Total coded events. \\
				H\_GOAL        & [18]  & Goals by home. \\
				A\_GOAL        & [19]  & Goals by away. \\
				H\_AUTOG       & [20]  & Own goals conceded by home. \\
				A\_AUTOG       & [21]  & Own goals conceded by away. \\
				D\_GOAL        & [22]  & Goal difference (home–away). \\
				H\_GOAL\_C     & [23]  & Cumulative goals by home. \\
				A\_GOAL\_C     & [24]  & Cumulative goals by away. \\
				H\_AUTOG\_C    & [25]  & Cumulative own goals by home. \\
				A\_AUTOG\_C    & [26]  & Cumulative own goals by away. \\
				H\_RESULT      & [27]  & Home result (W/D/L). \\
				H\_POINTS      & [28]  & Home points. \\
				A\_RESULT      & [29]  & Away result (W/D/L). \\
				A\_POINTS      & [30]  & Away points. \\
				H\_MPOINTS     & [31]  & Match home points (constant). \\
				A\_MPOINTS     & [32]  & Match away points (constant). \\
				M\_P           & [33]  & Maximum cumulative attainable points for league day. \\
				H\_P           & [34]  & Cumulative home points. \\
				A\_P           & [35]  & Cumulative away points. \\
				HAD\_RP        & [36]  & Difference in ranking points from last match. \\
				H\_P\_rel      & [37]  & Relative home points. \\
				A\_P\_rel      & [38]  & Relative away points. \\
				HAD\_RP\_rel   & [39]  & Relative difference in ranking points. \\
				
				\midrule
				\multicolumn{3}{@{}l}{\textit{Substitutions}}\\\midrule
				HC\_ACT        & [40]  & Home substitutions. \\
				AC\_ACT        & [41]  & Away substitutions. \\
				
				\midrule
				\multicolumn{3}{@{}l}{\textit{Tactical schemes}}\\\midrule
				H\_SCHEME\_1   & [42]  & Scheme 1 for home. \\
				A\_SCHEME\_1   & [43]  & Scheme 1 for away. \\
				HAD\_SCHEME\_1 & [44]  & Difference in scheme 1. \\
				H\_SCHEME\_2   & [45]  & Scheme 2 for home. \\
				A\_SCHEME\_2   & [46]  & Scheme 2 for away. \\
				HAD\_SCHEME\_2 & [47]  & Difference in scheme 2. \\
				H\_SCHEME\_3   & [48]  & Scheme 3 for home. \\
				A\_SCHEME\_3   & [49]  & Scheme 3 for away. \\
				HAD\_SCHEME\_3 & [50]  & Difference in scheme 3. \\
				
				\midrule
				\multicolumn{3}{@{}l}{\textit{Actions}}\\\midrule
				H\_CROSS       & [51]  & Crosses by home. \\
				A\_CROSS       & [52]  & Crosses by away. \\
				H\_SHOT        & [53]  & Shots by home. \\
				A\_SHOT        & [54]  & Shots by away. \\
				H\_GOALK       & [55]  & Goal kicks by home. \\
				A\_GOALK       & [56]  & Goal kicks by away. \\
				H\_OFFS        & [57]  & Offsides by home. \\
				A\_OFFS        & [58]  & Offsides by away. \\
				H\_CORNER      & [59]  & Corners by home. \\
				A\_CORNER      & [60]  & Corners by away. \\
				H\_CROSSCOR    & [61]  & Crosses→corner by home. \\
				A\_CROSSCOR    & [62]  & Crosses→corner by away. \\
				H\_CROSSNOCOR  & [63]  & Crosses¬→corner by home. \\
				A\_CROSSNOCOR  & [64]  & Crosses¬→corner by away. \\

			\end{longtable*}
		\end{minipage}
		\hspace*{4cm}
		\begin{minipage}[t]{0.475\textwidth}
			\centering\scriptsize
			\begin{longtable*}{@{}lp{0.15\linewidth}p{0.8\linewidth}@{}}
				\toprule
				\textbf{Variable} & \textbf{Col.} & \textbf{Description} \\
				\midrule
				\endfirsthead
				\multicolumn{3}{@{}l}{\dots continued}\\\midrule
				\textbf{Variable} & \textbf{Col.} & \textbf{Description} \\
				\midrule
				\endhead
				\midrule\multicolumn{3}{r}{\dots continued}\\
				\endfoot
				\bottomrule
				\endlastfoot
				
				\multicolumn{3}{@{}l}{\textit{Disciplinary}}\\\midrule
				HR\_YEL       & [65]  & Yellow cards to home. \\
				AR\_YEL       & [66]  & Yellow cards to away. \\
				HR\_RED       & [67]  & Red cards to home. \\
				AR\_RED       & [68]  & Red cards to away. \\
				HR\_FREEKICK    & [69]  & Free kicks to home. \\
				AR\_FREEKICK    & [70]  & Free kicks to away. \\
				HR\_PENALTYK    & [71]  & Penalties to home. \\
				AR\_PENALTYK    & [72]  & Penalties to away. \\
				HR\_FOUL        & [73]  & Fouls by home. \\
				AR\_FOUL        & [74]  & Fouls by away. \\
				
				HR\_FOUL\_A\_FREEK  & [75]  & Home fouls→away free kick. \\
				HR\_FOUL\_A\_PENAL  & [76]  & Home fouls→away penalty. \\
				AR\_FOUL\_H\_FREEK  & [77]  & Away fouls→home free kick. \\
				AR\_FOUL\_H\_PENAL  & [78]  & Away fouls→home penalty. \\
				
				\midrule
				\multicolumn{3}{@{}l}{\textit{Cumulative actions}}\\\midrule
				H\_CROSS\_C    & [79]  & Cumulative crosses by home. \\
				A\_CROSS\_C    & [80]  & Cumulative crosses by away. \\
				HAD\_CROSS\_C  & [81]  & Difference in cumulative crosses. \\
				H\_SHOT\_C     & [82]  & Cumulative shots by home. \\
				A\_SHOT\_C     & [83]  & Cumulative shots by away. \\
				HAD\_SHOT\_C   & [84]  & Difference in cumulative shots. \\
				H\_GOALK\_C    & [85]  & Cumulative goal kicks by home. \\
				A\_GOALK\_C    & [86]  & Cumulative goal kicks by away. \\
				HAD\_GOALK\_C  & [87]  & Difference in cumulative goal kicks. \\
				H\_OFFS\_C     & [88]  & Cumulative offsides by home. \\
				A\_OFFS\_C     & [89]  & Cumulative offsides by away. \\
				HAD\_OFFS\_C   & [90]  & Difference in cumulative offsides. \\
				H\_CORNER\_C   & [91]  & Cumulative corners by home. \\
				A\_CORNER\_C   & [92]  & Cumulative corners by away. \\
				HAD\_CORNER\_C & [93]  & Difference in cumulative corners. \\
				H\_CROSSCOR\_C & [94]  & Cumulative crosses→corner by home. \\
				A\_CROSSCOR\_C & [95]  & Cumulative crosses→corner by away. \\
				HAD\_CROSSCOR\_C & [96] & Difference in cumulative crosses→corner. \\
				H\_CROSSNOCOR\_C & [97] & Cumulative crosses¬→corner by home. \\
				A\_CROSSNOCOR\_C & [98] & Cumulative crosses¬→corner by away. \\
				HAD\_CROSSNOCOR\_C & [99] & Difference in cumulative crosses¬→corner. \\
				H\_CORNERNOCROSS & [100] & Corners¬→cross by home. \\
				A\_CORNERNOCROSS & [101] & Corners¬→cross by away. \\
				H\_CORNERNOCROSS\_C & [102] & Cumulative corners¬→cross by home. \\
				A\_CORNERNOCROSS\_C & [103] & Cumulative corners¬→cross by away. \\
				HAD\_CORNERNOCROSS\_C & [104] & Difference in cumulative corners¬→cross. \\
				HR\_YEL\_C     & [105] & Cumulative yellow cards to home. \\
				AR\_YEL\_C     & [106] & Cumulative yellow cards to away. \\
				HADR\_YEL\_C   & [107] & Difference in cumulative yellow cards. \\
				HR\_RED\_C     & [108] & Cumulative red cards to home. \\
				AR\_RED\_C     & [109] & Cumulative red cards to away. \\
				HADR\_RED\_C   & [110] & Difference in cumulative red cards. \\
				HR\_FREEKICK\_C & [111] & Cumulative free kicks to home. \\
				AR\_FREEKICK\_C & [112] & Cumulative free kicks to away. \\
				HADR\_FREEKICK\_C & [113] & Difference in cumulative free kicks. \\
				HR\_PENALTYK\_C & [114] & Cumulative penalties to home. \\
				AR\_PENALTYK\_C & [115] & Cumulative penalties to away. \\
				HADR\_PENALTYK\_C & [116] & Difference in cumulative penalties. \\
				HR\_FOUL\_C    & [117] & Cumulative fouls by home. \\
				AR\_FOUL\_C    & [118] & Cumulative fouls by away. \\
				HADR\_FOUL\_C  & [119] & Difference in cumulative fouls. \\
				HR\_FOUL\_A\_FREEK\_C & [120] & Cumulative home fouls→away free kick. \\
				AR\_FOUL\_H\_FREEK\_C & [121] & Cumulative away fouls→home free kick. \\
				HADR\_FOUL\_FREEK\_C & [122] & Difference in cumulative fouls→free kick. \\
				HR\_FOUL\_A\_PENAL\_C & [123] & Cumulative home fouls→away penalty. \\
				AR\_FOUL\_H\_PENAL\_C & [124] & Cumulative away fouls→home penalty. \\
				HADR\_FOUL\_PENAL\_C & [125] & Difference in cumulative fouls→penalty. \\
				HR\_FOUL\_2\_C & [126] & Cumulative second‐category fouls by home. \\
				AR\_FOUL\_2\_C & [127] & Cumulative second‐category fouls by away. \\
				HADR\_FOUL\_2\_C & [128] & Difference in cumulative second‐category fouls. \\
				
				\midrule
				\multicolumn{3}{@{}l}{\textit{Season ranking points}}\\\midrule
				RPOINTS\_S1$_{team}$ & [129–148] & Season 1 cumulative ranking points per team. \\
				RPOINTS\_S2$_{team}$ & [149–168] & Season 2 cumulative ranking points per team. \\
				RPOINTS\_S3$_{team}$ & [169–188] & Season 3 cumulative ranking points per team. \\
				
				\midrule
				\multicolumn{3}{@{}l}{\textit{Total ranking points}}\\\midrule
				TRPs\_2011$_{team}$  & [189–208] & Total cumulative ranking points in 2011 season per team. \\
				TRPs\_2012$_{team}$  & [209–228] & Total cumulative ranking points in 2012 season per team. \\
				TRPs\_2013$_{team}$  & [229–247] & Total cumulative ranking points in 2013 season per team. \\
			\end{longtable*}
		\end{minipage}
	}
	\caption{Codebook for the variables in the full balanced panel dataset.}
	\label{tab:codebook_variables}
	\end{table}
	
	
	\begin{table}[!p]
	\resizebox{\textwidth}{!}{%
		\begin{tabular}{@{}lp{0.15\linewidth}p{0.6\linewidth}@{}}
			\toprule
			\textbf{Variable Name} & \textbf{Col.} & \textbf{Description} \\
			\midrule
			ID & [1] & Match ID \\
			SEAS & [2] & Season \\
			LDAY & [3] & League day \\
			YEAR & [4] & Year of the match \\
			MONTH & [5] & Month of the match \\
			DAY & [6] & Day of the month of the match \\
			HOME & [7] & Home team name \\
			AWAY & [8] & Away team name \\
			COH & [9] & Coach name (home) \\
			COA & [10] & Coach name (away) \\
			REF & [11] & Referee name \\
			Y1 & [12] & $y_{i}^{(1)}$: Goal difference \\
			Y2 & [13] & $y_{i}^{(2)}$: Binary outcome if home team wins \\
			Y3 & [14] & $y_{i}^{(3)}$: Trivariate outcome (win/draw/loss) \\
			X1I--X3I & [15--17] & $s_{i,1},s_{i,2},s_{i,3}$: Init. diff. schemes \\
			X1Ih--X3Ih & [18--20] & $s_{i,1}^{H},s_{i,2}^{H},s_{i,3}^{H}$: Initial schemes home \\
			X1Ia--X3Ia & [21--23] & $s_{i,1}^{A},s_{i,2}^{A},s_{i,3}^{A}$: Initial schemes away \\
			X1wI--X3wI & [24--26] & $\tilde{s}_{i,1},\tilde{s}_{i,2},\tilde{s}_{i,3}$: Initial weighted difference schemes \\
			X1wIh--X3wIh & [27--29] & $\tilde{s}_{i,1}^{H},\tilde{s}_{i,2}^{H},\tilde{s}_{i,3}^{H}$: Initial weighted home schemes \\
			X1wIa--X3wIa & [30--32] & $\tilde{s}_{i,1}^{A},\tilde{s}_{i,2}^{A},\tilde{s}_{i,3}^{A}$: Initial weighted away schemes \\
			X1F--X3F & [33--35] & $s_{i,1}^{*},s_{i,2}^{*},s_{i,3}^{*}$: Final difference schemes \\
			X1Fh--X3Fh & [36--38] & $s_{i,1}^{*H},s_{i,2}^{*H},s_{i,3}^{*H}$: Final home schemes \\
			X1Fa--X3Fa & [39--41] & $s_{i,1}^{*A},s_{i,2}^{*A},s_{i,3}^{*A}$: Final away schemes\\
			X1wF--X3wF & [42--44] & $\tilde{s}_{i,1},\tilde{s}_{i,2},\tilde{s}_{i,3}$: Final weighted difference schemes \\
			X1wFh--X3wFh & [45--47] & $\tilde{s}_{i,1}^{H},\tilde{s}_{i,2}^{H},\tilde{s}_{i,3}^{H}$: Final weighted home schemes \\
			X1wFa--X3wFa & [48--50] & $\tilde{s}_{i,1}^{A},\tilde{s}_{i,2}^{A},\tilde{s}_{i,3}^{A}$: Final weighted away schemes \\
			W1--W8 & [51--58] & $w_{i,1}\dots w_{i,8}$: Team action diff. \\
			W1W--W8W & [59--66] & $\tilde{w}_{i,1}\dots \tilde{w}_{i,8}$: Weighted team action diff. \\
			W1h--W8h & [67--74] & $w_{i,1}^H\dots w_{i,8}^H$: Team action home \\
			W1Wh--W8Wh & [75--82] & $\tilde{w}_{i,1}^H\dots \tilde{w}_{i,8}^H$: Weighted action home \\
			W1a--W8a & [83--90] & $w_{i,1}^A\dots w_{i,8}^A$: Team action away \\
			W1Wa--W8Wa & [91--98] & $\tilde{w}_{i,1}^A\dots \tilde{w}_{i,8}^A$: Weighted action away \\
			Z1--Z5 & [99--103] & $z_{i,1}\dots z_{i,5}$: Ref. decision diff. \\
			Z1W--Z5W & [104--108] & $\tilde{z}_{i,1}\dots \tilde{z}_{i,5}$: Weighted ref. decision diff. \\
			Z1h--Z5h & [109--113] & $z_{i,1}^H\dots z_{i,5}^H$: Ref. against home \\
			Z1a--Z5a & [114--118] & $z_{i,1}^A\dots z_{i,5}^A$: Ref. against away \\
			Z1Wh--Z5Wh & [119--123] & $\tilde{z}_{i,1}^H\dots \tilde{z}_{i,5}^H$: Weighted home ref. \\
			Z1Wa--Z5Wa & [124--128] & $\tilde{z}_{i,1}^A\dots \tilde{z}_{i,5}^A$: Weighted away ref. \\
			MET, met\_1, met\_2 & [129--131] & Extra Time (total $c_{i,2}$, 1st and 2nd half)  \\
			DATE\_RANK & [132] & $k_{i,1}^{d}$: Ranked calendar day \\
			RELATIVE\_DATE & [133] & $k_{i,2}^{d}$: Continuous calendar effect \\
			EVENTS & [134] & Total event count \\
			EXR1--EXR13 & [135--137] & $c_{i,1}, c_{i,1}^2, c_{i,1}^3$: Stadium filling index \\
			H\_Atalanta--H\_Verona & [138--163] & $\bk_{i}^{T}$: Home team dummies \\
			SEASON\_2011--SEASON\_2013 & [164--166] & $k_{i}^{(11)},k_{i}^{(12)},k_{i}^{(13)}$: Season fixed effects \\
			DUM\_EXTR, DUM\_P, DUM\_N & [167--169] & $k_{i,1}^{e},k_{i,2}^{e},k_{i,3}^{e}$: Extreme result dummies \\
			RP\_LH, RP\_LA, RP\_LHAD & [170--172] & Ranking points (previous match): home, away, differenced ($c_{i,3}$) \\
			RP\_HOME\_REL\_P, RP\_AWAY\_REL\_P, RP\_LHAD\_REL & [173--175] & Relative ranking points previous match for home, away, and their difference ($c_{i,4}$) \\
			NEVENTS1, NEVENTS2 & [176--177] & Event counts for 1st and 2nd half \\
			ETW1, ETW2 & [178--179] & Weights from events for 1st and 2nd half \\
			MET\_1W, MET\_2W, METW & [180--182] & Weighted extra time for 1st, 2nd half, and total ($\tilde{c}_{i,2}$)\\
			\bottomrule
		\end{tabular}%
	}
	\caption{Coded variables in the cross-sectional dataset.}\label{tab:coded_var}
	\end{table}
	
	\begin{table}
	\centering
	\captionsetup{size=small} 
	\caption{Association statistics between match covariates and three outcome variables. The first two columns report Pearson’s product--moment correlation $\rho_{\mathrm{PM}}$ and its $p$-value for the discrete numeric outcome $y^{(1)}$. The next four columns display the point-biserial correlation $\rho_{\mathrm{PB}}$, the $\chi^2$ test statistic, and their corresponding $p$-values for the binary outcome $y^{(2)}$. The final two columns present the Kruskal--Wallis (KW) test statistic and $p$-value for the trichotomous ordinal outcome $y^{(3)}$. All covariates represent coach, team, and referee features.}\label{tab:SM_cor}
	\footnotesize 
	\resizebox{0.5\textwidth}{!}{%
		\begin{tabular}{crlrlrlrl}
			\toprule 
			\multicolumn{1}{c}{ } & \multicolumn{2}{c}{$y^{(1)}$} & \multicolumn{4}{c}{$y^{(2)}$} & \multicolumn{2}{c}{$y^{(3)}$} \\
			\cmidrule(l{3pt}r{3pt}){2-3} \cmidrule(l{3pt}r{3pt}){4-7} \cmidrule(l{3pt}r{3pt}){8-9}
			Variables & $\rho_{PM}$ & $p$-val. & $\rho_{PB}$ & $p$-val. & $\chi^2$ & $p$-val. & KW & $p$-val.\\
			\midrule
			$s_1$ & 0.10 & 0.00 & 0.07 & 0.02 & 12.11 & 0.27 & 9.94 & 0.01\\
			$s_2$ & 0.10 & 0.00 & 0.07 & 0.02 & 12.19 & 0.23 & 10.17 & 0.01\\
			$s_3$ & 0.10 & 0.00 & 0.07 & 0.02 & 34.35 & 0.11 & 9.57 & 0.01\\
			\addlinespace
			$s_1^{*}$ & -0.26 & 0.00 & -0.29 & 0.00 & 106.65 & 0.00 & 114.60 & 0.00\\
			$s_2^{*}$ & -0.16 & 0.00 & -0.20 & 0.00 & 57.99 & 0.00 & 55.72 & 0.00\\
			$s_3^{*}$ & -0.26 & 0.00 & -0.30 & 0.00 & 252.66 & 0.00 & 112.68 & 0.00\\
			\addlinespace
			$w_1$ & -0.08 & 0.01 & -0.13 & 0.00 & 106.68 & 0.02 & 21.19 & 0.00\\
			$w_2$ & -0.06 & 0.03 & -0.08 & 0.01 & 41.73 & 0.05 & 6.61 & 0.04\\
			$w_3$ & -0.07 & 0.02 & -0.08 & 0.01 & 16.66 & 0.57 & 8.05 & 0.02\\
			$w_4$ & -0.07 & 0.01 & -0.13 & 0.00 & 105.13 & 0.01 & 21.16 & 0.00\\
			$w_5$ & -0.04 & 0.20 & -0.05 & 0.09 & 25.05 & 0.28 & 3.22 & 0.20\\
			$w_6$ & 0.21 & 0.00 & 0.15 & 0.00 & 79.42 & 0.00 & 35.78 & 0.00\\
			$w_7$ & 0.04 & 0.21 & 0.07 & 0.01 & 34.72 & 0.50 & 6.59 & 0.04\\
			$w_8$ & 0.08 & 0.01 & 0.09 & 0.00 & 36.63 & 0.01 & 8.80 & 0.01\\
			\addlinespace
			$z_1$ & -0.16 & 0.00 & -0.15 & 0.00 & 36.44 & 0.00 & 31.82 & 0.00\\
			$z_2$ & -0.24 & 0.00 & -0.22 & 0.00 & 58.39 & 0.00 & 62.70 & 0.00\\
			$z_3$ & -0.01 & 0.83 & -0.02 & 0.46 & 39.97 & 0.20 & 0.76 & 0.68\\
			$z_4$ & -0.19 & 0.00 & -0.16 & 0.00 & 31.29 & 0.00 & 32.63 & 0.00\\
			$z_5$ & -0.02 & 0.48 & -0.04 & 0.22 & 39.47 & 0.21 & 1.52 & 0.47\\
			\bottomrule
		\end{tabular}
	}
	\end{table}
	
	\begin{table}
	\centering
	\captionsetup{size=small} 
	\caption{Model specifications for model selection. Blocks A–J itemize the ten core covariate families (see definitions in Table~\ref{tab:codebook}, Appendix); interactions K–M add all pairwise products among Coach schemes (A), Team actions (B), and Referee actions (C).  The no‐interaction setting yields 184\,320 unique specifications for Gaussian and Logit models, and 92\,160 for Ordered‑Logit (intercept always included).  Each interaction block likewise contributes the same count. $^{1}$Counts refer to unweighted variables only, without interaction - for each interaction we obtain the same count of models. The model selection exercise is also repeated for weighted variables. }
	\label{tab:mod_blocks_extended}
	\resizebox{0.85\textwidth}{!}{%
		\begin{tabular}{@{}llp{0.65\linewidth}c@{}}
			\toprule
			Group & Block & Description  & \# Choices \\
			\midrule
			A & Coach schemes       & 
			($s_{i,1},\,s_{i,1}^{*}$), 
			($s_{i,2},\,s_{i,2}^{*}$), 
			($s_{i,3},\,s_{i,3}^{*}$);
			($s_{i,1}^{H},\,s_{i,1}^{A},\,s_{i,1}^{*H},\,s_{i,1}^{*A}$), 
			($s_{i,2}^{H},\,s_{i,2}^{A},\,s_{i,2}^{*H},\,s_{i,2}^{*A}$), 
			($s_{i,3}^{H},\,s_{i,3}^{A},\,s_{i,3}^{*H},\,s_{i,3}^{*A}$) 
			& 6  \\
			
			B & Team actions        & 
			($w_{i,1},\,w_{i,2},\,w_{i,6},\,w_{i,7},\,w_{i,8}$), 
			($w_{i,3},\,w_{i,4},\,w_{i,5},\,w_{i,6},\,w_{i,7},\,w_{i,8}$);\\
			&                     & 
			($w_{i,1}^H,\,w_{i,2}^H,\,w_{i,6}^H,\,w_{i,7}^H,\,w_{i,8}^H,\,
			w_{i,1}^A,\,w_{i,2}^A,\,w_{i,6}^A,\,w_{i,7}^A,\,w_{i,8}^A$),\\
			&                     & 
			($w_{i,3}^H,\,w_{i,4}^H,\,w_{i,5}^H,\,w_{i,6}^H,\,w_{i,7}^H,\,w_{i,8}^H,\,
			w_{i,3}^A,\,w_{i,4}^A,\,w_{i,5}^A,\,w_{i,6}^A,\,w_{i,7}^A,\,w_{i,8}^A$) & 4  \\
			
			C & Referee actions     & 
			($z_{i,1},\,z_{i,2},\,z_{i,3},\,z_{i,4}$), 
			($z_{i,3},\,z_{i,4},\,z_{i,5}$);\\
			&                     & 
			($z_{i,1}^H,\,z_{i,2}^H,\,z_{i,3}^H,\,z_{i,4}^H,\,
			z_{i,1}^A,\,z_{i,2}^A,\,z_{i,3}^A,\,z_{i,4}^A$),\\
			&                     & 
			($z_{i,3}^H,\,z_{i,4}^H,\,z_{i,5}^H,\,
			z_{i,3}^A,\,z_{i,4}^A,\,z_{i,5}^A$) & 4  \\
			
			D & Extreme‐results      & ($k^e_1$), ($k^e_2,\,k^e_3$) & 2  \\
			E & Home‐advantage       & Intercept excluded vs.\ included & 2  \\
			F & Seasonal effects     & No effects; ($k^y_1, k^y_2, k^y_3$); discrete season & 3  \\
			G & Calendar effects     & No effects; ($k^d_1$); ($(k^d_1)^2$); ($k^d_2$); ($(k^d_2)^2$) & 5  \\
			
			J & Stadium‐filling index& No effects; ($c_1,c_1^2,c_1^3$) & 2  \\
			I & Extra‐time duration  & ($c_2$), ($\tilde c_2$) & 2  \\
			H & Team effects         & No effects; team FEs; ($c_3$); ($c_3^2$); ($c_4$); ($c_4^2$); ($c_3^H,c_3^A$); ($c_4^H, c_4^A$) & 8  \\
			
			\midrule
			K & A $\times$ B interactions   & All pairwise products of A and B & $6\times4=24$ \\
			L & A $\times$ C interactions   & All pairwise products of A and C & $6\times4=24$ \\
			M & B $\times$ C interactions   & All pairwise products of B and C & $4\times4=16$ \\
			
			\midrule
			& Configurations per model$^1$ 
			& Gaussian, dependent $y^{(1)}$      & 184\,320 \\
			&                                      & Logit, dependent $y^{(2)}$          & 184\,320 \\
			&                                      & Ordered Logit dependent $y^{(3)}$  & 92\,160  \\
			\bottomrule
	\end{tabular}}
	\end{table}
	
	
	\clearpage
	
	\begin{table}[!p]
	\centering
	\footnotesize
	\caption{Baseline specifications for Model 1, $y^{(1)}$, with robust standard errors (SEs). Model 1 with SE sandwich estimator HC3 using the following coach variables (columns): $s_1$, $s_2$, $s_3$, $\tilde{s}_1$, $\tilde{s}_3$. Reported values are coefficients with associated $p$-values in parentheses. Number of observations $= 1, 139$.}
	\label{tab:regression_summary}
	\resizebox{0.9\textwidth}{!}{
		\begin{tabular}{lrrrrr}
			\toprule
			\textbf{Variable} & $s_1$ & $s_2$ & $s_3$ & $\tilde{s}_1$ & $\tilde{s}_3$ \\
			\midrule
			\textbf{Coach} \\
			$s_1$ & 0.29 (0.000) & - & - & 0.29 (0.000) & - \\
			$s_1^{*}$ & -0.25 (0.000) & - & - & - & - \\
			$s_2$ & - & 0.30 (0.000) & - & - & - \\
			$s_2^{*}$ & - & -0.26 (0.000) & - & - & - \\
			$s_3$ & - & - & 8.92 (0.000) & - & 8.90 (0.000) \\
			$s_3^{*}$ & - & - & -7.53 (0.000) & - & - \\
			$\tilde{s}_1^{*}$ & - & - & - & -0.24 (0.000) & - \\
			$\tilde{s}_3^{*}$ & - & - & - & - & -7.21 (0.000) \\
			\midrule
			\textbf{Team} \\
			$w_1$ & -0.02 (0.000) & -0.01 (0.000) & -0.01 (0.000) & - & - \\
			$w_2$ & -0.02 (0.050) & -0.02 (0.050) & -0.02 (0.040) & - & - \\
			$w_6$ & 0.05 (0.000) & 0.05 (0.000) & 0.05 (0.000) & - & - \\
			$w_7$ & 0.03 (0.000) & 0.03 (0.000) & 0.03 (0.000) & - & - \\
			$w_8$ & 0.03 (0.060) & 0.03 (0.070) & 0.03 (0.070) & - & - \\
			$\tilde{w}_1$ & - & - & - & -0.01 (0.000) & -0.01 (0.000) \\
			$\tilde{w}_2$ & - & - & - & -0.02 (0.050) & -0.02 (0.040) \\
			$\tilde{w}_6$ & - & - & - & 0.05 (0.000) & 0.04 (0.000) \\
			$\tilde{w}_7$ & - & - & - & 0.03 (0.000) & 0.03 (0.000) \\
			$\tilde{w}_8$ & - & - & - & 0.03 (0.060) & 0.03 (0.070) \\
			\midrule
			\textbf{Referee} \\
			$z_1$ & -0.06 (0.010) & -0.06 (0.010) & -0.06 (0.010) & - & - \\
			$z_2$ & -0.65 (0.000) & -1.09 (0.000) & -0.62 (0.000) & - & - \\
			$z_3$ & 0.01 (0.120) & 0.01 (0.140) & 0.01 (0.140) & - & - \\
			$z_4$ & -0.34 (0.000) & -0.33 (0.000) & -0.33 (0.000) & - & - \\
			$\tilde{z}_1$ & - & - & - & -0.06 (0.010) & -0.06 (0.010) \\
			$\tilde{z}_2$ & - & - & - & -0.63 (0.000) & -0.60 (0.000) \\
			$\tilde{z}_3$ & - & - & - & -0.01 (0.120) & -0.01 (0.140) \\
			$\tilde{z}_4$ & - & - & - & 0.34 (0.000) & 0.33 (0.000) \\
			\midrule
			\textbf{Controls} \\
			$\bar{c}_1$ & 0.47 (0.860) & 0.45 (0.870) & 0.36 (0.890) & 0.43 (0.870) & 0.32 (0.900) \\
			$\bar{c}_1^2$ & -0.45 (0.930) & -0.44 (0.930) & -0.28 (0.960) & -0.38 (0.940) & -0.20 (0.970) \\
			$\bar{c}_1^3$ & -0.24 (0.930) & -0.22 (0.940) & -0.31 (0.910) & -0.28 (0.920) & -0.36 (0.900) \\
			\addlinespace
			$\tilde{c}_2$ & -0.04 (0.120) & -0.04 (0.120) & -0.04 (0.130) & -0.04 (0.120) & -0.04 (0.130) \\
			\addlinespace
			$c_4$ & 1.46 (0.000) & 1.47 (0.000) & 1.47 (0.000) & 1.46 (0.000) & 1.48 (0.000) \\
			\midrule
			\textbf{Fixed Effects} \\
			$k^{(11)}$ & 0.57 (0.230) & 0.56 (0.250) & 0.58 (0.240) & 0.58 (0.220) & 0.58 (0.230) \\
			$k^{(12)}$ & 0.49 (0.310) & 0.48 (0.330) & 0.49 (0.320) & 0.49 (0.300) & 0.49 (0.320) \\
			$k^{(13)}$ & 0.52 (0.280) & 0.50 (0.310) & 0.51 (0.310) & 0.53 (0.270) & 0.51 (0.300) \\
			\addlinespace
			$k_{2}^{d}$ & -0.06 (0.670) & -0.04 (0.760) & -0.05 (0.720) & -0.06 (0.680) & -0.05 (0.730) \\
			\addlinespace
			$k_{2}^{e}$ & 3.96 (0.000) & 3.96 (0.000) & 3.96 (0.000) & 3.96 (0.000) & 3.97 (0.000) \\
			$k_{3}^{e}$ & -4.72 (0.000) & -4.83 (0.000) & -4.83 (0.000) & -4.73 (0.000) & -4.84 (0.000) \\
			\midrule
			\textbf{Model Fit} \\
			\midrule
			AIC & 3821 & 3808 & 3817 & 3822 & 3818 \\
			BIC & 3937 & 3924 & 3933 & 3938 & 3934 \\
			Deviance & 1834 & 1813 & 1827 & 1836 & 1829 \\
			$\bar{R}^2$ & 0.44 & 0.44 & 0.44 & 0.44 & 0.44 \\
			\bottomrule
		\end{tabular}
	}
	\end{table}

	\begin{table}[h!]
	\centering
	\scriptsize
	\caption{Baseline specification for Model 2, $y^{(2)}$, with robust standard errors (SEs). Model 2 with SE sandwich estimator HC3 using the following coach variables (columns): $s_1$, $s_2$, $s_3$, $\tilde{s}_1$, $\tilde{s}_3$. Reported values are coefficients with associated $p$-values in parentheses. Number of observations $= 1, 139$.}
	\label{tab:logit_regression_summary}
	\resizebox{0.9\textwidth}{!}{
		\begin{tabular}{lrrrrr}
			\toprule
			\textbf{Variable} & $s_1$ & $s_2$ & $s_3$ & $\tilde{s}_1$ & $\tilde{s}_3$  \\
			\midrule
			\textbf{Coach} \\
			$s_1$ & 0.52 (0.000) & - & - & 0.52 (0.000) & - \\
			$s_1^{*}$ & -0.50 (0.000) & - & - & - & - \\
			$s_2$ & - & 0.54 (0.000) & - & - & - \\
			$s_2^{*}$ & - & -0.52 (0.000) & - & - & - \\
			$s_3$ & - & - & 15.95 (0.000) & - & 15.96 (0.000) \\
			$s_3^{*}$ & - & - & -15.04 (0.000) & - & - \\
			$\tilde{s}_1^{*}$ & - & - & - & -0.48 (0.000) & - \\
			$\tilde{s}_3^{*}$ & - & - & - & - & -14.44 (0.000) \\
			\midrule
			\textbf{Team} \\
			$w_1$ & -0.03 (0.000) & -0.03 (0.000) & -0.03 (0.000) & - & - \\
			$w_2$ & -0.02 (0.340) & -0.02 (0.330) & -0.02 (0.310) & - & - \\
			$w_6$ & 0.08 (0.000) & 0.07 (0.000) & 0.07 (0.000) & - & - \\
			$w_7$ & 0.06 (0.000) & 0.06 (0.000) & 0.06 (0.000) & - & - \\
			$w_8$ & 0.05 (0.100) & 0.04 (0.130) & 0.05 (0.120) & - & - \\
			$\tilde{w}_1$ & - & - & - & -0.03 (0.000) & -0.03 (0.000) \\
			$\tilde{w}_2$ & - & - & - & -0.02 (0.340) & -0.02 (0.310) \\
			$\tilde{w}_6$ & - & - & - & 0.07 (0.000) & 0.07 (0.000) \\
			$\tilde{w}_7$ & - & - & - & 0.05 (0.000) & 0.05 (0.000) \\
			$\tilde{w}_8$ & - & - & - & 0.05 (0.100) & 0.04 (0.120) \\
			\midrule
			\textbf{Referee} \\
			$z_1$ & -0.10 (0.020) & -0.11 (0.020) & -0.11 (0.020) & - & - \\
			$z_2$ & -1.25 (0.000) & -2.10 (0.000) & -1.20 (0.000) & - & - \\
			$z_3$ & 0.00 (0.950) & 0.00 (0.960) & 0.00 (0.960) & - & - \\
			$z_4$ & -0.50 (0.010) & -0.47 (0.010) & -0.47 (0.010) & - & - \\
			$\tilde{z}_1$ & - & - & - & -0.10 (0.020) & -0.10 (0.020) \\
			$\tilde{z}_2$ & - & - & - & -1.23 (0.000) & -1.17 (0.000) \\
			$\tilde{z}_3$ & - & - & - & 0.00 (0.950) & 0.00 (0.960) \\
			$\tilde{z}_4$ & - & - & - & 0.49 (0.010) & 0.46 (0.010) \\
			\midrule
			\textbf{Controls} \\
			$\bar{c}_1$ & 0.23 (0.970) & 0.29 (0.960) & 0.09 (0.990) & 0.17 (0.980) & 0.01 (1.000) \\
			$\bar{c}_1^2$ & -0.16 (0.990) & -0.42 (0.970) & 0.03 (1.000) & -0.04 (1.000) & 0.17 (0.990) \\
			$\bar{c}_1^3$ & -0.50 (0.940) & -0.29 (0.960) & -0.56 (0.930) & -0.56 (0.930) & -0.63 (0.920) \\
			$\tilde{c}_2$ & -0.09 (0.040) & -0.09 (0.030) & -0.09 (0.040) & -0.09 (0.040) & -0.09 (0.040) \\
			$c_4$ & 2.00 (0.000) & 2.03 (0.000) & 2.03 (0.000) & 2.01 (0.000) & 2.03 (0.000) \\
			\midrule
			\textbf{Fixed Effects} \\
			$k^{(11)}$ & 0.17 (0.880) & 0.16 (0.880) & 0.18 (0.870) & 0.18 (0.870) & 0.19 (0.860) \\
			$k^{(12)}$ & 0.22 (0.840) & 0.20 (0.860) & 0.22 (0.840) & 0.23 (0.830) & 0.23 (0.840) \\
			$k^{(13)}$ & 0.31 (0.770) & 0.28 (0.800) & 0.29 (0.790) & 0.32 (0.760) & 0.30 (0.780) \\
			$k_{2}^{d}$ & 0.01 (0.970) & 0.04 (0.870) & 0.04 (0.890) & 0.01 (0.970) & 0.04 (0.890) \\
			$k_{2}^{e}$ & 14.39 (0.000) & 14.38 (0.000) & 14.39 (0.000) & 14.40 (0.000) & 14.41 (0.000) \\
			$k_{3}^{e}$ & -13.57 (0.000) & -13.91 (0.000) & -13.92 (0.000) & -13.58 (0.000) & -13.93 (0.000) \\
			\midrule
			\textbf{Model Fit} \\
			\midrule
			AIC & 1165 & 1159 & 1164 & 1164 & 1163 \\
			BIC & 1275 & 1270 & 1274 & 1275 & 1274 \\
			Deviance & 1120 & 1115 & 1120 & 1120 & 1119 \\
			$\bar{R}^2_N$ & 0.44 & 0.44 & 0.44 & 0.44 & 0.44\\
			Accuracy & 0.75 & 0.75 & 0.75 & 0.75 & 0.75 \\
			Sensitivity & 0.76 & 0.77 & 0.77 & 0.76 & 0.77 \\
			Specificity & 0.73 & 0.72 & 0.73 & 0.73 & 0.73 \\
			F1 & 0.76 & 0.77 & 0.77 & 0.76 & 0.77 \\
			\bottomrule
		\end{tabular}
	}
	\end{table}
	
	\begin{table}[h!]
	\centering
	\scriptsize
	\caption{Baseline specification for Model 3, $y^{(3)}$, with bootstrapped standard errors ($B=1000$) using the following coach variables (columns): $s_1$, $s_2$, $s_3$, $\tilde{s}_1$, $\tilde{s}_3$. Reported values are coefficients with associated $p$-values in parentheses. Number of observations $= 1, 139$. Baseline case $k_{i}^{(11)}=1$.}
	\label{tab:baseline_ologit}
	\resizebox{0.9\textwidth}{!}{
		\begin{tabular}{lrrrrr}
			\toprule
			\textbf{Variable} & $s_1$ & $s_2$ & $s_3$ & $\tilde{s}_1$ & $\tilde{s}_3$ \\
			\midrule
			\textbf{Coach} \\
			$s_1$ & 0.52 (0.000) & - & - & 0.52 (0.000) & - \\
			$s_1^{*}$ & -0.49 (0.000) & - & - & - & - \\
			$s_2$ & - & 0.53 (0.000) & - & - & - \\
			$s_2^{*}$ & - & -0.51 (0.000) & - & - & - \\
			$s_3$ & - & - & 15.81 (0.000) & - & 15.77 (0.000) \\
			$s_3^{*}$ & - & - & -14.91 (0.000) & - & - \\
			$\tilde{s}_1^{*}$ & - & - & - & -0.47 (0.000) & - \\
			$\tilde{s}_3^{*}$ & - & - & - & - & -14.26 (0.000) \\
			\midrule
			\textbf{Team} \\
			$w_1$ & -0.03 (0.001) & -0.02 (0.001) & -0.02 (0.001) & - & - \\
			$w_2$ & -0.04 (0.079) & -0.04 (0.066) & -0.04 (0.060) & - & - \\
			$w_6$ & 0.07 (0.000) & 0.07 (0.000) & 0.07 (0.000) & - & - \\
			$w_7$ & 0.04 (0.011) & 0.04 (0.008) & 0.04 (0.009) & - & - \\
			$w_8$ & 0.05 (0.056) & 0.05 (0.071) & 0.05 (0.069) & - & - \\
			$\tilde{w}_1$ & - & - & - & -0.03 (0.001) & -0.02 (0.001) \\
			$\tilde{w}_2$ & - & - & - & -0.04 (0.079) & -0.04 (0.059) \\
			$\tilde{w}_6$ & - & - & - & 0.07 (0.000) & 0.07 (0.000) \\
			$\tilde{w}_7$ & - & - & - & 0.04 (0.012) & 0.04 (0.009) \\
			$\tilde{w}_8$ & - & - & - & 0.05 (0.055) & 0.04 (0.068) \\
			\midrule
			\textbf{Referee} \\
			$z_1$ & -0.10 (0.015) & -0.10 (0.013) & -0.10 (0.013) & - & - \\
			$z_2$ & -1.20 (0.000) & -2.03 (0.000) & -1.14 (0.000) & - & - \\
			$z_3$ & 0.01 (0.642) & 0.00 (0.731) & 0.00 (0.732) & - & - \\
			$z_4$ & -0.51 (0.001) & -0.48 (0.002) & -0.47 (0.002) & - & - \\
			$\tilde{z}_1$ & - & - & - & -0.10 (0.016) & -0.10 (0.014) \\
			$\tilde{z}_2$ & - & - & - & -1.17 (0.000) & -1.11 (0.000) \\
			$\tilde{z}_3$ & - & - & - & -0.01 (0.640) & -0.00 (0.730) \\
			$\tilde{z}_4$ & - & - & - & 0.50 (0.001) & 0.47 (0.002) \\
			\midrule
			\textbf{Controls} \\
			$\bar{c}_1$ & 0.77 (0.888) & 1.14 (0.842) & 1.05 (0.855) & 0.70 (0.898) & 0.81 (0.887) \\
			$\bar{c}_1^2$ & -1.11 (0.911) & -1.86 (0.857) & -1.63 (0.875) & -0.97 (0.922) & -1.17 (0.910) \\
			$\bar{c}_1^3$ & -0.16 (0.977) & 0.29 (0.960) & 0.14 (0.980) & -0.24 (0.966) & -0.11 (0.985) \\
			$\tilde{c}_2$ & -0.03 (0.380) & -0.03 (0.409) & -0.03 (0.442) & -0.03 (0.397) & -0.03 (0.460) \\
			$c_4$ & 2.23 (0.000) & 2.27 (0.000) & 2.27 (0.000) & 2.24 (0.000) & 2.27 (0.000) \\
			\midrule
			\textbf{Fixed Effects} \\
			$k^{(12)}$ & -0.05 (0.766) & -0.05 (0.745) & -0.05 (0.761) & -0.05 (0.765) & -0.05 (0.760) \\
			$k^{(13)}$& 0.06 (0.717) & 0.03 (0.851) & 0.03 (0.853) & 0.06 (0.726) & 0.03 (0.862) \\
			$k_{2}^{d}$ & -0.12 (0.596) & -0.09 (0.699) & -0.10 (0.666) & -0.12 (0.598) & -0.10 (0.666) \\
			$k_{2}^{e}$& 13.14 (0.000) & 13.11 (0.000) & 13.12 (0.000) & 13.16 (0.000) & 13.14 (0.000) \\
			$k_{3}^{e}$& -14.45 (0.000) & -14.84 (0.000) & -14.86 (0.000) & -14.45 (0.000) & -14.85 (0.000) \\
			\midrule
			\textbf{Thresholds} \\
			0$|$1 & -1.61 (0.105) & -1.53 (0.134) & -1.52 (0.139) & -1.61 (0.103) & -1.55 (0.131) \\
			1$|$3 & 0.10 (0.924) & 0.18 (0.864) & 0.18 (0.861) & 0.09 (0.927) & 0.15 (0.883) \\
			\midrule
			AIC & 1895.09 & 1889.73 & 1893.03 & 1895.60 & 1893.63 \\
			BIC & 2010.97 & 2005.61 & 2008.90 & 2011.47 & 2009.50 \\
			Deviance & 1849.09 & 1843.73 & 1847.03 & 1849.60 & 1847.63 \\
			$\bar{R}^2_N$ & 0.45 & 0.45&0.45&0.45&0.45\\
			Accuracy & 0.63 & 0.63 &0.63&0.63&0.63\\
			Sensitivity & 0.65, 0.22, 0.84 &0.65, 0.24, 0.83&0.66, 0.23, 0.84&0.65, 0.22, 0.84&0.66, 0.22, 0.84\\
			Specificity & 0.86, 0.88, 0.66 &0.87, 0.87, 0.67&0.87, 0.88, 0.66&0.86, 0.88, 0.66&0.86, 0.88, 0.66\\
			F1 & 0.64, 0.28, 0.75 &0.65, 0.30, 0.75&0.65, 0.29, 0.75&0.65, 0.28, 0.75&0.65, 0.28, 0.75\\
			Brant Test ($p$) & 0.750 & 0.770 & 0.760 & 0.740 & 0.750 \\
			\bottomrule
		\end{tabular}
	}
	\end{table}
	
	\begin{table}
	\centering
	\tiny
	\captionsetup{size=footnotesize}
	\caption{Rank-1 AIC- and BIC-selected models for Models~1--3 with weighted in-game actions including block-interactions. Significance levels in parentheses. Model~1 uses robust HC3 SE. Below, $p$-values for SW (Shapiro--Wilk), KS (Kolmogorov--Smirnov), JB (Jarque--Bera), BP (Breusch--Pagan), and RR (Ramsey RESET). Model~2 also has robust HC3 SE, with p-values for HL (Hosmer--Lemeshow). Model~3 employs nonparametric bootstrap ($B=1000$) for the SEs and $p$-values also on Brant (proportional-odds) test. S = significant, NS = not significant; ``--'' = excluded variable; if Intercept, baseline is Juventus. For Models~2--3, $R_{MF}^2$ = McFadden's $R^2$; for Model~3, $R_{N}^2$ = Nagelkerke's $R^2$. For Model 3, standard $p$-values for LI = Lipsitz. ACCU = Accuracy, SENS = Sensitivity, SPEC = Specificity, PREC = Precision.}
	\label{tab:unified_combined}
	\resizebox{0.90\textwidth}{!}{
		\begin{tabular}{lcccccc}
			\toprule
			\textbf{Model} & \multicolumn{2}{c}{Model 1} & \multicolumn{2}{c}{Model 2} & \multicolumn{2}{c}{Model 3} \\
			\textbf{Variable} & AIC & BIC & AIC & BIC & AIC & BIC \\
			\midrule
			(Intercept) & -- & 0.49 (0.002) & -- & -- &  -- & -- \\
			\addlinespace
			\textbf{Coach} &&&&&& \\
			$s_2^{H}$ & 0.32 (0.000) & -- & -- & -- & -- & -- \\
			$s_2^{A}$ & -0.23 (0.000) & -- & -- & -- & -- & -- \\
			$\tilde{s}_2^{*H}$& -0.26 (0.000) & -- & -- & -- & -- & -- \\
			$\tilde{s}_2^{*A}$& 0.24 (0.000) & -- & -- & -- & -- & -- \\
			$s_2$ & -- & 0.29 (0.000) & 0.55 (0.000) & 0.53 (0.000) & 0.53 (0.000) & 0.51 (0.000) \\
			$\tilde{s}_2^{*}$ & -- & -0.25 (0.000) & -0.51 (0.000) & -0.50 (0.000) & -0.50 (0.000) & -0.48 (0.000) \\
			\addlinespace
			\textbf{Team} &&&&&& \\
			$\tilde{w}_1$ & 0.13 (0.172) & -0.02 (0.000) & -0.04 (0.000) & -0.03 (0.000) & -0.03 (0.000) & -0.03 (0.000) \\
			$\tilde{w}_2$ & -0.86 (0.001) & -0.02 (0.045) & -0.02 (0.428) & -0.02 (0.300) & -0.03 (0.225) & -0.04 (0.048) \\
			$\tilde{w}_6$ & 0.27 (0.049) & 0.04 (0.000) & 0.07 (0.000) & 0.07 (0.000) & 0.07 (0.000) & 0.06 (0.000) \\
			$\tilde{w}_7$ & -0.02 (0.936) & 0.03 (0.002) & 0.06 (0.001) & 0.06 (0.003) & 0.05 (0.001) & 0.04 (0.009) \\
			$\tilde{w}_8$ & 0.08 (0.812) & 0.02 (0.116) & 0.07 (0.046) & 0.04 (0.196) & 0.06 (0.035) & 0.04 (0.089) \\
			\addlinespace
			\textbf{Referee} &&&&&& \\
			$\tilde{z}_1$ & -0.05 (0.054) & -0.06 (0.010) & -- & -0.11 (0.014) & -0.09 (0.055) & -0.10 (0.010) \\
			$\tilde{z}_2$ & -1.04 (0.000) & -1.08 (0.000) & -- & -2.10 (0.000) & -2.15 (0.000) & -2.01 (0.000) \\
			$\tilde{z}_3$ & -0.01 (0.500) & -0.01 (0.186) & -- & 0.00 (0.931) & 0.00 (0.935) & -0.00 (0.793) \\
			$\tilde{z}_4$ & 0.30 (0.000) & 0.33 (0.000) & -- & 0.45 (0.011) & 0.42 (0.011) & 0.46 (0.003) \\
			$\tilde{z}_1^H$ & -- & -- & -0.22 (0.001) & -- & -- & -- \\
			$\tilde{z}_2^H$ & -- & -- & -2.16 (0.000) & -- & -- & -- \\
			$\tilde{z}_3^H$ & -- & -- & -0.01 (0.731) & -- & -- & -- \\
			$\tilde{z}_4^H$ & -- & -- & -0.41 (0.196) & -- & -- & -- \\
			$\tilde{z}_1^A$ & -- & -- & -0.04 (0.537) & -- & -- & -- \\
			$\tilde{z}_2^A$ & -- & -- & 2.36 (0.000) & -- & -- & -- \\
			$\tilde{z}_3^A$ & -- & -- & 0.01 (0.526) & -- & -- & -- \\
			$\tilde{z}_4^A$ & -- & -- & 0.47 (0.037) & -- & -- & -- \\
			\addlinespace
			\textbf{Controls} &&&&&& \\
			$\bar{c}_{1}$ & -0.26 (0.924) & -- & -5.09 (0.431) & -- & -5.58 (0.344) & -- \\
			$\bar{c}_{1}^2$ & -1.19 (0.814) & -- & 7.29 (0.538) & -- & 6.86 (0.504) & -- \\
			$\bar{c}_{1}^3$ & -0.23 (0.938) & -- & -5.55 (0.421) & -- & -5.04 (0.380) & -- \\
			
			$c_{2}$ & -- & -- & -- & -- & 0.01 (0.730) & -0.03 (0.468) \\
			$\tilde{c}_{2}$ & -- & -0.03 (0.198) & -0.03 (0.478) & -0.06 (0.000) & -- & -- \\
			
			$c_{3}$ & -- & 0.03 (0.000) & -- & -- & -- & 0.06 (0.000) \\
			
			\addlinespace
			
			\textbf{Fixed Effects} &&&&&& \\
			Team FE & NS & -- & S & -- & S & -- \\
			
			$k_{1}^{e}$ & -- & -- & -- & 0.63 (0.337) & -- & -- \\
			$k_{2}^{e}$ & 4.10 (0.000) & 4.09 (0.000) & 14.12 (0.000) & -- & 12.98 (0.000) & 13.06 (0.000) \\
			$k_{3}^{e}$ & -4.65 (0.000) & -5.31 (0.000) & -13.90 (0.000) & -- & -14.35 (0.000) & -15.43 (0.000) \\
			\addlinespace
			
			\textbf{Interactions} &&&&&& \\
			$\bs \times \bw^\top$ & NS & -- & -- & -- & -- & -- \\
			
			\addlinespace
			\textbf{Thresholds} &&&&&& \\
			0|1 & -- & -- & -- & -- & -6.25 (0.000) & -1.54 (0.000) \\
			1|3 & -- & -- & -- & -- & -4.41 (0.000) & 0.20 (0.386) \\
			\midrule
			\textbf{Goodness-of-fit} &&&&&& \\
			\midrule
			AIC & 3778 & 3782 & 1123 & 1131 & 1841 & 1852 \\
			BIC & 4106 & 3868 & 1359 & 1201 & 2063 & 1937 \\
			DEV & 1641 & 1791 & 1029 & 1103 & 1753 & 1818 \\
			\addlinespace
			$\bar{R}^2$ & 0.48 & 0.43 & -- & -- & -- & -- \\
			$R_{MF}^2$ & -- & -- & 0.35 & 0.30 & 0.27 & 0.25 \\
			$R_{N}^2$ & -- & -- & 0.51 & 0.45 & 0.50 & 0.46 \\
			\addlinespace
			ACCU & -- & -- & 0.78 & 0.76 & 0.65 & 0.64 \\
			SENS & -- & -- & 0.80 & 0.77 & 0.68, 0.28, 0.84 & 0.66, 0.26, 0.84 \\
			SPEC & -- & -- & 0.76 & 0.74 & 0.87, 0.85, 0.73 & 0.88, 0.86, 0.69 \\
			PREC & -- & -- & 0.80 & 0.77 & 0.67, 0.39, 0.73 & 0.67, 0.39, 0.70 \\
			F1 & -- & -- & 0.80 & 0.77 & 0.67, 0.33, 0.78 & 0.66, 0.31, 0.76 \\
			\midrule
			\textbf{Residual Diagnostics Tests ($p$-values)} &&&&&& \\
			\midrule
			SW & 0.06 & 0.23 & -- & -- & -- & -- \\
			KS & 0.59 & 0.52 & -- & -- & -- & -- \\
			JB & 0.14 & 0.24 & -- & -- & -- & -- \\
			BP & 0.07 & 0.00 & -- & -- & -- & -- \\
			RR & 0.14 & 0.30 & 0.15 & 0.73 & -- & -- \\
			HL & -- & -- & 0.75 & 0.53 & 0.37 & 0.70 \\
			LI & -- & -- & -- & -- & 0.63 & 0.56 \\
			Brant & -- & -- & -- & -- & 1.00 & 0.67 \\
			\bottomrule
		\end{tabular}
	}
	\end{table}
	
	\begin{table}[!h]
	\centering
	\caption{Model averaging results for Model 1, 2, and 3 using Akaike weights for the top AIC-ranked models (Sets 1 and Set 2). ``--'' = excluded variable; Coefficients with $p$-values $>0.15$ are omitted. Team FE marked as ``S'' if at least one team effect is significant, ``NS'' if not significant, and ``--'' if excluded.}
	\label{tab:modelavg2}
	\resizebox{0.9\textwidth}{!}{
		\begin{tabular}{l cc cc cc | cc cc cc}
			\toprule
			Sets & \multicolumn{6}{c}{Set 1} & \multicolumn{6}{c}{Set 2} \\
			\cmidrule(lr){2-7} \cmidrule(lr){8-13}
			Models & \multicolumn{2}{c}{Model 1} & \multicolumn{2}{c}{Model 2} & \multicolumn{2}{c}{Model 3} 
			& \multicolumn{2}{c}{Model 1} & \multicolumn{2}{c}{Model 2} & \multicolumn{2}{c}{Model 3} \\
			\cmidrule(lr){2-3} \cmidrule(lr){4-5} \cmidrule(lr){6-7} \cmidrule(lr){8-9} \cmidrule(lr){10-11} \cmidrule(lr){12-13}
			Variables & Est. & $(p$-val$)$ & Est. & $(p$-val$)$ & Est. & $(p$-val$)$ & Est. & $(p$-val$)$ & Est. & $(p$-val$)$ & Est. & $(p$-val$)$ \\
			\midrule
			$s_2$                 & 0.29 & (0.000) & 0.53 & (0.000) & 0.41 & (0.032) & -- & -- & -- & -- & -- & -- \\
			$\tilde{s}_2^{*}$         & -0.24 & (0.000) & -0.49 & (0.000) & -0.39 & (0.029) & -- & -- & -- & -- & -- & -- \\
			$s_2^{H}$              & -- & -- & -- & -- & -- & -- & 0.35 & (0.000) & 0.59 & (0.000) & 0.51 & (0.000) \\
			$s_2^{A}$              & -- & -- & -- & -- & -- & -- & -0.24 & (0.000) & -0.50 & (0.000) & -0.49 & (0.000) \\
			$\tilde{s}_2^{*H}$      & -- & -- & -- & -- & -- & -- & -0.28 & (0.000) & -0.52 & (0.000) & -0.46 & (0.000) \\
			$\tilde{s}_2^{*A}$      & -- & -- & -- & -- & -- & -- & 0.21 & (0.000) & 0.49 & (0.000) & 0.48 & (0.000) \\
			\addlinespace
			$\tilde{w}_1$        & -0.01 & (0.099) & -- & -- & -- & -- & -0.01 & (0.125) & -0.03 & (0.068) & -0.03 & (0.083) \\
			$\tilde{w}_6$        & 0.04 & (0.000) & 0.07 & (0.000) & 0.06 & (0.002) & 0.04 & (0.000) & 0.07 & (0.000) & 0.07 & (0.000) \\
			$\tilde{w}_7$        & 0.03 & (0.006) & 0.06 & (0.003) & 0.05 & (0.020) & 0.03 & (0.005) & 0.06 & (0.004) & 0.05 & (0.002) \\
			$\tilde{w}_8$        & 0.03 & (0.096) & 0.05 & (0.109) & 0.05 & (0.072) & 0.02 & (0.141) & 0.05 & (0.128) & 0.06 & (0.037) \\
			\addlinespace
			$\tilde{z}_1$         & -0.05 & (0.076) & -- & -- & -- & -- & -0.04 & (0.130) & -- & -- & -0.08 & (0.105) \\
			$\tilde{z}_1^H$       & -- & -- & -0.20 & (0.044) & -- & -- & -- & -- & -0.21 & (0.016) & -- & -- \\
			$\tilde{z}_2$         & -0.97 & (0.001) & -- & -- & -1.44 & (0.105) & -0.89 & (0.016) & -- & -- & -1.99 & (0.001) \\
			$\tilde{z}_2^H$       & -- & -- & -1.81 & (0.019) & -- & -- & -- & -- & -1.96 & (0.002) & -- & -- \\
			$\tilde{z}_2^A$       & -- & -- & 1.92 & (0.018) & -- & -- & -- & -- & 2.11 & (0.001) & -- & -- \\
			$\tilde{z}_4$         & 0.28 & (0.015) & -- & -- & -- & -- & 0.26 & (0.046) & -- & -- & 0.40 & (0.040) \\
			\addlinespace
			$k_{2}^{e}$           & 4.12 & (0.000) & 9.95 & (0.094) & 11.53 & (0.000) & 4.05 & (0.000) & 12.67 & (0.001) & 10.90 & (0.000) \\
			$k_{3}^{e}$    & -4.79 & (0.000) & -10.03 & (0.121) & -12.76 & (0.001) & -5.04 & (0.000) & -13.18 & (0.005) & -13.08 & (0.000) \\
			\addlinespace
			Team FE & \multicolumn{2}{c}{--} & \multicolumn{2}{c}{--} & \multicolumn{2}{c}{S} & \multicolumn{2}{c}{--} & \multicolumn{2}{c}{--} & \multicolumn{2}{c}{S} \\
			\midrule
			No. Models ($R$)  & \multicolumn{2}{c}{152} & \multicolumn{2}{c}{353} & \multicolumn{2}{c}{148} 
			& \multicolumn{2}{c}{318} & \multicolumn{2}{c}{42} & \multicolumn{2}{c}{32} \\
			\bottomrule
		\end{tabular}
	}
	\end{table}

	
	\begin{table}
	\centering
	\footnotesize
	\begin{tabular}{clcccc}
		\toprule
		&& \multicolumn{2}{c}{\textbf{AIC}} & \multicolumn{2}{c}{\textbf{BIC}} \\
		\cmidrule(lr){3-4} \cmidrule(lr){5-6}
		\textbf{Response} & \textbf{Estimator} & $\mathrm{CI}_{2.5}$ & $\mathrm{CI}_{97.5}$ & $\mathrm{CI}_{2.5}$ & $\mathrm{CI}_{97.5}$\\
		\midrule
		&&\multicolumn{2}{c}{(\(s_2^{H} \))}&\multicolumn{2}{c}{(\(s_2 \))}\\
		\textbf{$y^{(1)}$} & Classical  &  0.2267 & 0.4116 &  0.2365 & 0.3505  \\
		& BCa &  0.2319 & 0.4188 &  0.2332 & 0.3487  \\
		& BCa $\mathrm{L}_1/\mathrm{L}_2$ & 0.2415 & 0.3855 & 0.2299 & 0.3453\\
		& Model Avg. & 0.2630  & 0.4281 & 0.2290 & 0.3466\\ 
		\midrule
		&&\multicolumn{2}{c}{(\(s_2 \))}&\multicolumn{2}{c}{(\(s_2 \))}\\
		\textbf{$y^{(2)}$} & Classical &  0.4151 & 0.6847  &  0.4056 & 0.6494 \\
		& BCa &  0.3870 & 0.6508  &  0.4000 & 0.6429  \\
		& BCa $\mathrm{L}_1/\mathrm{L}_2$ &  0.3937 & 0.6376  &  0.4034 & 0.6382  \\
		& Model Avg. & 0.3145 & 0.7370 & 0.2856 & 0.7590 \\ 
		\midrule
		&&\multicolumn{2}{c}{(\(s_2 \))}&\multicolumn{2}{c}{(\(s_2 \))}\\
		\textbf{$y^{(3)}$} & Classical &  0.4242 & 0.6413  &  0.4077 & 0.6089 \\
		& BCa &  0.4011 & 0.6269 & 0.4078 & 0.6015 \\
		& BCa $\mathrm{L}_1/\mathrm{L}_2$ &  0.4130 & 0.6320  &  0.4115 & 0.6044 \\
		& Model Avg. & 0.0355 & 0.7913 & 0.0885 & 0.7503 \\ 
		\bottomrule
	\end{tabular}
	\captionsetup{size=small} 
	\caption{Asymptotic, BCa bootstrap, and model averaging CIs for different estimators at the 95\% level for the initial scheme, (\(s_2^{H}\)), based on the best AIC and BIC-ranked specifications with weighted variables. Since in the best-AIC Model 1 for $y^{(1)}$ the specification includes the home initial (\(s_2^{H}\)), its CI is shown. All bootstrap CIs are computed using \(B=2000\) resamples. $\mathrm{L}_1/\mathrm{L}_2$ = regularized/penalized estimation. Avg.=averaging. Model Avg. for Model 3 uses $B=1000$ repetitions. 
	}
	\label{tab:boot2}
	\end{table}
	
	\clearpage
	\subsection{Figures}\label{SM:appfig}
	\noindent
	
	\begin{figure}
	\centering
	\begin{minipage}{0.45\textwidth}
		\centering
		\begin{subfigure}[b]{\linewidth}
			\includegraphics[width=\linewidth]{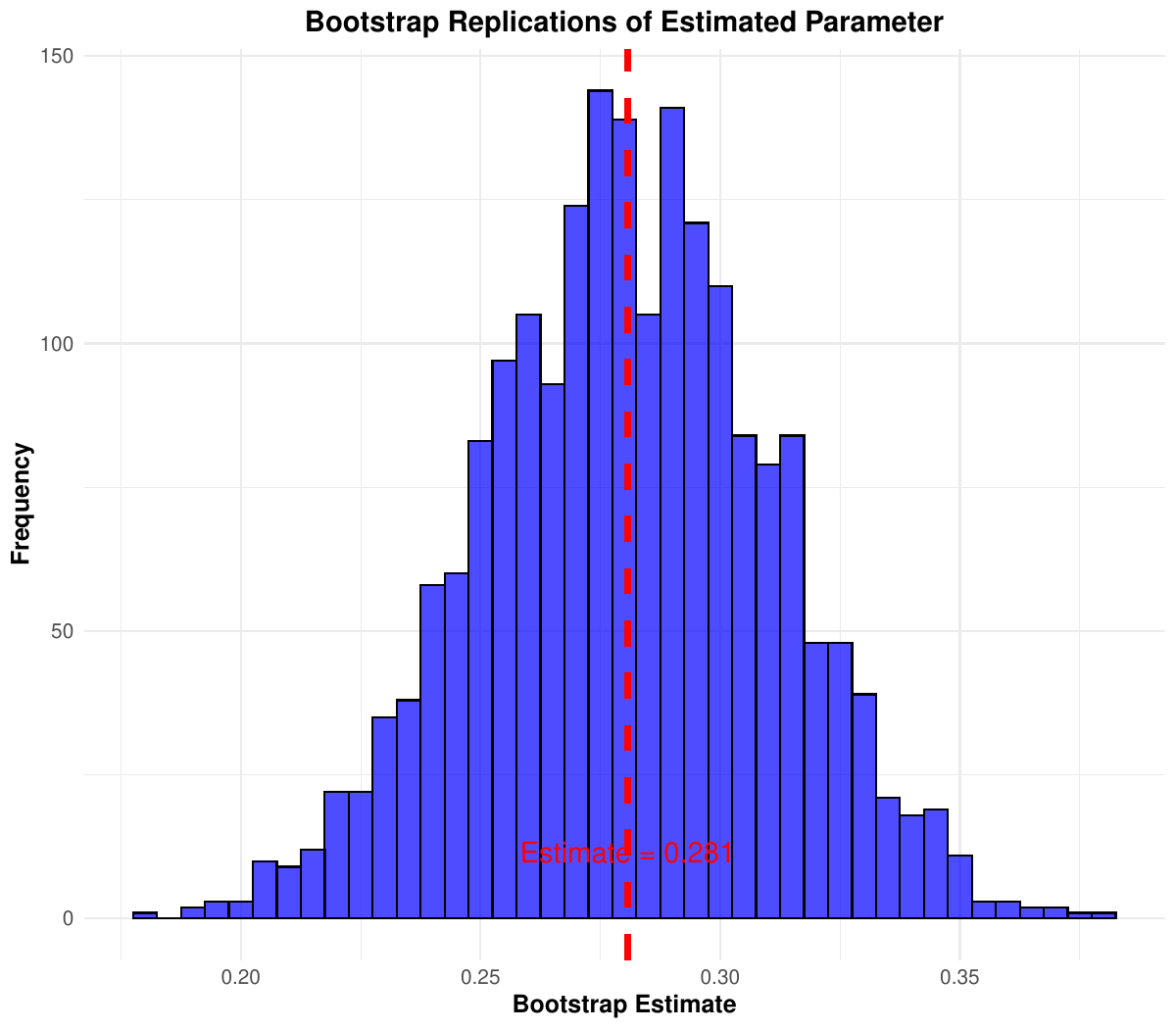}
			\caption{Best AIC for Model 1}
			\label{fig:SM_MR_AIC}
		\end{subfigure}
		\hfill
		\begin{subfigure}[b]{\linewidth}
			\includegraphics[width=\linewidth]{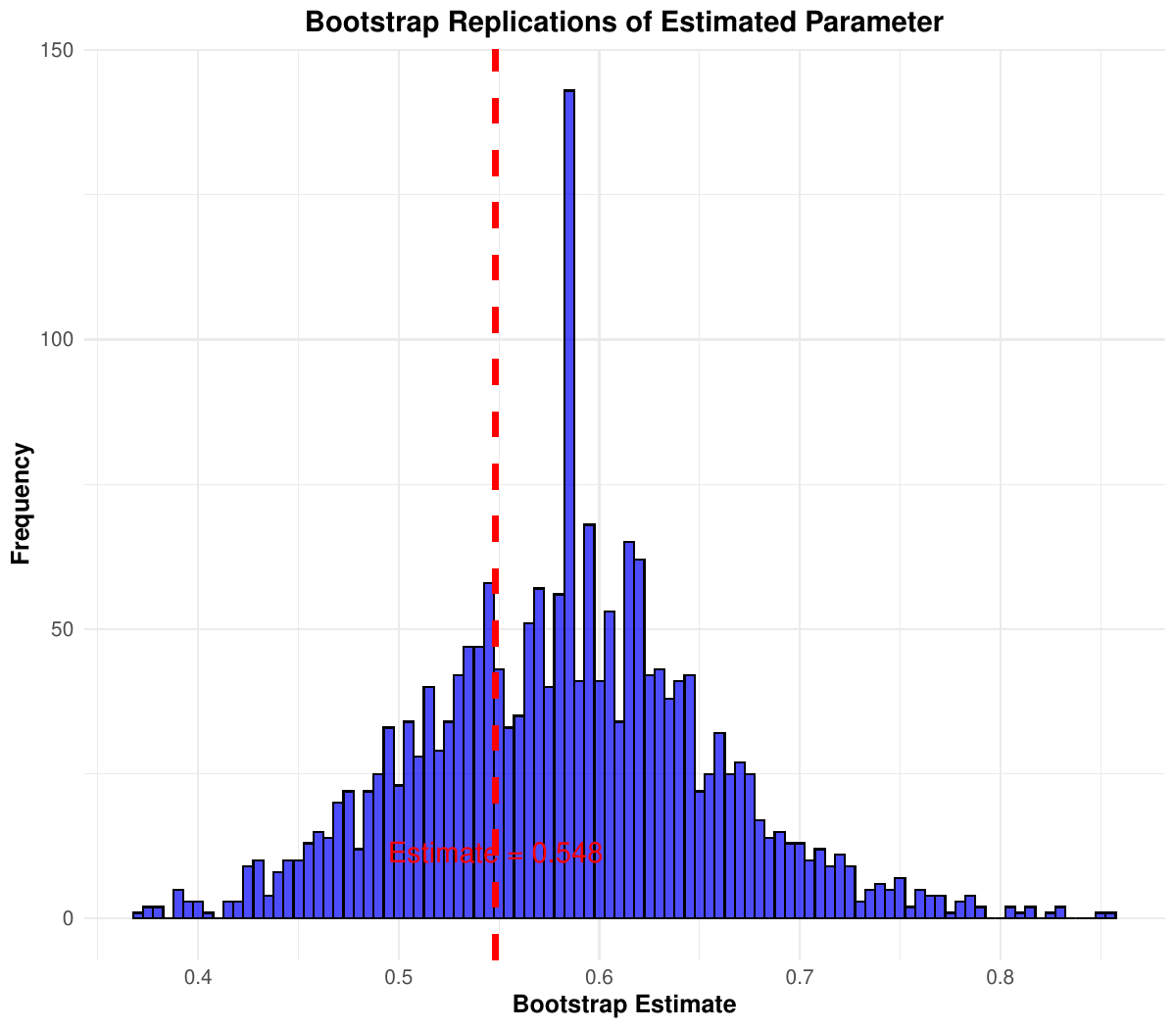}
			\caption{Best AIC for Model 2}
			\label{fig:SM_LR_AIC}
		\end{subfigure}
		\hfill
		\begin{subfigure}[b]{\linewidth}
			\includegraphics[width=\linewidth]{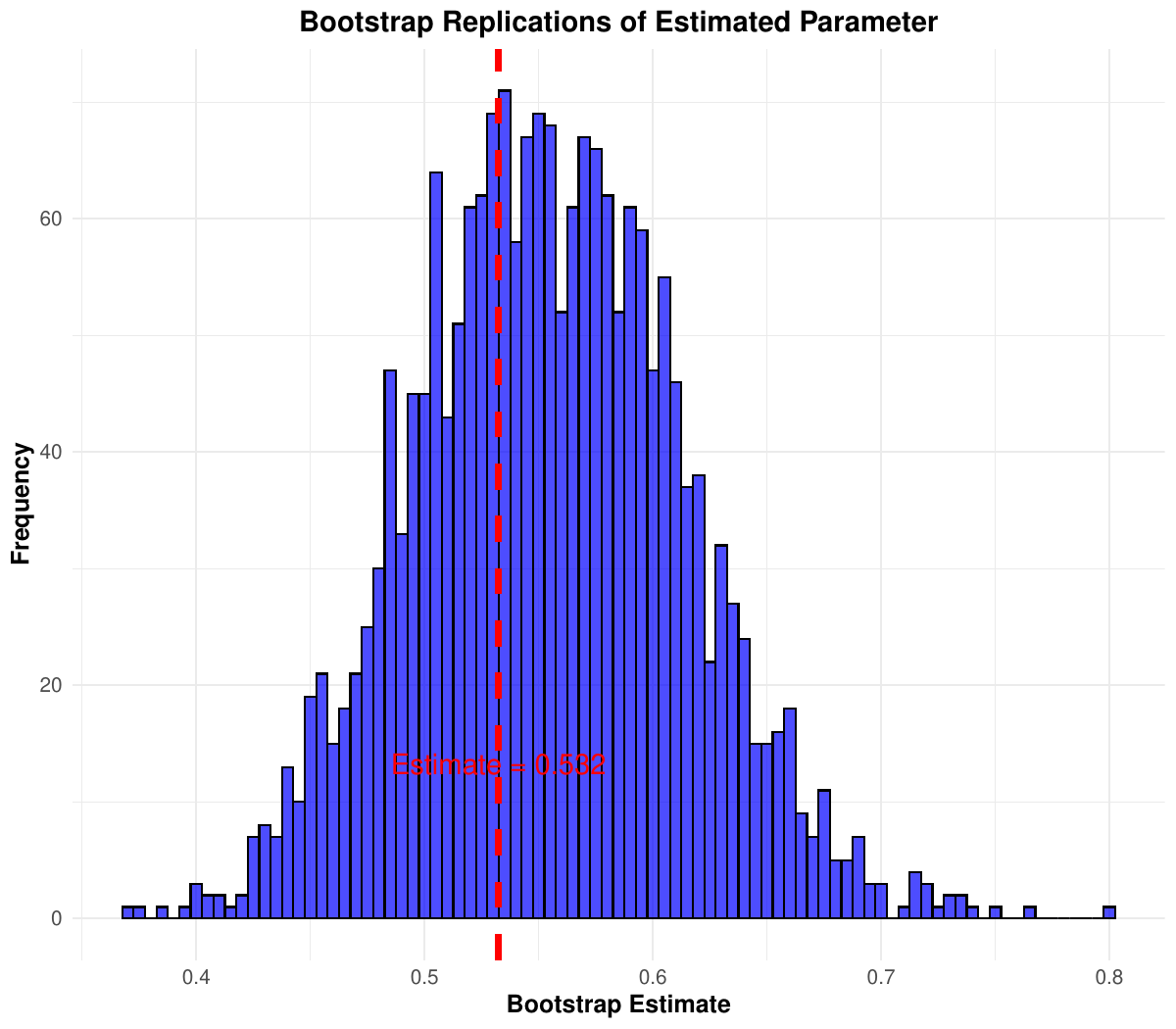}
			\caption{Best AIC for Model 3}
			\label{fig:SM_OR_AIC}
		\end{subfigure}
	\end{minipage}
	\begin{minipage}{0.45\textwidth}
		\centering
		\begin{subfigure}[b]{\linewidth}
			\includegraphics[width=\linewidth]{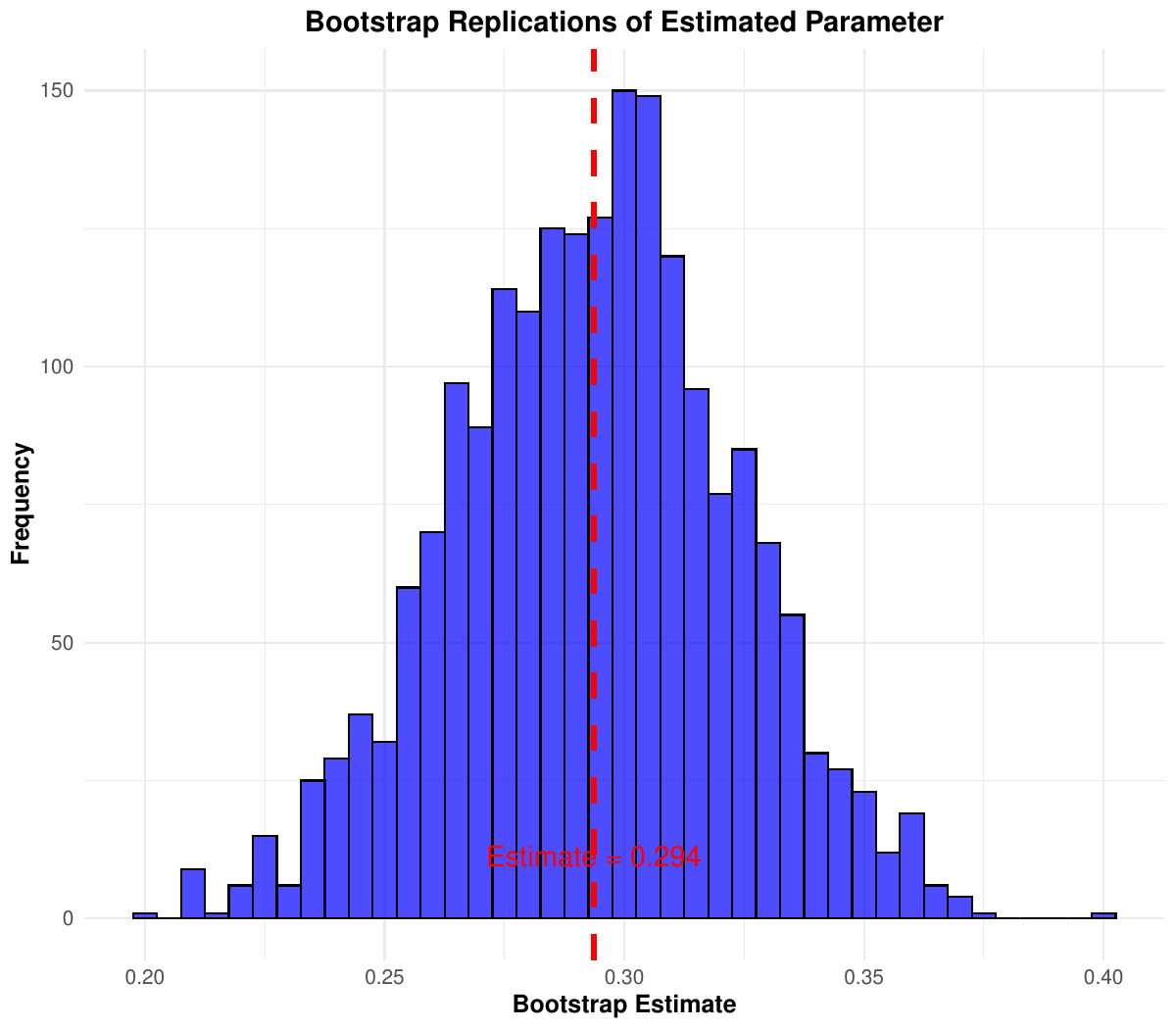}
			\caption{Best BIC for Model 1}
			\label{fig:SM_MR_BIC}
		\end{subfigure}
		\hfill
		\begin{subfigure}[b]{\linewidth}
			\includegraphics[width=\linewidth]{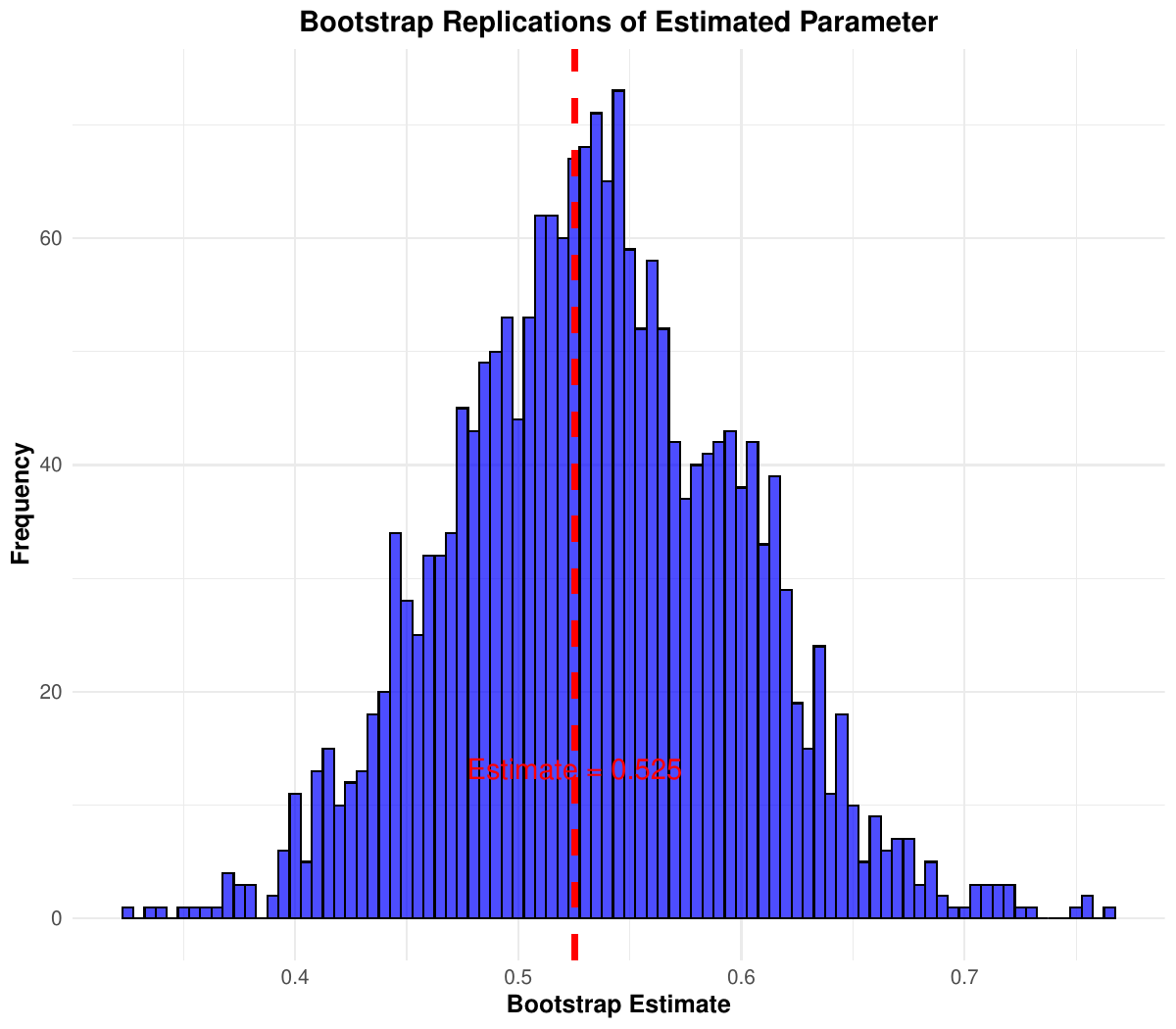}
			\caption{Best BIC for Model 2}
			\label{fig:SM_LR_BIC}
		\end{subfigure}
		\hfill
		\begin{subfigure}[b]{\linewidth}
			\includegraphics[width=\linewidth]{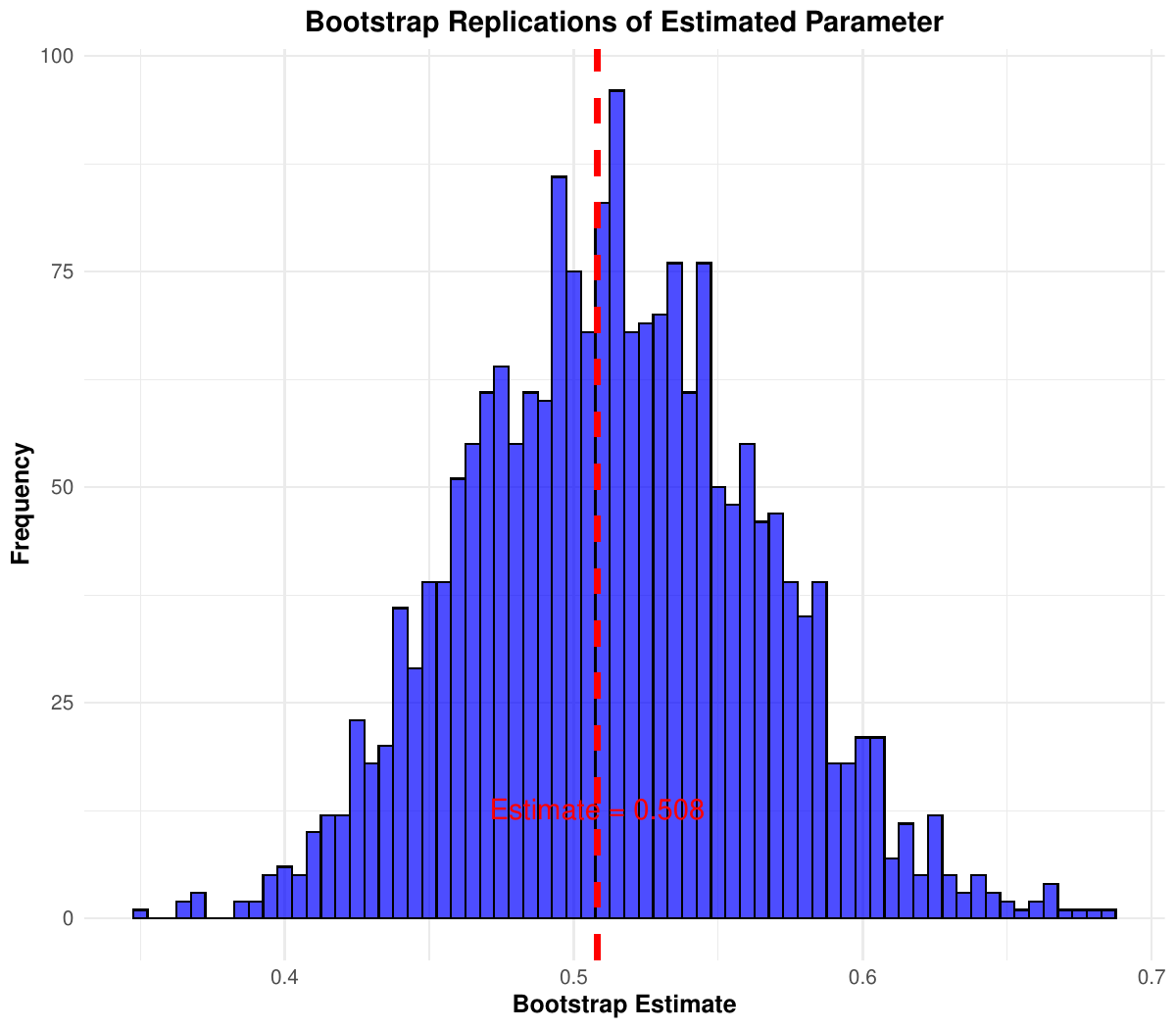}
			\caption{Best BIC for Model 3}
			\label{fig:SM_OR_BIC}
		\end{subfigure}
	\end{minipage}
	\caption{Comparative BCa distributions for AIC- (left) and BIC-selected models (right) of the estimated parameter for $s_2$. 
		(a) Model 1: slight downward bias, $\hat{z}_0=0.019$, $49.25\%$ of bootstrap replicates exceed $0.28$. 
		(b) Model 2: upward bias, $\hat{z}_0=-0.47$, $68.1\%$ of replicates above $0.55$ (peak explained by separability issues, substituting 104 over 2000 cases with the median value). 
		(c) Model 3: upward bias, $\hat{z}_0=-0.35$, $63.6\%$ of replicates are over $0.53$.
		(d) Model 1: slight upward bias, $\hat{z}_0=-0.029$, $51.15\%$ of bootstrap replicates exceed $0.29$. 
		(e) Model 2: upward bias, $\hat{z}_0=-0.147$, $55.85\%$ of replicates above $0.53$. 
		(f) Model 3: upward bias, $\hat{z}_0=-0.11$, $54.3\%$ of replicates are over $0.51$.
	}
	\label{fig:bca_dist_combined}
	\end{figure}
	
	
	\clearpage
	\section*{Author addresses}
	\begingroup
	\small
	\setlength{\parindent}{0pt}
	\setlength{\parskip}{0.35ex}
	\begin{tabular}{@{}p{0.26\linewidth}p{0.72\linewidth}@{}}
		Francesco Angelini &
		Department of Statistical Sciences “Paolo Fortunati”, University of Bologna,
		Via delle Belle Arti 41, 40126 Bologna, Italy. \emph{Email:}
		\href{mailto:francesco.angelini7@unibo.it}{francesco.angelini7@unibo.it} \\
		Massimiliano Castellani &
		Department of Statistical Sciences “Paolo Fortunati”, University of Bologna,
		Via delle Belle Arti 41, 40126 Bologna, Italy. \emph{Email:}
		\href{mailto:m.castellani@unibo.it}{m.castellani@unibo.it} \\
		Gery A. D\'\i az Rubio &
		Department of Statistical Sciences “Paolo Fortunati”, University of Bologna,
		Via delle Belle Arti 41, 40126 Bologna, Italy. \emph{Email:}
		\href{mailto:geryandre.diazrubio2@unibo.it}{geryandre.diazrubio2@unibo.it} \\
		Simone Giannerini &
		Department of Economics and Statistics, University of Udine, via Tomadini 30/a,
		33100 Udine, Italy. \emph{Email:}
		\href{mailto:simone.giannerini@uniud.it}{simone.giannerini@uniud.it} \\
		Greta Goracci &
		Faculty of Economics and Management, University of Bozen--Bolzano,
		Piazza Universit\`a 1, 39100 Bozen--Bolzano, Italy. \emph{Email:}
		\href{mailto:greta.goracci@unibz.it}{greta.goracci@unibz.it}
	\end{tabular}
	\endgroup
	
\end{document}